\documentclass[12pt,notitlepage]{revtex4-1}
\usepackage{graphicx}
\usepackage{epstopdf}
\usepackage{amsmath}

\begin{document}


\font\msytw=msbm10 scaled\magstep1
\font\msytww=msbm7 scaled\magstep1
\font\msytwww=msbm5 scaled\magstep1
\font\cs=cmcsc10
\font\ottorm=cmr8

\let\a=\alpha \let\b=\beta  \let\g=\gamma  \let\d=\delta \let\e=\varepsilon
\let\z=\zeta  \let\h=\eta   \let\th=\theta \let\k=\kappa \let\l=\lambda
\let\m=\mu    \let\n=\nu    \let\x=\xi     \let\p=\pi    \let\r=\rho
\let\s=\sigma \let\t=\tau   \let\f=\varphi \let\ph=\varphi\let\c=\chi
\let\ps=\psi  \let\y=\upsilon \let\o=\omega\let\si=\varsigma
\let\G=\Gamma \let\D=\Delta  \let\Th=\Theta\let\L=\Lambda \let\X=\Xi
\let\P=\Pi    \let\Si=\Sigma \let\F=\Phi    \let\Ps=\Psi
\let\O=\Omega \let\Y=\Upsilon

\def\PPP{{\cal P}}\def\EE{{\cal E}}\def\MM{{\cal M}} \def\VV{{\cal V}}
\def\CC{{\cal C}}\def\FF{{\cal F}} \def\HHH{{\cal H}}\def\WW{{\cal W}}
\def\TT{{\cal T}}\def\NN{{\cal N}} \def\BBB{{\cal B}}\def\III{{\cal I}}
\def\RR{{\cal R}}\def\LL{{\cal L}} \def\JJ{{\cal J}} \def\OO{{\cal O}}
\def\DD{{\cal D}}\def\AAA{{\cal A}}\def\GG{{\cal G}} \def\SS{{\cal S}}
\def\KK{{\cal K}}\def\UU{{\cal U}} \def\QQ{{\cal Q}} \def\XXX{{\cal X}}

\def\hh{{\bf h}} \def\HH{{\bf H}} \def\AA{{\bf A}} \def\qq{{\bf q}}
\def\BB{{\bf B}} \def\XX{{\bf X}} \def\PP{{\bf P}} \def\pp{{\bf p}}
\def\vv{{\bf v}} \def\xx{{\bf x}} \def\yy{{\bf y}} \def\zz{{\bf z}}
\def\aaa{{\bf a}}\def\bbb{{\bf b}}\def\hhh{{\bf h}}\def\II{{\bf I}}
\def\ii{{\bf i}}\def\jj{{\bf j}}\def\kk{{\bf k}}\def\bS{{\bf S}}
\def\mm{{\bf m}}\def\Vn{{\bf n}}\def\uu{{\bf u}}\def\tt{{\bf t}}

\def\RRR{\hbox{\msytw R}} \def\rrrr{\hbox{\msytww R}}
\def\rrr{\hbox{\msytwww R}} \def\CCC{\hbox{\msytw C}}
\def\cccc{\hbox{\msytww C}} \def\ccc{\hbox{\msytwww C}}
\def\NNN{\hbox{\msytw N}} \def\nnnn{\hbox{\msytww N}}
\def\nnn{\hbox{\msytwww N}} \def\ZZZ{\hbox{\msytw Z}}
\def\zzzz{\hbox{\msytww Z}} \def\zzz{\hbox{\msytwww Z}}


\def\\{\hfill\break}
\def\={:=}
\let\io=\infty
\let\0=\noindent\def\pagina{{\vfill\eject}}
\def\media#1{{\langle#1\rangle}}
\let\dpr=\partial
\def\sign{{\rm sign}}
\def\const{{\rm const}}
\def\tende#1{\,\vtop{\ialign{##\crcr\rightarrowfill\crcr\noalign{\kern-1pt
    \nointerlineskip} \hskip3.pt${\scriptstyle #1}$\hskip3.pt\crcr}}\,}
\def\otto{\,{\kern-1.truept\leftarrow\kern-5.truept\to\kern-1.truept}\,}
\def\defin{{\buildrel def\over=}}
\def\wt{\widetilde}
\def\wh{\widehat}
\def\to{\rightarrow}
\def\la{\left\langle}
\def\ra{\right\rangle}
\def\qed{\hfill\raise1pt\hbox{\vrule height5pt width5pt depth0pt}}
\def\Val{{\rm Val}}
\def\ul#1{{\underline#1}}
\def\lis{\overline}
\def\V#1{{\bf#1}}
\def\be{\begin{equation}}
\def\ee{\end{equation}}
\def\bea{\begin{eqnarray}}
\def\eea{\end{eqnarray}}
\def\nn{\nonumber}
\def\pref#1{(\ref{#1})}
\def\ie{{\it i.e.}}
\def\lb{\label}
\def\eg{{\it e.g.}}
\def\sl{{\displaystyle{\not}}}
\def\Tr{\mathrm{Tr}}
\def\BBBB{\hbox{\msytw B}}
\def\bbb{\hbox{\msytww B}}
\def\TTT{\hbox{\msytw T}}
\def\dd{\delta}
\def\bT{{\bf T}}
\def\mod{{\rm mod}}
\def\der{{\rm d}}
\def\bs{\backslash}

\newtheorem{lemma}{Lemma}[section]
\newtheorem{theorem}{Theorem}[section]
\newtheorem{proposition}{Proposition}[section]
\newtheorem{oss}{Remark}


\title{The ground state construction of the two-dimensional Hubbard model on 
the honeycomb lattice}

\author{Alessandro Giuliani}
\affiliation{Dipartimento di Matematica, Universit\`a di Roma Tre\\ 
L.go S. L. Murialdo 1, 00146 Roma, Italy}
\begin{abstract}
In these lectures I consider the half-filled two-dimensional (2D) Hubbard model
on the honeycomb lattice and I review the rigorous construction of its ground 
state properties by making use of constructive fermionic Renormalization Group 
methods.
\end{abstract}

\maketitle

\renewcommand{\thesection}{\arabic{section}}


\section{Introduction}\label{sec1}
\setcounter{equation}{0}
\renewcommand{\theequation}{\ref{sec1}.\arabic{equation}}

There are very few quantum interacting systems whose ground state properties 
(thermodynamic functions, reduced density matrices, etc.) can be computed
without approximations. Among these, the Luttinger and the Thirring model 
\cite{To,Lu,ML,Th,J} (one-dimensional spinless relativistic fermions), 
the one-dimensional Hubbard model \cite{LW,EFGKK} (non-relativistic lattice 
fermions with spin), the Lieb-Liniger model \cite{LL,L} (one-dimensional bosons
with repulsive delta interactions) and the BCS model \cite{BCS,RS,DPS}
($d$-dimensional spinning fermions with mean field interactions, $d\ge 1$); 
the construction of the ground states of these systems is based on some 
remarkable exact solutions, which make use of bosonization techniques and Bethe
ansatz. Unfortunately, in most cases these exact solutions do not allow to 
compute the long-distance decay of the $n$-particles correlation functions 
in a closed form (with some notable exceptions, namely the Luttinger and 
Thirring models, see \cite{K,Ma}). Moreover, even the computation of the 
thermodynamic functions crucially relies on a very special choice of the 
particle-particle interaction; as soon as the interparticle potential is 
slightly modified the exact solvability of these models is destroyed and no 
conclusion on the new ``perturbed'' system can be drawn from their solution. 
This is, of course, very annoying and unsatisfactory from a physical point of
view.

There is another rigorous powerful method that allows in a few cases to fully 
construct the ground state of a system of interacting particles,
known as Renormalization Group (RG). This method, whenever it works, has the 
advantage that it provides full information on the zero or low temperatures 
state of the system, including correlations and that, typically, it is robust 
under small modifications of the interparticle potential; even better, it 
usually gives a very precise meaning to what ``small modifications'' means: 
it allows one to classify perturbations into ``relevant'' and ``irrelevant'' 
and to show that the addition of a small irrelevant perturbation does not 
change the asymptotic behavior of correlations. Unfortunately, it only works in
the weak coupling regime and it has only been successfully applied to a limited
number of interacting quantum systems. Most of the available results on the
ground (or thermal) states of interacting quantum systems obtained via RG
concern one-dimensional (1D) weakly interacting fermions, e.g., ultraviolet 
$O(N)$ models with $N\ge 2$ \cite{GK,FMRS}, ultraviolet QED-like models 
\cite{Le} and non-relativistic lattice systems \cite{BGPS,GeM,BM}. In more than
one dimension, most of the result derived by rigorous RG techniques concern the
finite temperature properties of two-dimensional (2D) fermionic systems above 
the BCS critical temperature \cite{BG,FMRT,DR1,DRprl,BGM1,BGM2}. Two 
remarkable exceptions are the Fermi liquid construction by Feldman, Kn\"orrer 
and Trubowitz \cite{FKT}, which concerns zero temperature properties of a 
system of 2D interacting fermions with highly asymmetric Fermi surface, and the
ground state construction of the short range half-filled 2D Hubbard model on 
the honeycomb lattice by Giuliani and Mastropietro \cite{GM10,GMprb}, which 
will be reviewed here. 

The 2D Hubbard model on the honeycomb lattice is a basic model for describing 
{\it graphene}, a newly discovered material consisting of a one-atom thick 
layer of graphite \cite{N}, see \cite{CGPNG} for a comprehensive and up-to-date review of its 
low temperature properties. One of the most remarkable features of graphene is 
that at half-filling the Fermi surface is highly degenerate and it consists of 
just two isolated points. This makes the infrared properties of the system very
peculiar: in the absence of interactions, it behaves in the same way as a system of 
non-interacting $(2+1)$-dimensional Dirac particles \cite{W}. Therefore, the interacting system is a 
sort of $(2+1)$-dimensional QED, with some peculiar differences that make its study new and 
non-trivial \cite{Se, Ha}. 

The goal of these lectures is to give a self-contained proof of the analyticity of the ground state
energy of the Hubbard model on the 2D honeycomb lattice at half filling and
weak coupling via constructive RG methods. A simple extension of the proof of convergence of the
series for the specific ground 
state energy presented below allows one to construct the whole set of reduced density matrices 
at weak coupling (see \cite{GM10}): it turns out that the off-diagonal elements of these matrices
decay to zero at infinity, with the same decay exponents as the non-interacting system; in this
sense, the construction presented below rigorously exclude the presence of long range order
 in the ground state, and the absence of anomalous critical exponents (in other words,
the interacting system is in the same universality class as the non-interacting one). Let me also 
mention that more sophisticated extensions of the methods exposed here also allowed us to: (i) give an 
order by order construction of the ground state correlations of the same model in the presence of 
electromagnetic interactions \cite{GMP1,GMP2}; (ii) predict a possible mechanism for the 
spontaneous generation of the Peierls-Kekul\'e instability \cite{GMP2}; (iii) prove the universality of 
the optical conductivity \cite{GMP3}. 

The plan of these lectures is the following:
\begin{itemize}
\item In Section \ref{sec2}, I  introduce the model and state the main result.
\item In Section \ref{sec3}, I review the non-interacting case.
\item In Section \ref{sec4}, I describe the formal series expansion for the 
ground state energy, I explain how to conveniently re-express it in terms of Grassmann 
functional integrals and estimate by naive power-counting the generic 
$N$-th order in perturbation theory, so identifying two main issues in the convergence of the 
series: a combinatorial problem, related to the large number of Feynman graphs contributing at 
a generic perturbative order, and a divergence problem, related to the (very mild) ultraviolet 
singularity of the propagator and to its (more serious) infrared singularity. 
\item In Section \ref{sec5}, I describe a way to reorganize and estimate the
perturbation theory, via the so-called {\it determinant expansion}, that allows one to prove 
convergence of the series, for any fixed choice of the ultraviolet and infrared cut-offs.
\item In Section \ref{sec6}, I describe how to resum the determinant expansion in order
to cure the mild ultraviolet divergences appearing in perturbation theory, via a multiscale 
expansion, whose result is conveniently expressed in terms of {\it Gallavotti-Nicol\`o} trees 
\cite{GN,G}.
\item In Section \ref{sec7}, I describe how to resum and cure the infrared divergences
and conclude the proof of the main result.
\item Finally, in Section \ref{sec8}, I draw the conclusions. A few technical aspects of the proof are 
described in the Appendixes
\end{itemize}

The material presented in this lectures is mostly taken from \cite{GM10}. Some technical proofs 
concerning the determinant and the tree expansions are taken from the review 
\cite{GeM}. Other reviews of the constructive RG methods discussed here, which the reader
may find useful to consult, are \cite{BGbook,G,M2,Sa}.

\section{The Model and the Main Results}\label{sec2}
\setcounter{equation}{0}
\renewcommand{\theequation}{\ref{sec2}.\arabic{equation}}

The grandcanonical
Hamiltonian of the 2D Hubbard model on the honeycomb lattice at half
filling in second quantized form is given by:
\bea
&&H_\L=-t\sum_{\substack{\vec x\in \L_A\\ i=1,2,3}}\sum_{\s=\uparrow\downarrow} \Big(
a^+_{\vec x,\s} b_{\vec x+\vec \d_i,\s}^-+
b_{\vec x+\vec \d_i,\s}^+ a^-_{\vec x,\s} \Big)
+\label{1.1}\\
&&+U \sum_{\vec x\in \L_A}
\big(a^+_{\vec x,\uparrow} a^-_{\vec x,\uparrow} -\frac12\big)\big(
a^+_{\vec x,\downarrow}
a^-_{\vec x,\downarrow}-\frac12\big)+ U \sum_{\vec x\in \L_B}\big(
b_{\vec x,\uparrow}^+b_{\vec x,
\uparrow}^--\frac12\big)\big(b_{\vec x,\downarrow}^+
b_{\vec x,\downarrow}^--\frac12\big)\nn\eea
where:
\begin{enumerate}
\item $\L_A=\L$ is a periodic triangular lattice, defined as
$\L=\BBBB/L\BBBB$, where $L\in\NNN$ and $\BBBB$ is the infinite triangular
lattice with basis
${\vec l_1}=\frac12(3,\sqrt{3})$, ${\vec l_2}=\frac12(3,-\sqrt{3})$. $\L_B=\L_A+\vec\d_i$ is obtained by translating $\L_A$ by a nearest neighbor vector $\vec\d_i$, $i=1,2,3$, where
\be {\vec \d_1}=(1,0)\;,\quad {\vec \d_2}=\frac12
(-1,\sqrt{3})\;,\quad{\vec \d_3}=\frac12(-1,-\sqrt{3})\;.\label{1.2}\ee
The honeycomb lattice we are interested in 
is the union of the two triangular sublattices $\L_A$ and $\L_B$, see Fig.\ref{fig2az}
\begin{figure}[htp]
\centering
\includegraphics[width=.75\textwidth]{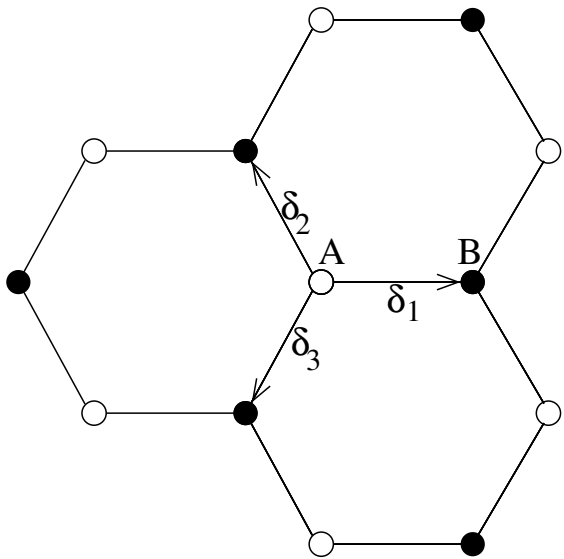}
\caption{\label{fig2az}
A portion of the honeycomb lattice $\L$. The white and black dots correspond 
to the sites of the $\L_A$ and $\L_B$ triangular sublattices, respectively. These two 
sublattices are one the translate of the other. They are connected by nearest neighbor vectors
$\vec\d_1,\vec\d_2,\vec\d_3$ that, in our units, are of unit length.}
\end{figure}
\item 
$a_{\vec x,\s}^\pm$ are creation or annihilation
fermionic operators with spin index $\s=\uparrow\downarrow$ and
site index $\vec x\in\L_A$, satisfying periodic boundary conditions
in $\vec x$. Similarly, 
$b_{\vec x,\s}^\pm$ are creation or
annihilation
fermionic operators with spin index $\s=\uparrow\downarrow$ and
site index $\vec x\in\L_B$,
satisfying periodic boundary conditions in $\vec x$.
\item $U$ is the strength of the
on--site density--density interaction;
it can be either positive or negative.
\end{enumerate}
Note that the terms in the second line of Eq.(\ref{1.1})
can be rewritten as the sum of a truly quartic term in the creation/annihilation operators (the density-density interaction), plus a quadratic term (a chemical potential term, of the form $-\m N$,
with $\m=U/2$ the chemical potential and $N$ the total particles number operator), plus a constant
(which plays no role in the thermodynamic properties of the system). 
The Hamiltonian (\ref{1.1}) is hole-particle symmetric,
i.e., it is invariant under the exchange $a^\pm_{\vec x,\s}\otto
a^\mp_{\vec x,\s}$, $b^\pm_{\vec x+\vec\d_1,\s}\otto-b^\mp_{\vec
x+ \vec\d_1,\s}$. This invariance implies in particular that, if
we define the average density of the system to be
$\r=(2|\L|)^{-1}\media{N}_{\b,\L}$, with $N=\sum_{\vec
x,\s}(a^+_{\vec x,\s} a^-_{\vec x,\s}+b^+_{\vec x+\vec\d_1,\s}
b^-_{\vec x+\vec\d_1,\s})$ the total particle number operator and
$\media{\cdot}_{\b,\L} =\Tr\{e^{-\b H_\L}\cdot\}/\Tr\{e^{-\b
H_\L}\}$ the average with respect to the (grandcanonical) Gibbs
measure at inverse temperature $\b$, one has $\r\equiv1$, for any
$|\L|$ and any $\b$;  in other words, the grand-canonical  Hamiltonian Eq.(\ref{1.1}) 
describes the system at half-filling, for all $U,\b,\L$. Let 
\be f_\b(U)=-\frac1\b\lim_{|\L|\to\io}|\L|^{-1}\log\Tr\{e^{-\b H_\L}\}\;.\ee
be the specific free energy of the system and 
$e(U)=\lim_{\b\to\io}f_\b(U)$ the specific ground state energy. We will prove the following 
Theorem.

\begin{theorem}\label{thm1}
There exists a constant $U_0>0$ such that, if $|U|\le U_0$, the specific
free energy $f_\b(U)$ of the 2D Hubbard model on the honeycomb lattice at half filling is an
analytic function of $U$, uniformly in $\b$ as
$\b\to\io$, and so is the specific ground state energy $e(U)$.
\end{theorem}

The proof is based on RG methods, which will be reviewed below. A straightforward extension
of the proof of Theorem \ref{thm1} allows one to prove that the correlation functions (i.e., the 
off-diagonal elements of the reduced density matrices of the system) are analytic functions
of $U$ and they decay to zero at infinity with the same decay exponents as in the non-interacting 
($U=0$) case, see \cite{GM10}. This rigorously excludes the presence of LRO in the ground state
and proves that the interacting system is in the same universality class as the non-interacting 
system. 

\section{The non-interacting system}\label{sec3}
\setcounter{equation}{0}
\renewcommand{\theequation}{\ref{sec3}.\arabic{equation}}

Let us begin by reviewing the construction of the finite and zero temperature states for the 
non-interacting ($U=0$) case. 
In this case the Hamiltonian of interest reduces to
\be H^0_\L=
-t\sum_{\substack{\vec x\in \L\\ i=1,2,3}}\sum_{\s=\uparrow\downarrow} \Big(
a^+_{\vec x,\s} b_{\vec x+\vec \d_i,\s}^-+
b_{\vec x+\vec \d_i,\s}^+ a^-_{\vec x,\s} \Big)\;,\label{A.1}\ee
with $\L$, $a^{\pm}_{\vec x,\s}$, $b^{\pm}_{\vec x+\vec\d_i,\s}$ defined
as in items (1)--(4) after (\ref{1.1}). We aim at computing the spectrum of $H^0_\L$ by 
diagonalizing the right hand side (r.h.s.) of (\ref{A.1}). To this purpose, we pass 
to Fourier space. 
We identify the periodic triangular lattice $\L$ with the set of vectors of the infinite triangular lattice
within a ``box" of size $L$, i.e., 
\be \L= \{n_1 {\vec l_1}+n_2 {\vec l_2}\ :\ 0\le n_1,n_2\le L-1\} \;,
\label{A.2a}\ee
with $\vec l_{1}=\frac12(3,\sqrt3)$ and $\vec l_2=\frac12(3,-\sqrt3)$.
The reciprocal lattice $\L^*$ is the set of vectors $\vec K$ such
that $e^{i\vec K \vec x}=1$, if $\vec x\in \L$.
A basis $\vec G_1,\vec G_2$ for $\L^*$ can be obtained by the inversion
formula:
\be\begin{pmatrix}G_{11}&G_{12}\\ G_{21}&G_{22}\end{pmatrix}=2\p\begin{pmatrix}l_{11}&l_{21}\\
l_{12}&l_{22}\end{pmatrix}^{\!\!\!-1}\;,\label{A.3a}\ee
which gives
\be {\vec G_1}=\frac{2\pi}{3}(1,\sqrt{3})\;,\qquad {\vec G_2}=\frac{2\pi}{3}(1,-\sqrt{3})\;.\label{A.4a}\ee
We call $\BBB_L$ the set of quasi-momenta $\vec k$ of the form
\be \vec k=\frac{m_1}{L}{\vec G_1}+\frac{m_2}{L}{\vec G_2}\;,\qquad m_1,m_2
\in\ZZZ\;,\label{A.5a}\ee
identified modulo $\L^*$; this means that $\BBB_L$ can be identified
with the vectors $\vec k$ of the form (\ref{1.2}) and restricted to the
{\it first Brillouin zone}:
\be \BBB_L=\{\vec k=\frac{m_1}{L}{\vec G_1}+
\frac{m_2}{L}{\vec G_2}\ :\ 0\le m_1,m_2\le L-1\}\;.
\label{A.6a}\ee
Given a periodic function $f:\L\to\RRR$, its Fourier transform is
defined as
\be  f(\vec x)=\frac{1}{|\L|}
\sum_{\vec k\in \BBB_L} e^{i\vec k\vec x} \hat f(\vec k)\;,\label{A.7a}\ee
which can be inverted into
\be \hat f(\vec k)=\sum_{\vec x\in \L}
e^{-i\vec k\vec x} f(\vec x)\;,\qquad \vec k\in\BBB_L\;,\label{A.8}\ee
where we used the identity
\be \sum_{\vec x\in\L} e^{i\vec k\vec x}= |\L| \d_{\vec k, \vec 0}\label{A.9}
\ee
and $\d$ is the periodic Kronecker delta function over $\L^*$.

\begin{figure}\begin{center}
\includegraphics[width=0.85\textwidth]{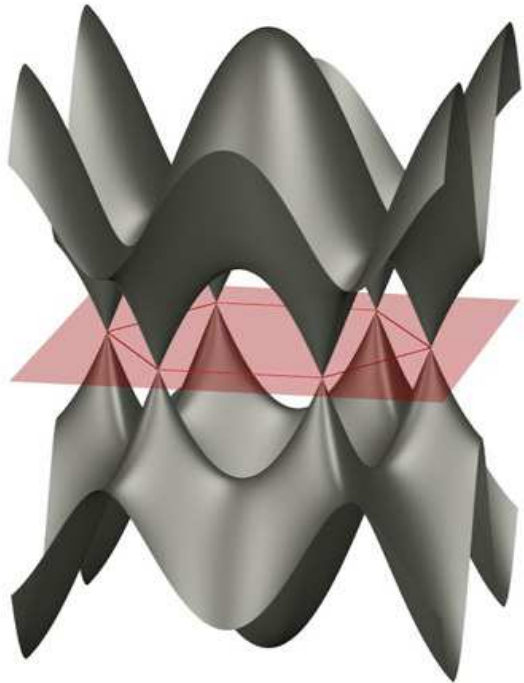}\end{center}
\caption{\label{fig1az}
A sketch of the energy bands of the free electron gas with nearest neighbor hopping 
on the honeycomb lattice. The red plane corresponds to the Fermi energy at half-filling. It cuts the 
bands at a discrete set of points, known as the Fermi points or Dirac points. From the picture, it seems that there are six distinct Fermi points. However, after identification of the points modulo 
vectors of the reciprocal lattice, it turns out that only two of them are independent.}
\end{figure}

We now associate to the set of creation/annihilation operators
$a^{\pm}_{\vec x,\s}$, $b^{\pm}_{\vec x+\vec\d_i,\s}$ the corresponding set
of operators in momentum space:
\be a^\pm_{\vec x,\s}=\frac{1}{|\L|} \sum_{\vec k\in \BBB_L} e^{\pm
i\vec k\vec x} \hat a^\pm_{\vec k,\s}\;,\qquad\qquad b^\pm_{\vec x+\vec
\d_1,\s}=\frac{1}{|\L|} \sum_{\vec k\in \BBB_L} e^{\pm i\vec k\vec x}
\hat b^\pm_{\vec k,\s} \;.\label{A.10}\ee
Using (\ref{A.7a})--(\ref{A.9}), we find that
\be \hat a^\pm_{\vec k,\s}=\sum_{\vec x\in\L}e^{\mp i\vec k\vec x}
a^\pm_{\vec x,\s}\;,\qquad\qquad \hat b^\pm_{\vec k,\s}
=\sum_{\vec x\in\L}e^{\mp i\vec k\vec x}
b^\pm_{\vec x+\vec\d_1,\s}
\label{A.11}\ee
are fermionic creation/annihilation operators, {\it periodic over $\L^*$}, satisfying
\be \{a^\e_{\vec k,\s},a^{\e'}_{\vec k',\s'}\}=|\L|\d_{\vec k,\vec k'}
\d_{\e,-\e'}\d_{\s,\s'}\;,\qquad \qquad \{b^\e_{\vec k,\s},
b^{\e'}_{\vec k',\s'}\}=|\L|\d_{\vec k,\vec k'}
\d_{\e,-\e'}\d_{\s,\s'}\label{A.12}\ee
and $\{a^\e_{\vec k,\s},b^{\e'}_{\vec k',\s'}\}=0$ \cite{footnote}. 
With the previous definitions, we can rewrite
\bea H^0_\L&=&-t\sum_{\substack{\vec x\in \L\\
i=1,2,3}}\sum_{\s=\uparrow\downarrow} (a^+_{\vec x,\s} b_{\vec
x+\vec \d_i,\s}^-+ b_{\vec
x+\vec \d_i,\s}^+a^-_{\vec x,\s} )=\label{A.13}\\
&=&-\frac{v_0}{|\L|}\sum_{\substack{\vec k\in \BBB_L\\\s=\uparrow\downarrow}}
\begin{matrix} \big(\hat a^+_{\vec k,\s}, \hat b^+_{\vec k,\s}\big)\cr\hskip2.truecm\end{matrix}\!\!
\begin{pmatrix} 0 && \O^*(\vec k)\cr \O(\vec k)&& 0\end{pmatrix}\!\!
\begin{pmatrix} \hat a^-_{\vec k,\s}\cr\hat b^-_{\vec k,\s}\nn\end{pmatrix}\eea
%
%
with $v_0=\frac32t$ the unperturbed Fermi velocity (if the hopping strength $t$ is chosen 
to be the one measured in real graphene, $v_0$ turns out to be approximately 300 times 
smaller than the speed of light) and 
\be \O({\vec k})=\frac23\sum_{i=1}^3 e^{i(\vec \d_i-\vec \d_1) \vec k}=\frac23\Big[1+2
e^{-i \frac32 k_1}\cos(\frac{\sqrt{3}}2 k_2)\Big]\label{A.14}\ee
the complex {\it dispersion relation}. By looking at the second line of Eq.(\ref{A.13}), 
one realizes that the natural creation/annihilation operator is a 2D spinor of components 
$a$ and $b$:
\be \hat \Psi^+_{\vec k,\s}=(\hat a^+_{\vec k ,\s},\hat b^+_{\vec k,\s})\;,\qquad \hat \Psi^-_{\vec k,\s}=
\begin{pmatrix} \hat a^-_{\vec k,\s}\\ \hat b^-_{\vec k,\s}\end{pmatrix}\;,\label{A.14qq}\ee
whose real space counterparts read
\be \Psi^+_{\vec x,\s}=(a^+_{\vec x ,\s}, b^+_{\vec x+\vec\d_1,\s})\;,\qquad \Psi^-_{\vec x,\s}=
\begin{pmatrix} \hat a^-_{\vec x,\s}\\ \hat b^-_{\vec x+\vec\d_1,\s}\end{pmatrix}\;.\label{A.14qqq}\ee
In order to fully diagonalize the theory, one needs to perform the diagonalization of the 
$2\times2$ quadratic form in the second line of Eq.(\ref{A.13}), which can be realized by the 
$\vec k$-dependent unitary transformation 
\be U_{\vec k}=\begin{pmatrix} \frac1{\sqrt2} & \frac1{\sqrt2}\frac{\O^*(\vec k)}{|\O(\vec k)|}\\
-\frac1{\sqrt2}\frac{\O(\vec k)}{|\O(\vec k)|}& \frac1{\sqrt2}\end{pmatrix}\;,\label{A.14qazz}\ee
in terms of which, defining 
\be  \hat \Phi^-_{\vec k,\s}=\begin{pmatrix}\hat \a^-_{\vec k,\s}\\ \hat \b^-_{\vec k,\s}\end{pmatrix}:=U
\hat \Psi^-_{\vec k,\s}=
\frac1{\sqrt2}\begin{pmatrix} \hat a_{\vec k,\s}^-+\frac{\O^*({\vec k})}{|\O({\vec k})|}\hat b_{\vec k,\s}^-
\\ -\frac{\O(\vec k)}{|\O(\vec k)|}\hat a_{\vec k,\s}^-+\hat b^-_{\vec k,\s}\end{pmatrix}\;,
\label{A.14ads}\ee
we can rewrite
\bea H^0_\L&=&-\frac{v_0}{|\L|}\sum_{\substack{\vec k\in\BBB_L\\\s=\uparrow\downarrow}}
\hat \Psi^+_{\vec k,\s}U^\dagger_{\vec k}\, U_{\vec k}\begin{pmatrix}0 & \O^*(\vec k)\\ \O(\vec k) & 0 \end{pmatrix}
U^\dagger_{\vec k}\,U_{\vec k}\hat \Psi^-_{\vec k,\s}=\nn\\
&=&-\frac{v_0}{|\L|}\sum_{\substack{\vec k\in\BBB_L\\\s=\uparrow\downarrow}}
\hat \Phi^+_{\vec k,\s} \begin{pmatrix} |\O(\vec k)| &0 \\ 0 & - |\O(\vec k)|\end{pmatrix}
\hat \Phi^-_{\vec k,\s}=\label{A.16}\\
&=&-\frac{v_0}{|\L|}\sum_{\substack{\vec k\in\BBB_L\\\s=\uparrow\downarrow}}
\big(|\O({\vec k})|\hat \a^+_{\vec k,\s}
\hat \a_{\vec k,\s}^--|\O({\vec k})|\hat \b^+_{\vec k,\s}\hat \b_{\vec k,\s}^-
\big)\;.\nn\eea
The two energy bands $\pm v_0|\O(\vec k)|$ are plotted in Fig.\ref{fig1az}.
They cross the Fermi energy $E_F=0$ at the {\it Fermi points} $\vec k=\vec p_F^{\,\o}$, 
$\o=\pm$, with
\be \vec p_{F}^{\,\o}=(\frac{2\pi}{3},\o\frac{2\pi}{3\sqrt{3}})\;,\label{A.18}\ee
close to which the complex dispersion relation vanishes linearly:
\be \O(\vec p_F^{\,\o}+\vec k')=ik_1'+\o k_2'+O(|\vec k'|^2)\;,\label{A.18a}\ee
resembling in this sense the relativistic dispersion relation of $(2+1)$-dimensional Dirac fermions.
From Eq.(\ref{A.16}) it is apparent that the ground state of the system consists of a Fermi sea 
such that all the negative energy states (the ``$\a$-states") are filled and all the positive energy 
states (the ``$\b$-states") are empty. The specific ground state energy $e_{0,\L}$ is
\be e_{0,\L}=-\frac{2v_0}{|\L|}\sum_{\vec k\in\BBB_L}|\O(\vec k)|\;,\label{A.18b}\ee
from which we find that  the specific ground state energy in the thermodynamic limit is
\be e(0)=-2v_0\int_{\BBB}\frac{d\vec k}{|\BBB|}|\O(\vec k)|\;,\label{A.18c}\ee
where $\BBB\= \{\vec k=t_1\vec G_1+t_2\vec G_2\ :\ t_i\in[0,1)\}$ and $|\BBB|=8\p^2/(3\sqrt3)$.
Similarly, the finite volume specific free energy 
$f_{0,\L}^\b:=-(\b|\L|)^{-1}\log\Tr\{e^{-\b H^0_\L}\}$ is
\be f_{0,\L}^\b=-\frac{2}{\b|\L|}\sum_{\vec k\in\BBB_L}\log\Big[
\big(1+e^{\b v_0|\O(\vec k)|}\big)\big(1+e^{-\b v_0|\O(\vec k)|}\big)\Big]\;,\label{A.18d}\ee
from which 
\be f_\b(0)=-\frac{2}{\b}\int_{\BBB}\frac{d\vec k}{|\BBB|}\log\Big[
\big(1+e^{\b v_0|\O(\vec k)|}\big)\big(1+e^{-\b v_0|\O(\vec k)|}\big)\Big]\;.\label{A.18e}\ee
Besides these thermodynamic functions, it is also useful to compute the {\it Schwinger 
functions} of the free gas, in terms of which, in the next section, we will write down the perturbation 
theory for the interacting system.  These are defined as follows.  
Let $x_0\in[0,\b)$ be an {\it imaginary time}, let $\xx:=(x_0,\vec x)\in[0,\b)\times\L$ and let us 
consider the time-evolved operator $\Psi^\pm_{\xx,\s}= e^{H
x_0}\Psi^\pm_{\vec x,\s} e^{-H x_0}$, where $\Psi^\pm_{\vec x,\s}$ is the two-components spinor defined in Eq.(\ref{A.14qqq}); in the following we shall denote its components by 
$\Psi^\pm_{\xx,\s,\r}$, with $\r\in\{1,2\}$, $\Psi^\pm_{\xx,\s,1}=a^\pm_{\xx,\s}$
and $\Psi^\pm_{\xx,\s,2}=b^\pm_{\xx+\dd_1,\s}$ (here $\dd_1=(0,\vec\d_1)$). 
We define the $n$-points Schwinger functions
at finite volme and finite temperature as:
\be S_n^{\b,\L}(\xx_1,\e_1,\s_1,\r_1;\ldots;\xx_n,\e_n,\s_n,\r_n)
=\media{\bT\{
\Psi^{\e_1}_{\xx_1,\s_1,\r_1}\cdots \Psi^{\e_n}_{\xx_n,\s_n,\r_n}\}}_{\b,\L}
\label{1.3}\ee
where: $\xx_i\in[0,\b]\times\L$, $\s_i=\uparrow\downarrow$,
$\e_i=\pm$, $\r_i=1,2$ and
$\bT$ is the operator of fermionic time ordering, acting on a
product of fermionic fields as:
\be \bT(
\Psi^{\e_1}_{\xx_1,\s_1,\r_1}\cdots \Psi^{\e_n}_{\xx_n,\s_n,\r_n})= (-1)^\p
\Psi^{\e_{\p(1)}}_{\xx_{\p(1)},\s_{\p(1)},\r_{\p(1)}}\cdots
\Psi^{\e_{\p(n)}}_{
\xx_{\p(n)},\s_{\p(n)},\r_{\p(n)}}\label{1.4}\ee
where $\p$ is a permutation of $\{1,\ldots,n\}$, chosen in such a
way that $x_{\p(1)0}\ge\cdots\ge x_{\p(n)0}$, and $(-1)^\p$ is its
sign. [If some of the time coordinates are
equal each other, the arbitrariness of the definition is solved by
ordering each set of operators with the same time coordinate so
that creation operators precede the annihilation operators.]
Taking the limit $\L\to\io$ in (\ref{1.3}) we get the finite
temperature $n$-point Schwinger functions, denoted by
$S^\b_n(\xx_1,\e_1,\s_1,\r_1; \ldots;\xx_n,\e_n,\s_n,\r_n) $,
which describe the properties of the infinite volume system at
finite temperature. Taking the $\b\to\io$ limit of the finite
temperature Schwinger functions, we get the zero temperature
Schwinger functions, simply denoted by
$S_n(\xx_1,\e_1,\s_1,\r_1;\ldots; \xx_n,\e_n,\s_n,\r_n)$, which
by definition characterize the properties of the {\it thermal ground state} 
of (\ref{1.1}) in the thermodynamic limit.

In the non-interacting case, i.e., if $H_\L=H^0_\L$, the Hamiltonian is quadratic in the 
creation/annihilation operators. Therefore, the $2n$-point
Schwinger functions satisfy the Wick rule, i.e.,
\bea&& \media{{\bf T}\{\Psi_{\xx_1,\s_1,\r_1}^-\Psi_{\yy_1,\s_1',\r_1'}^+\cdots 
\Psi_{\xx_n,\s_n,\r_n}^-\Psi_{\yy_n,\s_n',\r_n'}^+}_{\b,\L}
=\det G\;,\nn\\
&& G_{ij}=\d_{\s_i\s_j'}\media{{\bf T}\{\Psi^-_{\xx_i,\s_i,\r_i}
\Psi^+_{\yy_j,\s'_j,\r'_j}\}}_{\b,\L}
\;.
\label{A.2}\eea
Moreover, every $n$--point
Schwinger function $S_n^{\b,\L}(\xx_1,\e_1,\s_1,\r_1;\ldots;\xx_n,\e_n,\s_n,
\r_n)$ with $\sum_{i=1}^n\e_i\neq 0$
is identically zero. Therefore, in order to construct the whole set of
Schwinger functions of $H^0_\L$, it is enough to compute the $2$--point
function $S_0^{\b,\L}(\xx-\yy)=\media{{\bf T}\{\Psi^-_{\xx,\s,\r}
\Psi^+_{\yy,\s,\r'}\}}_{\b,\L}$. This can be easily reconstructed from the 
2-point function of the $\a$-fields and $\b$-fields, see Eq.(\ref{A.14ads}). 

Let $\vec x\in\L$, $\a_{\vec x,\s}^\pm=
|\L|^{-1}\sum_{\vec k\in\BBB_L} e^{\pm i\vec k\vec x}\hat \a_{\vec
k,\s}$ and $\b_{\vec x,\s}^\pm= |\L|^{-1}\sum_{\vec k\in\BBB_L}
e^{\pm i\vec k\vec x}\hat \a_{\vec k,\s}$; if $\xx=(x_0,\vec x)$, let
$\a_{\xx,\s}^\pm=e^{H^0_\L x_0}\a_{\vec x,\s}^\pm
e^{-H^0_\L x_0}$ and $\b_{\xx,\s}^\pm=e^{H^0_\L x_0}\b_{\vec
x,\s}^\pm e^{-H^0_\L x_0}$.  A straightforward computation
shows that, if $-\b<x_0-y_0\le \b$,
\bea &&\media{{\bf T}\{\a_{\xx,\s}^-\a_{\yy,\s'}^+\}}_{\b,\L}= \frac{\d_{\s,\s'}}{|\L|}
\sum_{\vec k\in\BBB_L}e^{-i\vec k( \vec x-\vec y)}\cdot\\
&&\qquad\qquad\cdot\Big[\openone\big(x_0-y_0>0\big)\frac{e^{v_0(x_0-y_0) |\O({\vec
k})|}} {1+e^{v_0\b|\O({\vec k})|}}-\openone\big(x_0-y_0\le
0\big)\frac{e^{v_0(x_0-y_0+\b) |\O({\vec k})|}} {1+e^{v_0\b|\O({\vec k})|}}
\Big]\;,\label{A.19aa}\nn\eea
\bea &&\media{{\bf T}\{\b_{\xx,\s}^-\b_{\yy,\s'}^+\}}_{\b,\L}= \frac{\d_{\s,\s'}}{|\L|}
\sum_{\vec k\in\BBB_L}e^{-i\vec k( \vec x-\vec y)}\cdot\\
&&\qquad\qquad\cdot\Big[\openone\big(x_0-y_0>0\big)\frac{e^{-v_0(x_0-y_0) |\O({\vec
k})|}} {1+e^{-v_0\b|\O({\vec k})|}}-\openone\big(x_0-y_0\le
0\big)\frac{e^{-v_0(x_0-y_0+\b) |\O({\vec k})|}} {1+e^{-v_0\b|\O({\vec k})|}}
\Big]\label{A.19}\nn\eea
and $\media{{\bf T}\{\a_{\xx,\s}^-\b_{\yy,\s'}^+\}}_{\b,\L}=
\media{{\bf T}\{\b_{\xx,\s}^-\a_{\yy,\s'}^+\}}_{\b,\L}=0$. A priori, 
Eqs.(\ref{A.19aa}) and (\ref{A.19}) are defined only for
$-\b<x_0-y_0\le \b$, but we can extend them periodically over the 
whole real axis; the periodic extension of the propagator is 
continuous in the time variable for $x_0-y_0\not\in\b \ZZZ$, and it has
jump discontinuities at the points $x_0-y_0\in\b\ZZZ$. 
Note that at $x_0-y_0=\b n$, the difference between the right and left 
limits is equal to
$(-1)^n\d_{\vec x,\vec y}$, so that the propagator is discontinuous only
at $\xx-\yy=\b\ZZZ\times \vec 0$. If we define $\BBB_{\b,L}:=\BBB_\b\times\BBB_L$, 
with $\BBB_\b=\{k_0=\frac{2\pi}{\b}(n_0+\frac{1}{2})
\;:\;n_0\in\ZZZ\}$, then 
for $\xx-\yy\not\in\b\ZZZ\times \vec 0$ we can write
\bea &&\media{{\bf T}\{\a_{\xx,\s}^-\a_{\yy,\s'}^+\}}_{\b,\L}=
\frac{\d_{\s,\s'}}{\b|\L|}\sum_{\kk\in\BBB_{\b,L}}e^{-i\kk(\xx-\yy)}
\frac1{-ik_0-v_0|\O({\vec k})|}\;, \label{A.19a1}\\
&&\media{{\bf
T}\{\b_{\xx,\s}^-\b_{\yy,\s'}^+\}}_{\b,\L}=\frac{\d_{\s,\s'}}
{\b|\L|}\sum_{\kk\in\BBB_{\b,L}}e^{-i\kk(\xx-\yy)}
\frac1{-ik_0+ v_0|\O({\vec k})|}\;. \label{A.19a2}\eea
If we now re-express
$\a_{\xx,\s}^\pm$ and $\b^\pm_{\xx,\s}$ in terms of
$a^\pm_{\xx,\s}$ and $b^\pm_{\xx+\dd_1,\s}$, using Eq.(\ref{A.14ads}),
we find that, for $\xx-\yy\not\in\b\ZZZ\times\vec 0$:
\bea S_0^{\b,\L}
(\xx-\yy)_{\r,\r'} &\=& S_{2}^{\b,\L}(\xx,\s,-,\r;\yy,\s,+,\r')\Big|_{U=0}
=\nn\\
&=&\frac{1}{\b|\L|}
\sum_{\kk\in\BBB_{\b,L}}\frac{e^{-i\kk\cdot(\xx-\yy)}}{k_0^2+v_0^2|\O(\vec
k)|^2} \begin{pmatrix}i k_0 & -v_0\O^*(\vec k) \\ -v_0\O(\vec k) &
ik_0\end{pmatrix}_{\r,\r'}\label{1.5}\eea
Finally, if $\xx-\yy=(0^-,\vec 0)$:
\be S_0^{\b,\L}(0^-,\vec 0)=-\frac12+\frac{1}{\b|\L|}
\sum_{\kk\in\BBB_{\b,L}} \frac{1}{k_0^2+v_0^2|\O(\vec k)|^2}
\begin{pmatrix}0&-v_0\O^*(\vec k)\\ -v_0\O(\vec k)&0\end{pmatrix}\;.\label{A.20}\ee

\section{Perturbation theory and Grassmann integration}\label{sec4}
\setcounter{equation}{0}
\renewcommand{\theequation}{\ref{sec4}.\arabic{equation}}

\begin{figure}\begin{center}
\includegraphics[width=0.85\textwidth]{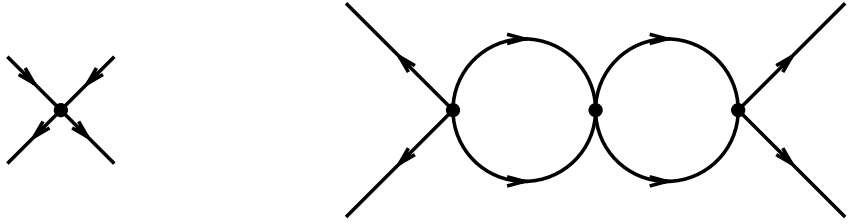}\end{center}
\caption{\label{fey} The four-legged graph element (left); an example of a Feynman 
diagram of order 3 (right).}
\end{figure}

Let us now turn to the interacting case. The first step is to derive a formal 
perturbation theory for the specific free energy and ground state energy. In other words, we want 
to find rules to compute the generic perturbative order in $U$ of $f_{\b,\L}:=
-(\b|\L|)^{-1}\log\Tr\{e^{-\b H_{\L}}\}$. We write $H_\L=H^0_\L+V_\L$, with $V_\L$ the operator
in the second line of Eq.(\ref{1.1}) and we use Trotter's product formula 
\be e^{-\b H_{\L}}=\lim_{n\to\infty}\Big[e^{-\b H^0_\L/n}(1-\frac{\b}{n}V_\L)\Big]^n\label{s4.1}\ee
so that, defining $V_\L(t):=e^{tH^0_\L}V_\L e^{-tH^0_\L}$, 
\bea && \frac{\Tr\{e^{-\b H_{\L}}\}}{\Tr\{e^{-\b H^0_\L}\}}=\label{s4.2}\\
&&\qquad=1+\sum_{N\ge 1}(-1)^N
\int_0^\b dt_1\int_0^{t_1}dt_2\cdots \int_0^{t_{N-1}}
dt_N \frac{\Tr\{e^{-\b H^0_\L}V_\L(t_1)\cdots V_\L(t_N)\}}{\Tr\{e^{-\b H^0_\L}\}}\;.\nn\eea
Using the fermionic time-ordering operator defined in Eq.(\ref{1.4}), we can rewrite Eq.(\ref{s4.2})
as
\be \frac{\Tr\{e^{-\b H_{\L}}\}}{\Tr\{e^{-\b H^0_\L}\}}=
1+\sum_{N\ge 1}\frac{(-1)^N}{N!}\media{\bT\{(V_{\b,\L}(\Psi))^N\}}^0_{\b,\L}\;,\label{s4.3}\ee
where $\media{\cdot}^0_{\b,\L}=\Tr\{e^{-\b H^0_\L}\cdot\}/\Tr\{e^{-\b H^0_\L}\}$, 
\be V_{\b,\L}(\Psi):=U\sum_{\r=1,2}\int_{(\b,\L)} d\xx \big(\Psi_{\xx,\uparrow,\r}^+
\Psi_{\xx,\uparrow,\r}^--\frac12\big)\big(\Psi_{\xx,\downarrow,\r}^+
\Psi_{\xx,\downarrow,\r}^--\frac12\big)\;,\label{s4.4}\ee
and $\int_{(\b,\L)} d\xx$ must be interpreted as $\int_{(\b,\L)} d\xx=
\int_{-\b/2}^{\b/2} dx_0\sum_{\vec x\in \L}$.
Note that the $N$-th term in the sum in the r.h.s. of Eq.(\ref{s4.3}) can be computed by 
using the Wick rule (\ref{A.2}) and the explicit expression for the 2-point function 
Eqs.(\ref{1.5})-(\ref{A.20}). It is straightforward to check that the ``Feynman rules" needed 
to compute $\media{\bT\{(V_{\b,\L}(\Psi))^N\}}^0_{\b,\L}$ are the following: 
(i) draw $N$ graph elements consisting of 
4-legged vertices, with the vertex associated to two labels $\xx_i$ and $\r_i$, $i=1,\ldots,N$, and 
the four legs associated to two exiting fields (with labels $(\xx_i,\uparrow,\r_1)$ and 
$(\xx_i,\downarrow,\r_i)$) and two entering fields (with labels $(\xx_i,\uparrow,\r_i)$ and 
$(\xx_i,\downarrow,\r_i)$), respectively; 
(ii) pair the fields in all possible ways, in such a way that every pair consists of one entering and
one exiting field, with the same spin index, see Fig.\ref{fey};
(iii) associate to every pairing a sign, corresponding to 
the sign of the permutation needed to bring every pair of contracted fields next to each other; 
(iv) associate to every paired pair of fields $[\Psi^-_{\xx_i,\s_i,\r_i},\Psi^+_{\xx_j,\s_j,\r_j}]$ 
an oriented line connecting the $i$-th with the $j$-th vertex, with orientation from $j$ to $i$;
(v) associate to every oriented line $[j\to i]$ a value equal to 
\be g_{\r_i,\r_j}(\xx_i-\xx_j):=\frac{1}{\b|\L|}
\sum_{\kk\in\BBB^{(M)}_{\b,L}}e^{-i\kk\cdot(\xx_i-\xx_j)}\frac{\c_0(2^{-M}|k_0|)}{k_0^2+v_0^2
|\O(\vec k)|^2} \begin{pmatrix}i k_0 & -v_0\O^*(\vec k) \\ -v_0\O(\vec k) &
ik_0\end{pmatrix}_{\r_i,\r_j}\label{s4.4a}\ee
where $\BBB^{(M)}_{\b,L}=\BBB_\b^{(M)}\times\BBB_L$, $\BBB_\b^{(M)}=\BBB_\b\cap 
\{k_0\ :\ \c_0(2^{-M}|k_0|)>0\}$ and $\c_0(t)$ is a smooth compact support function that 
is equal to 1 for (say) $|t|\le 1/3$ and equal to 0 for $|t|\ge 2/3$; (vi) associate to every pairing 
(i.e., to every Feynman graph) a value, equal to the product of the sign of the pairing times 
$U^N$ times the product of the values of all the oriented
lines; (vii) integrate over $\xx_i$ and sum over $\r_i$ the value of each pairing, then sum over 
all pairings; (viii) finally, take the $M\to\infty$ limit: 
the result is equal to $\media{\bT\{(V_{\b,\L}(\Psi))^N\}}^0_{\b,\L}$. Note that the $M\to\infty$ limit 
of the {\it propagator} $g(\xx)$ is equal to $S_0^{\b,\L}(\xx)$ if $\xx\neq\V0$, while 
$\lim_{M\to\infty}g(\V0)=S_0^{\b,\L}(0^-,\vec 0)+\frac12$, see Eq.(\ref{A.20}): the difference 
between $\lim_{M\to\infty}g(\xx)$ and $S_0^{\b,\L}(\xx)$ takes into account the $-\frac12$ terms in
the definition of $V_{\b,\L}(\Psi)$. \\

An algebraically convenient way to re-express Eq.(\ref{s4.3}) is in terms of 
{\it Grassmann integrals}. Consider the set ${\cal A}_{M,\b,L}=\{\hat\psi^\pm_{\kk,\s,\r}\}_{ \kk \in
\BBB_{\b,L}^{(M)}}^{\s=\uparrow\downarrow,\ \r=1,2}$, where the {\it Grassmann variables}
$\hat\psi^\pm_{\kk,\s,\r}$ satisfy by the definition the anticommutation rules 
$\{\hat\psi^\e_{\kk,\s,\r},$ $ \hat\psi^{\e'}_{\kk',\s',\r'}\}=0$.
In particular, the square of a Grassmann 
variable is zero and the only non-trivial Grassmann monomials 
are at most linear in each variable. Let the Grassmann algebra generated by ${\cal A}_{M,\b,L}$
be the set of all polynomials obtained by linear combinations of such  
non-trivial monomials. Let us also define the Grassmann
integration $\int
\big[\prod_{\kk\in\BBB^{(M)}_{\b,L}}\prod_{\s=\uparrow\downarrow}^{\r=1,2}
d\hat\psi_{\kk,\s,\r}^+ d\hat\psi_{\kk,\s,\r}^-\big]$ as
the linear operator on the Grassmann algebra such that, given a
monomial $Q( \hat\psi^-, \hat\psi^+)$ in the variables
$\hat\psi^\pm_{\kk,\s,\r}$, its action on $Q(\hat\psi^-,
\hat\psi^+)$ is $0$ except in the case $Q(\hat\psi^-,
\hat\psi^+)=\prod_{\kk\in\BBB_{\b,L}^{(M)}}
\prod_{\s=\uparrow\downarrow}^{\r=1,2} \hat\psi^-_{\kk,\s,\r}
\hat\psi^+_{\kk,\s,\r}$, up to a permutation of the variables. In
this case the value of the integral is determined, by using the
anticommutation properties of the variables, by the condition
\be \int \Big[\prod_{\kk\in\BBB_{\b,L}^{(M)}}\prod_{\s=\uparrow\downarrow}^{\r=1,2}
d\hat\psi_{\kk,\s,\r}^+
d\hat\psi_{\kk,\s,\r}^-\Big]\prod_{\kk\in\BBB_{\b,L}^{(M)}}
\prod_{\s=\uparrow\downarrow}^{\r=1,2}
\hat\psi^-_{\kk,\s,\r} \hat\psi^+_{\kk,\s,\r}=1\label{2.1}\ee
Defining the free propagator matrix $\hat g_\kk$ as
\be \hat g_\kk=\c_0(2^{-M}|k_0|)
\begin{pmatrix}-i k_0 & -v_0\O^*(\vec k) \\ -v_0\O(\vec k) & -i k_0\end{pmatrix}^{-1}
\label{2.2}\ee
and the ``Gaussian integration'' $P_M(d\psi)$ as
\bea P_M(d\psi) &=& \Big[\prod_{\kk\in\BBB_{\b,L}^{(M)}}
^{\s=\uparrow\downarrow}\frac{-\b^2|\L|^2\,[\c_0(2^{-M}|k_0|)]^{2}}
{k_0^2+v_0^2|\O(\vec k)|^2}
d\hat\psi_{\kk,\s,1}^+
d\hat\psi_{\kk,\s,1}^-d\hat\psi_{\kk,\s,2}^+
d\hat\psi_{\kk,\s,2}^-\Big]\cdot\nn\\&&\hskip3.truecm
\cdot\;\exp \Big\{-(\b|\L|)^{-1}
\sum_{\kk\in\BBB_{\b,L}^{(M)}}^{\s=\uparrow\downarrow}
\hat\psi^{+}_{\kk,\s}\,{\hat g}_\kk^{-1}\hat\psi^{-}_{\kk,\s}
\Big\}\;,\label{2.3}\eea
it turns out that
\be\int P(d \psi) \hat \psi^-_{\kk_1,\s_1}\hat \psi^+_{\kk_2,\s_2} =
\b|\L|\d_{\s_1,\s_2}\d_{\kk_1,\kk_2} \hat g_{\kk_1}
\;,\label{2.4}\ee
while the average of an arbitrary monomial in the Grassmann variables with respect 
to $P_M(d\psi)$ is given by the fermionic Wick rule with propagator equal to the r.h.s. of 
Eq.(\ref{2.4}). Using these definitions and the Feynman rules described above, we can rewrite
Eq.(\ref{s4.3}) as
\be \frac{\Tr\{e^{-\b H_{\L}}\}}{\Tr\{e^{-\b H^0_\L}\}}=\lim_{M\to\infty}
\int P_M(d\psi) e^{-\VV(\psi)}\;,\label{s4.5}\ee
where 
\be \VV(\psi)
=U\sum_{\r=1,2}\int_{(\b,\L)} d\xx \, \psi^+_{\xx,\uparrow,\r}
\psi^-_{\xx,\uparrow,\r}\psi^+_{\xx,\downarrow,\r}
\psi^-_{\xx,\downarrow,\r}\;,\label{s4.6}\ee
\be\psi_{\xx,\s,\r}^{\pm}=\frac{1}{\b|\L|}\sum_{\kk\in\BBB_{\b,L}^{(M)}}
e^{\pm i\kk\xx}\hat\psi^\pm_{\kk,\s,\r}\;,\qquad \xx\in(-\b/2,\b/2]\times\L
\label{2.6}\ee
and the exponential $e^{-\VV(\psi)}$ in the r.h.s. of Eq.(\ref{s4.5}) must be identified with its 
Taylor series in $U$ (which is finite for every finite $M$, due to the anticommutation rules of the
Grassmann variables and the fact that the Grassmann algebra is finite for every finite $M$). 
Apriori, Eq.(\ref{s4.5}) must be understood as an equality between formal power series in $U$. 
However, it can be given a non-perturbative meaning, provided that we can prove 
the convergence of the Grassmann functional integral in the r.h.s., as shown by the following Proposition. 

\begin{proposition}\label{prop1}
Let
\be F_{\b,\L}^{(M)}:=-\frac1{\b|\L|}\log\int P_M(d\psi)\big( e^{-\VV(\psi)}\big)\label{s4.7}\ee
and let $\b$ and $|\L|$ be sufficiently large. Assume that there exists $U_0>0$
such that $ F_{\b,\L}^{(M)}$ is analytic in the complex domain $|U|\le U_0$ and is uniformly 
convergent as $M\to\infty$. Then, if $|U|\le U_0$,
\be f_{\b,\L}=
-\frac{2}{\b|\L|}\sum_{\vec k\in\BBB_L}\log\big(2+2\cosh(\b v_0|\O(\vec k)|)\big)
+\lim_{M\to\infty}F_{\b,\L}^{(M)}\;.\label{s4.8}\ee
\end{proposition}

{\bf Proof.} We need to prove that 
\be \frac{\Tr\{e^{-\b H_\L}\}}{\Tr\{e^{-\b H_{0,\L}}\}}=\exp\big\{-\b|\L|
\lim_{M\to \io}F_{\b,\L}^{(M)}\big\}\label{appB.0}\ee
under the given analyticity assumptions on $F_{\b,\L}^{(M)}$.
The first key remark is that, if $\b,\L$ are finite, the left hand side of 
Eq.(\ref{appB.0}) is a priori well defined and analytic on the whole complex 
plane. In fact, by the Pauli principle, the Fock space generated by the fermion
operators $a^\pm_{\vec x,\s}, b^\pm_{\vec x+\vec \d_1,\s}$, with
$\vec x\in\L, \s=\uparrow\downarrow$, is finite dimensional.
Therefore, writing $H_\L=H^0_\L+V_\L$, with $H^0_\L$ and $V_\L$
two bounded operators, we see that $\Tr\{e^{-\b H_\L}\}$ is an
entire function of $U$, simply because $e^{-\b
H_\L}$ converges
in norm over the whole complex plane: 
\bea ||e^{-\b H_\L}||\le \sum_{n=0}^\io \frac{\b^n}{n!}
\big(||H_{\L}^0||+ ||
V_\L||\big)^n&=&\sum_{k=0}^\io \frac{\b^k  ||V_\L||^k}{k!}\sum_{n\ge k} 
\frac{\b^{n-k}||H_\L^0||^{n-k}}{(n-k)!}=\nn\\
&=&e^{\b||H^0_\L||+\b\,||V_\L||}\;,\eea
where the norm $||\cdot||$ is, e.g., the Hilbert-Schmidt norm $||A||=\sqrt{
\Tr(A^\dagger A)}$. 

On the other hand, by assumption, $F_{\b,\L}^{(M)}$ is analytic in 
$|U|\le U_0$, with $U_0$ independent of $\b,\L,M$, and uniformly 
convergent as $M\to\io$. Hence, by Weierstrass theorem, the limit 
$F_{\b,\L}=\lim_{M\to\io}F_{\b,\L}^{(M)}$ is analytic in $|U|\le U_0$
and its Taylor coefficients coincide with the limits as $M\to\io$
of the Taylor coefficients of $F_{\b,\L}^{(M)}$. Moreover, 
$\lim_{M\to\io}e^{-\b|\L|F_{\b,\L}^{(M)}}=e^{-\b|\L|F_{\b,\L}}$, 
again by Weierstrass theorem.

As discussed above, the Taylor coefficients of $e^{-\b|\L|F_{\b,\L}}$ 
coincide with the Taylor coefficients of $\Tr\{e^{-\b H_\L}\}/
\Tr\{e^{-\b H_{0,\L}}\}$: therefore, 
$\Tr\{e^{-\b H_\L}\}/\Tr\{e^{-\b H_{0,\L}}\}= e^{-\b|\L|F_{\b,\L}}$
in the complex region $|U|\le U_0$, simply because the l.h.s. is entire
in $U$, the r.h.s. is analytic in $|U|\le U_0$ and the Taylor coefficients
at the origin of the two sides are the same. Taking logarithms at both sides 
proves Eq.(\ref{s4.8}).\qed

\vskip.4truecm

By Proposition \ref{prop1}, the Grassmann integral Eq.(\ref{s4.7}) can be used to compute the free
energy of the original Hubbard model, provided that the r.h.s. of 
Eq.(\ref{s4.7}) is analytic in a domain that is uniform in $M,\b,\L$ and that it converges to 
a well defined analytic function uniformly as $M\to\infty$. The rest of these notes are devoted 
to the proof of this fact. We start from Eq.(\ref{s4.7}), which can be rewritten as
\be F_{\b,\L}^{(M)}:=-\frac1{\b|\L|}\sum_{N\ge 1}\frac{(-1)^N}{N!}\EE^T(\VV;N)\;,
\label{s4.9}\ee
where the {\it truncated expectation} $\EE^T$ is defined as 
\be \EE^T(\VV; N):=\frac{\dpr^N}{\dpr\l^N}\log\int P_M(d\psi) e^{\l\VV(\psi)}\big|_{\l=0}\;.
\label{s4.10}\ee
More in general,
\be \EE^T(\VV_1,\ldots,\VV_N):=\frac{\dpr^N}{\dpr\l_1\cdots\dpr\l_N}\log\int P_M(d\psi) 
e^{\l_1\VV_1(\psi)+\cdots+\l_N\VV_N(\psi)}\big|_{\l_i=0}
\label{s4.11}\ee
and $\EE^T(\VV_1,\ldots,\VV_N)\big|_{\VV_i=\VV}=\EE^T(\VV;N)$. It can be checked by induction
that the truncated expectation is related to the simple expectation $\EE(X(\psi))=
\int P_M(d\psi) X(\psi)$ by 
\be  \EE(\VV_1\cdots\VV_N)=\sum_{m=1}^N\ \sum_{(Y^1,\ldots, Y^m)}
\EE^T(\VV_{j^1_1},\ldots,\VV_{j^1_{|Y^1|}})\cdots
\EE^T(\VV_{j^m_1},\ldots,\VV{j^m_{|Y^m|}})\;,\label{s4.12}\ee
where the second sum in the r.h.s. runs over partitions of $\{1,\ldots,N\}$ of multiplicity $m$,
i.e., over $m$-ples of disjoint sets such that $\cup_{i=1}^m Y^i=\{1,\ldots,N\}$,
with $Y^i=\{j^i_1,\ldots,j^i_{|Y_i|}\}$. Note that 
$\EE(\VV^N)=\EE(\VV_1,\ldots,\VV_N)\big|_{\VV_i=\VV}$ can be computed as a sum of 
Feynman diagrams whose values are determined by the same Feynman rules described after 
Eq.(\ref{s4.4}) (with the exception of rule (viii): of course, since $\EE(X)=\int P_M(d\psi)X$,
$M$ should be temporarily kept fixed in the computation); we shall write 
\be \EE(\VV^N)=\sum_{\GG\in\G_N}\widehat\Val(\GG)\;,\label{s4.13}\ee
where $\G_N$ is the set of all Feynman diagrams with $N$ vertices, 
constructed with the rules described above; $\widehat\Val(\GG)$ includes the integration over 
the space-time labels $\xx_i$  and the sum over the component labels $\r_i$: if $\GG\in\G^T_N$,
we shall write symbolically 
\be \widehat\Val(\GG)=\s_\GG U^N\sum_{\r_1,\ldots,\r_N}\int d\xx_1\cdots d\xx_N
\prod_{\ell\in\GG}\d_{\s(\ell),\s'(\ell)}g_{\r(\ell),\r'(\ell)}(\xx(\ell)-\xx'(\ell))\;,\label{s4.13a}\ee
where $\s_\GG$ is the sign of the permutation associated to the graph $\GG$ and
we denoted by $(\xx(\ell), \s(\ell),\r(\ell))$ and $(\xx'(\ell),\s'(\ell), \r'(\ell))$ 
the labels of the two vertices, which the line $\ell$ exits from and enters in, respectively. 
Using Eqs.(\ref{s4.12})-(\ref{s4.13}), it can be proved by induction that 
\be  \EE^T(\VV; N)=\sum_{\GG\in\G^T_N}\widehat\Val(\GG)\;,\label{s4.14}\ee
where $\G^T_N\subset\G_N$ is the set of {\it connected} Feynman diagrams 
with $N$ vertices. Combining Eq.(\ref{s4.9}) with Eq.(\ref{s4.14}) we finally have a 
formal power series expansion for the specific free energy of our model (more precisely, of its
ultraviolet regularization associated to the imaginary-time ultraviolet cutoff $\c_0(2^{-M}|k_0|)$).
The Feynman rules for computing $\widehat\Val(\GG)$ allow us to derive a first {\it very naive}
upper bound on the $N$-th order contribution to $F_{\b,\L}^{(M)}$, that is to 
\be F_{\b,\L}^{(M;N)}:=-\frac1{\b|\L|}\frac{(-1)^N}{N!}\EE^T(\VV;N)\;.\label{s4.14a}\ee
We have:
\bea |F_{\b,\L}^{(M;N)}|&\le& \frac1{\b|\L|}\frac{1}{N!} \sum_{\GG\in\G^T_N}|\widehat\Val(\GG)|
\le\nn\\
&\le& \frac{|\G^T_N|}{N!} 2^N |U|^N ||g||_{\infty}^{N+1}\,||g||_{1}^{N-1}\;,\label{s4.15}\eea
where $|\G^T_N|$ is the number of connected Feynman diagrams of order $N$
and $\b|\L|(2|U|)^N||g||_{\infty}^{N+1}\,||g||_{1}^{N-1}$ is a uniform bound on the value of a generic 
connected Feynman diagram of order $N$. The bound is obtained as follows: given 
$\GG\in\G^T_N$, select an arbitrary ``spanning tree" in $\GG$, i.e. a loopless subset 
of $\GG$ that connects all the $N$ vertices; now: the integrals over the space-time
coordinates of the product of the propagators on the spanning tree can be bounded 
by $\b|\L|\,||g||_{1}^{N-1}$; the product of the remaining propagators can be bounded by 
$||g||_\io^{N+1}$; finally, the sum over the $\r_i$ labels is bounded by $2^N$. 
Using Eq.(\ref{s4.15}) and the facts that, for a suitable constant $C>0$:
(i)  $|\G^T_N|\le C^N (N!)^2$ (see, e.g., \cite[Appendix A.1.3]{GeM} for a proof of this fact), 
(ii) $||g||_{\infty}\le C M$, (iii) $||g||_{1}\le C \b$, we find:
\be |F_{\b,\L}^{(M;N)}|\le (2C^3|U|)^N N! \,M^{N+1}\b^{N-1}\;.\label{s4.16}\ee
\vskip.1truecm
{\bf Remark.} While the bound $||g||_{1}\le C \b$  (see Appendix \ref{app0} for a proof) is dimensionally optimal, 
the estimate $||g||_{\infty}\le C M$ could be improved to $||g||_{\infty}\le(\const.)$, at the cost 
of a more detailed analysis of the definition of $g(\xx)$, which shows that the apparent 
ultraviolet logarithmic divergence associated to the sum over $k_0$ in Eq.(\ref{s4.4a})
is in fact related to a jump singularity of $g(x_0,\vec 0)$ at $x_0=\b\ZZZ$. This can be proved 
along the lines of \cite[Section 2]{Be}. However, to the purpose of the present discussion,
the rough (and easier) bound $||g||_{\infty}\le C M$ is enough; see Appendix \ref{app0} for a proof.

\vskip.2truecm
The pessimistic bound Eq.(\ref{s4.16}) has two main problems: (i) a combinatorial problem, 
associated to the $N!$, which makes the r.h.s. of Eq.(\ref{s4.16}) not summable over $N$, not
even for finite $M$ and $\b$; (ii) a divergence problem, associated to 
the factor $M^{N+1}\b^{N-1}$, which diverges exponentially as $M\to\infty$ (i.e., as 
the ultraviolet regularization is removed) and as $\b\to\infty$ (i.e., as the temperature is sent to 0).
The combinatorial problem is solved by a smart reorganization of the perturbation theory, in the 
form of a determinant expansion, together with a systematic use of the Gram-Hadamard bound. 
The divergence problem is solved by systematic resummations of the series: we will first identify 
the class of contributions that produce ultraviolet or infrared divergences 
and then we show how to inductively resum them into a redefinition of the coupling constants of 
the theory; the inductive resummations are based on a multiscale integration of the theory: at the 
end of the construction, they will allow us to express the specific free energy in terms of modified 
Feynman diagrams, whose values are not affected anymore by ultraviolet or infrared divergences. 

\section{The determinant expansion}\label{sec5}
\setcounter{equation}{0}
\renewcommand{\theequation}{\ref{sec5}.\arabic{equation}}

Let us now show how to attack the first of two problems that arose at the end of previous section. 
In other words, let us show how to solve the combinatorial problem by reorganizing the 
perturbative expansion discussed above into a more compact and more convenient form.
In the previous section we discussed a Feynman diagram representation of the truncated 
expectation, see Eq.(\ref{s4.14}). A slightly more general version of 
Eq.(\ref{s4.14}) is the following. For a given set of indices $P=\{f_1,\ldots,f_{|P|}\}$, with 
$f_i=(\xx_i,\s_i,\r_i,\e_i)$, $\e_i\in\{+,-\}$, let 
\be \psi_P:=\prod_{f\in P}\psi^{\e(f)}_{\xx(f),\s(f),\r(f)}\;.\label{s5.1}\ee
Each field $\psi^{\e(f)}_{\xx(f),\s(f),\r(f)}$ can be represented as an oriented
half-line, emerging from the point $\xx(f)$ and carrying an arrow, pointing in the direction entering 
or exiting the point, depending on whether $\e(f)$ is equal to $-$ or $+$, respectively; moreover,
the half-line carries two labels, $\s(f)\in\{\uparrow,\downarrow\}$ and $\r(f)\in\{1,2\}$. 
Now, given $s$ set of indices $P_1,\ldots,P_s$, we can enclose the points $\xx(f)$ belonging 
to the set $P_j$, for some $j=1,\ldots,s$, in a box: in this way, assuming that all the points $\xx(f)$, 
$f\in\cup_iP_i$, are distinct, we obtain $s$ disjoint boxes. Given $\PPP:=\{P_1,\ldots,P_s\}$, 
we can associate to it the set $\G^T(\PPP)$ of connected Feynman diagrams, obtained by 
pairing the half-lines with consistent orientations, in such a way that the two half-lines of any 
connected pairs carry the same spin index, and in such a way that all the boxes are connected. 
Using a notation similar to Eq.(\ref{s4.13a}), we have:
\bea &&\EE^T(\psi_{P_1}, \ldots,\psi_{P_s})=
\sum_{\GG\in\G^T(\PPP)}\Val(\GG)\;,\nn\\
&&\Val(\GG)=\s_\GG \prod_{\ell\in\GG}\d_{\s(\ell),\s'(\ell)}
g_{\r(\ell),\r'(\ell)}(\xx(\ell)-\xx'(\ell))\;,\label{s5.2}\eea
A different a more compact representation for the truncated expectation, alternative to 
Eq.(\ref{s5.2}), is the following:
\be\EE^T(\psi_{P_1}, \ldots,\psi_{P_s})=
\sum_{T\in\bT(\PPP)}\a_T\prod_{\ell\in T}g_\ell
\int dP_{T}(\tt) \det G^{T}(\tt)\;,\label{s5.3}\ee
where:\begin{itemize}
\item
any element $T$ of the set $\bT(\PPP)$ is a set of lines forming an {\it anchored tree} between
the boxes $P_1,\ldots,P_{s}$, i.e., $T$ is a set
of lines that becomes a tree if one identifies all the points
in the same clusters;
\item $\a_T$ is a sign (irrelevant for the subsequent bounds);
\item $g_\ell$ is a shorthand for $\d_{\s(\ell),\s'(\ell)}\,g_{\r(\ell),\r'(\ell)}(\xx(\ell)-\xx'(\ell))$;
\item if $\tt=\{t_{i,i'}\in [0,1], 1\le i,i' \le s\}$, then $dP_{T}(\tt)$
is a probability measure with support on a set of $\tt$ such that
$t_{i,i'}=\uu_i\cdot\uu_{i'}$ for some family of vectors $\uu_i\in \RRR^n$ of
unit norm;
\item if $2n=\sum_{i=1}^s|P_{i}|$, then 
$G^{T}(\tt)$ is a $(n-s+1)\times (n-s+1)$ matrix, whose
elements are given by $G^{T}_{f,f'}=t_{i(f),i(f')}g_{\ell(f,f')}$, where:
$f, f'\not\in\cup_{\ell\in T}\{f^1_\ell,f^2_\ell\}$ and $f^1_\ell,f^2_\ell$ are 
the two field labels associated to the two (entering and exiting) half-lines contracted into $\ell$;
$i(f)\in\{1,\ldots,s\}$ is s.t. $f\in P_{i(f)}$; $g_{\ell(f,f')}$ is the propagator associated to the line 
obtained by contracting the two half-lines with indices $f$ and $f'$. 
\end{itemize}

If $s=1$ the sum over $T$ is empty, but we can still
use the Eq.(\ref{s5.3}) by interpreting the r.h.s.
as equal to $1$ if $P_{1}$ is empty and equal to $\det G^T({\bf 1})$ otherwise.
\\

The proof of the determinant representation is described in Appendix \ref{app1}; 
this representation is due to a fermionic reinterpretation 
of the interpolation formulas by Battle, Brydges and Federbush \cite{BF,B,BrF}, 
used originally by Gawedski-Kupianen \cite{GK} and by Lesniewski \cite{Le}, among others, 
to study certain $(1+1)$-dimensional fermionic Quantum Field Theories. Using 
Eq.(\ref{s5.3}) we get an alternative representation for the $N$-th order contribution to 
the specific free energy:
\bea F_{\b,\L}^{(M;N)}&=&-\frac1{\b|\L|}\frac{(-1)^N}{N!}\EE^T(\VV;N)=
-\frac1{\b|\L|}\frac{(-1)^N}{N!}U^N\!\!\!\!\sum_{\r_1,\ldots,\r_N}
\sum_{T\in\bT_N}\a_T\int d\xx_1\cdots d\xx_N\nn\\
&&\qquad\cdot\prod_{\ell\in T}\d_{\s(\ell),\s'(\ell)}\,g_{\r(\ell),\r'({\ell})}(\xx(\ell)-\xx'(\ell))
\int dP_{T}(\tt) \det G^{T}(\tt)\;.\label{s5.4}\eea
Using the fact that the number of anchored trees in $\bT_N$ is bounded by $C^N N!$ for a 
suitable constant $C$ (see, e.g., \cite[Appendix A.3.3]{GeM} for a proof of this fact), 
from Eq.(\ref{s5.4}) we get:
\be |F_{\b,\L}^{(M;N)}|\le (\const.)^N|U|^N\,||g||_1^{N-1}||\det G^T(\cdot)||_\io\;.\label{s5.5}\ee
In order to bound $\det G^T$, we use the {\it Gram-Hadamard inequality}, stating
that, if $M$ is a square matrix with elements $M_{ij}$ of the form
$M_{ij}=\media{A_i,B_j}$, where $A_i$, $B_j$ are vectors in a Hilbert space with
scalar product $\media{\cdot,\cdot}$, then
\be |\det M|\le \prod_i ||A_i||\cdot ||B_i||\;.\label{s5.6}\ee
where $||\cdot||$ is the norm induced by the scalar product. See \cite[Theorem A.1]{GeM}
for a proof of Eq.(\ref{s5.6}).
 
Let $\HHH=\RRR^n\otimes \HHH_0$, where $\HHH_0$ is the Hilbert space of the
functions ${\bf F} : [-\b/2,\b/2)\times\L\to \CCC^2$,
with scalar product $\media{{\bf F},{\bf G}}=\sum_{\r=1,2}\int d\zz\, F^*_\r(\zz)G_\r(\zz) $,
where $F_\r=[{\bf F}]_\r$, $G_\r=[{\bf G}]_\r$, $\r=1,2$, are the components of the vectors 
${\bf F}$ and ${\bf G}$. It is easy to verify that
\bea G^T_{f,f'}&=&t_{i(f),i(f')}\,\d_{\s(\ell),\s'(\ell)}\, g_{\r(f),\r(f')}(\xx(f)-\xx(f'))=\nn\\
&=&\media{\uu_{i(f)}\otimes{\bf e}_{\s(f)} \otimes
{\bf A}_{\xx(f),\r(f)},
\uu_{i(f')}\otimes {\bf e}_{\s(f')}\otimes{\bf B}_{\xx(f'),\r(f')}}\;,\label{s5.8}\eea
where: $\uu_i\in \RRR^n$, $i=1,\ldots,n$, are vectors such that
$t_{i,i'}=\uu_i\cdot\uu_{i'}$; ${\bf e}_\uparrow=(1,0)$, ${\bf e}_{\downarrow}=(0,1)$; ${\bf A}_{\xx,\r}$ and ${\bf B}_{\xx,\r}$ have components:
\bea &&[{\bf A}_{\xx,\r}(\zz)]_i=\frac{1}{\b|\L|}\sum_{\kk\in\BBB^{(M)}_{\b,L}} 
\frac{\sqrt{\c_0(2^{-M}|k_0|)}\,e^{-i\kk(\zz-\xx)}}{\big[k_0^2+v_0^2|\O(\vec k)|^2\big]^{1/4}}\d_{\r,i}\;,
\label{B.4}\\
&&[{\bf B}_{\xx,\r}(\zz)]_i=\frac1{\b|\L|}\sum_{\kk\in\BBB^{(M)}_{\b,L}} 
\,\frac{\sqrt{\c_0(2^{-M}|k_0|)}\,e^{-i\kk(\zz-\xx)}}{\big[k_0^2+v_0^2|\O(\vec k)|^2\big]^{3/4}}\begin{pmatrix}i k_0 & -v_0\O^*(\vec k)\cr
-v_0\O(\vec k)& i k_0\end{pmatrix}_{i,\r}\;,\nn\eea
so that
\be ||{\bf A}_{\xx,\r}||^2=||{\bf B}_{\xx,\r}||^2=\frac1{\b|\L|}\sum_{\kk\in\BBB^{(M)}_{\b,L}}
\frac{\c_0(2^{-M}|k_0|)}{\big[k_0^2+v_0^2|\O(\vec k)|^2\big]^{1/2}}\le C M\;,\label{B.5}\ee
for a suitable constant $C$. 
Using the Gram-Hadamard inequality, we find 
$||\det G^T||_\io\le (\const.)^N M^{N+1}$; substituting this result into Eq.(\ref{s5.5}),
we finally get:
\be |F_{\b,\L}^{(M;N)}|\le (\const.)^N|U|^N M^{N+1}\b^{N-1}\;,\label{s5.9}\ee
which is similar to Eq.(\ref{s4.16}), but for the fact that there is no $N!$ in the r.h.s.! In other words, 
using the determinant expansion, we recovered the same dimensional estimate as the one
obtained by the Feynman diagram expansion and we combinatorially gained a $1/N!$. 
The r.h.s. of Eq.(\ref{s5.9}) is now summable over $N$ for $|U|$ sufficiently small,
even though non uniformly in $M$ and $\b$. In the next section we will discuss how to 
systematically improve the dimensional bound by an iterative resummation method. 

\section{The multiscale integration: the ultraviolet regime}\label{sec6}
\setcounter{equation}{0}
\renewcommand{\theequation}{\ref{sec6}.\arabic{equation}}

In this section we begin to illustrate the multiscale integration of the fermionic 
functional integral of interest. This method will later allow us to perform iterative resummations
and to re-express the specific free energy in terms of a modified expansion, whose $N$-th order 
term is summable in $N$ and uniformly convergent as $M\to\infty$ and $\b\to-\io$, as desired.  

The first step in the computation of the partition function \be\X_{M,\b,\L}:=\int P_M(d\psi)
e^{-\VV(\psi)}\label{s6.0}\ee 
and of its logarithm is the integration of the ultraviolet degrees of freedom 
corresponding to the large 
values of $k_0$. We proceed in the following way. We decompose the free
propagator $\hat g_\kk$ into a sum of two propagators supported in
the regions of $k_0$ ``large'' and ``small'', respectively. The
regions of $k_0$ large and small are defined in terms of the smooth
support function $\chi_0(t)$ introduced after Eq.(\ref{s4.4a}); note that, by the very definition 
of $\c_0$, the supports
of $\chi_0\Big(\sqrt{k_0^2+|\vec k-\vec p_F^+|^2}\; \Big)$ and
$\chi_0\Big(\sqrt{k_0^2+|\vec k-\vec p_F^-|^2}\;\Big)$ are disjoint (here $|\cdot|$
is the euclidean norm over $\RRR^2/\L^*$). We define
\be f_{u.v.}(\kk)=1-
\chi_0\Big(\sqrt{k_0^2+|\vec k-\vec p_F^+|^2}\; \Big)
-\chi_0\Big(
\sqrt{k_0^2+|\vec k-\vec p_F^-|^2}\;\Big)\label{2.10}\ee
and $f_{i.r.}(\kk)=1-f_{u.v.}(\kk)$, 
so that we can rewrite $\hat g_\kk$ as:
\be \hat g_\kk=f_{u.v.}(\kk)\hat g_\kk+ f_{i.r.}(\kk)\hat g_\kk\defin
\hat g^{(u.v.)}(\kk)+\hat g^{(i.r.)}(\kk)\;.\label{2.11}\ee
We now introduce two independent set of Grassmann fields $\{
\psi^{(u.v.)\pm}_{\kk,\s,\r}\}$ and $\{
\psi^{(i.r.)\pm}_{\kk,\s,\r}\}$, with $\kk\in\BBB^{(M)}_{\b,L}$,
$\s=\uparrow\downarrow$, $\r=1,2$,
and the Gaussian integrations $P(d\psi^{(u.v.)})$ and
$P(d\psi^{(i.r.)})$ defined by
\bea &&\int P(d \psi^{(u.v.)}) \hat \psi^{(u.v.)-}_{\kk_1,\s_1}\hat
\psi^{(u.v.)+}_{ \kk_2,\s_2} = \b|\L|\d_{\s_1,\s_2}\d_{\kk_1,\kk_2}
\hat g^{(u.v.)}(\kk_1)\;,\nn\\
&&\int P(d \psi^{(i.r.)}) \hat \psi^{(u.v.)-}_{\kk_1,\s_1}\hat
\psi^{(i.r.)+}_{ \kk_2,\s_2} = \b|\L|\d_{\s_1,\s_2}\d_{\kk_1,\kk_2}
\hat g^{(i.r.)}(\kk_1)\;.\label{2.12}\eea
Similarly to $P_M(d\psi)$, the Gaussian integrations $P(d
\psi^{(u.v.)})$, $P(d \psi^{(i.r.)})$ also admit an explicit
representation analogous to (\ref{2.3}), with $\hat g_\kk$ replaced
by $\hat g^{(u.v.)}(\kk)$ or $\hat g^{(i.r.)}(\kk)$ and the sum
over $\kk$ restricted to the values in the support of $f_{u.v.}(\kk)$ or
$f_{i.r.}(\kk)$, respectively. The definition of Grassmann integration
implies the following identity (``addition principle''):
\be \int P(d\psi)e^{-\VV(\psi)}=\int P(d\psi^{(i.r.)})\int P(d\psi^{(u.v.)})
e^{-\VV(\psi^{(i.r.)}+\psi^{(u.v.)})}\label{2.13}\ee
so that we can rewrite the partition function as
\bea \Xi_{M,\b,\L}&=& e^{-\b|\L| F_{\b,\L}^{(M)}}=
\int P(d\psi^{(i.r.)})\exp\,\big\{\, \sum_{n\ge
1}\frac{1}{n!}\EE^T_{u.v.}(-\VV(\psi^{(i.r.)}+\cdot);n)\big\}\=\nn\\
&\=&
e^{-\b |\L| F_{0,M}}\int P(d\psi^{(i.r.)}) e^{-{\cal V}_0(\psi^{(i.r.)})}\;,
\label{2.14}\eea
where the {\it truncated expectation} $\EE^T_{u.v.}$ is defined, given
any polynomial $V_1(\psi^{(u.v.)})$ with coefficients depending on
$\psi^{(i.r.)}$, as
\be \EE^T_{u.v.}(V_1(\cdot);n)=\frac{\dpr^n}{\dpr\l^n}
\log\int P(d\psi^{(u.v.)})e^{\l
V_1(\psi^{(u.v.)})}\Big|_{\l=0}\label{2.15}\ee
and ${\cal V}_0$ is fixed by the condition ${\cal V}_0(0)=0$.
We will prove below that ${\cal V}_0$ can be written as
\bea && {\cal V}_0(\psi)=\sum_{n=1}^\io (\b|\L|)^{-2n}
\sum_{\s_1,\ldots,\s_n=\uparrow
\downarrow}\sum_{\r_1,\ldots,\r_{2n}=1,2}\sum_{\kk_1,\ldots,\kk_{2n}}
\Big[\prod_{j=1}^n\hat \psi^{(i.r.)+}_{\kk_{2j-1},\s_j,\r_{2j-1}}
\hat \psi^{(i.r.)-}_{\kk_{2j},\s_j,\r_{2j}}\Big]\;\cdot\nn\\
&&\hskip5.truecm\cdot\hat W_{M,2n,\ul\r}(\kk_1,\ldots,
\kk_{2n-1})\;\d(\sum_{j=1}^n(\kk_{2j-1}-\kk_{2j}))\;,\label{2.16}\eea
where $\ul\r=(\r_1,\ldots,\r_{2n})$ and we used the notation
\be \d(\kk)=\d(\vec k)\d(k_0)\;,\qquad \d(\vec k)=|\L|\sum_{n_1,n_2\in\zzz}
\d_{\vec k,n_1\vec G_1+n_2\vec G_2}\;,\qquad \d(k_0)=\b\d_{k_0,0}\;,
\label{2.16a}\ee
with $\vec G_1,\vec G_2$ a basis of $\L^*$.
The possibility of representing ${\cal V}_M$ in the form
(\ref{2.16}), with the {\it kernels} $\hat W_{M,2n,\ul\r}$ independent of the
spin indices $\s_i$, follows from a number of remarkable symmetries, discussed 
in Appendix \ref{app2}, see in particular symmetries (1)--(3) in Lemma \ref{lem2.4}. 
The regularity properties of the kernels are summarized in the 
following Lemma, which will be proved below. 
\\

{\bf Lemma \ref{sec6}.1} {\it 
The constant $F_{0,M}$ in (\ref{2.14}) and the kernels $\hat W_{M,2n,\ul\r}$ in
(\ref{2.16}) are
given by power series in $U$, convergent
in the complex disc $|U|\le U_0$, for $U_0$ small enough and independent of
$M,\b,\L$; after Fourier
transform, the $\xx$-space counterparts of the kernels $\hat W_{M,2n,\ul\r}$
satisfy the following bounds:
\be \int d\xx_1\cdots d\xx_{2n}\big|W_{M,2n,\ul\r}(\xx_1,\ldots,\xx_{2n})\big|
\le \b|\L|C^n |U|^{\max\{1,n-1\}}\label{2.17}\ee
for some constant $C>0$.
Moreover, the limits $F_0=\lim_{M\to\io}F_{0,M}$ and 
$W_{2n,\ul\r}
(\xx_1,\ldots,\xx_{2n})=\lim_{M\to\io}W_{M,2n,\ul\r}
(\xx_1,\ldots,\xx_{2n})$ exist and are reached uniformly in $M$, so that, 
in particular, the limiting functions are analytic in the same domain $|U|\le 
U_0$ and so are their $\b,|\L|\to\infty$ limits 
(that, with some abuse of notation, we shall denote by the same symbols).}\\

{\bf Remark.}
Once that the ultraviolet degrees of freedom have been integrated out, the 
remaining infrared problem (i.e., the computation of the Grassmann integral 
in the second line of Eq.(\ref{2.14})) is essentially independent of $M$, 
given the fact that the limit $W_{2n,\ul\r}$ of the kernles $W_{M,2n,\ul\r}$
is reached uniformly and that 
the limiting kernels are analytic and satisfy the same bounds as Eq.(\ref{2.17}). 
For this reason, in the infrared integration described in the next two 
sections, $M$ will not play any essential role and, whenever possible, we will 
simplify the notation by dropping the label $M$.
\\

Before we present the proof of Lemma \ref{sec6}.1, let us note that  the kernels 
$W_{M,2n,\ul\r}$ satisfy a number of non-trivial invariance properties. We will be particularly interested in the invariance
properties of the quadratic part $\hat W_{M,2,(\r_1,\r_2)}(\kk_1,\kk_2)$, which will
be used below to show that the structure of the
quadratic part of the new effective interaction has the same symmetries
as the free integration. The crucial properties that we will need are summarized in the following 
Lemma, which is proved in Appendix \ref{app2}.\\

{\bf Lemma \ref{sec6}.2.} {\it The kernel $\hat W_{2}(\kk):=\hat W_{M,2}(\kk,\kk)$, thought as a $2\times2$ matrix with components $ \hat W_{M,2,(i,j)}(\kk,\kk)$, 
satisfies the following symmetry properties:
\bea \hat W_{2}(\kk)&=&e^{i\vec k(\vec \d_1-\vec\d_2)\frac{\s_3}{2}} \hat W_{2}\big((k_0,
e^{i\frac{2\p}{3}\s_2}\vec k)\big)e^{-i\vec k(\vec \d_1-\vec\d_2)\frac{\s_3}{2}}= \hat W_{2}^*(-\kk)=
\hat W_{2}\big((k_0,k_1,-k_2)\big)\nn\\
&=&\s_1 \hat W_{2}\big((k_0,-k_1,k_2)\big)\s_1
=  \hat W_{2}^T\big((k_0,-\vec k)\big)=-\s_3 \hat W_{2}\big(
(-k_0,\vec k)\big)\s_3\;, \label{sym}\eea
where  $\s_1,\s_2,\s_3$ are the standard Pauli 
matrices. In particular, if $\hat W(\kk)=\lim_{\b,|\L|\to\infty}\hat W_{2}(\kk)$, 
in the vicinity of the Fermi point $\pp_F^\o=(0,\vec p_F^\o)$, with $\o=\pm$,
\be \hat W(\kk)=-\begin{pmatrix} iz_0k_0 & \d_0 \O^*(\vec k) \\
\d_0\O(\vec k) & iz_0k_0\end{pmatrix}+O(|\kk-\pp_F^\o|^2)\;,\label{sym2}\ee
for some real constants $z_0,\d_0$.}

\vskip.4truecm
Note that Eq.(\ref{sym2}) can be read by saying that, in the zero temperature
and thermodynamic limits,  the two-legged kernel has the same structure as 
the inverse of the free covariance, $\hat S_0^{-1}(\kk)$, modulo higher order terms in 
$\kk-\pp_F^\o$. This fact will be used in the next section to define a dressed infrared 
propagator $[\hat S_0^{-1}(\kk)+\hat W(\kk)]^{-1}$, with the same infrared singularity structure as the free one. We will come back to this point in more detail. For the moment, let us turn to the proof of Lemma \ref{sec6}.1, which illustrates the main RG strategy 
that will be also used below, in the more difficult infrared integration. \\

\begin{figure}[ht]
\includegraphics[height=.25\textwidth]{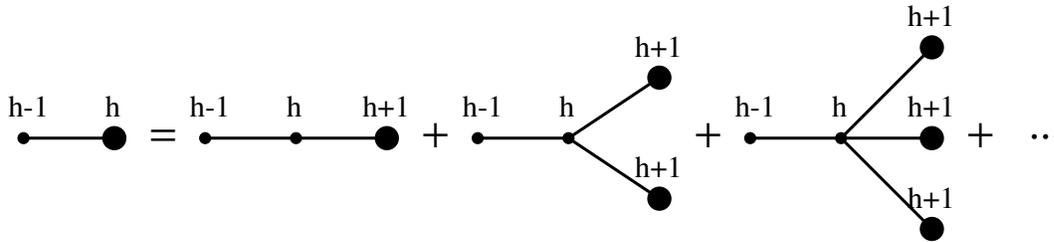}
\caption{The graphical representation of $\VV^{(h-1)}$.\label{fig6.1}}
\end{figure}

\begin{figure}[ht]
\includegraphics[height=.5\textwidth]{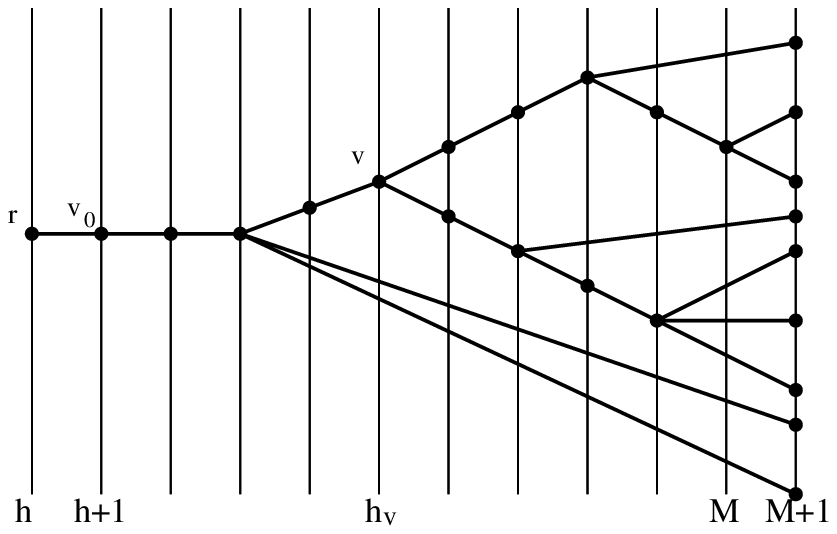}
\caption{A tree $\t\in\widetilde\TT_{M;h,n}$ with $n=9$: 
the root is on scale $h$ and the 
endpoints are on scale $M+1$.\label{fig6.2}}
\end{figure}

{\bf Proof of Lemma \ref{sec6}.1.}
Let us rewrite 
the Fourier transform of $\hat g^{(u.v.)}(\kk)$ as
\be g^{(u.v.)}(\xx)=\sum_{h=1}^{M} g^{(h)}(\xx)\;,\label{B.1}\ee
where
\be g^{(h)}(\xx)=\frac{1}{\b|\L|}\sum_{\kk\in\BBB^{(M)}_{\b,L}} 
 e^{-i\kk\xx}\,\frac{f_{u.v.}(\kk)H_h(k_0)}{k_0^2+v_0^2
|\O(\vec k)|^2}\begin{pmatrix}i k_0 & -v_0\O^*(\vec k) \\ -v_0\O(\vec k) &
ik_0\end{pmatrix}\;,\label{B.2}\ee
with $H_1(k_0)=\c_0(2^{-1}|k_0|)$ and, if $h\ge 2$, $H_h(k_0)=
\c_0(2^{-h}|k_0|)-
\c_0(2^{-h+1}|k_0|)$. Note that $[g^{(h)}({\bf 0})]_{\r\r}=0$, $\r=1,2$,  
and, for any integer $K\ge 0$, $g^{(h)}(\xx)$ satisfies the bound
\be ||g^{(h)}(\xx)||\le \frac{C_K}{1+ (2^h |x_0|_\b+|\vec x|_\L)^K}
\;,\label{B.3}\ee
where  $|x_0|_\b=\min_{n\in\zzzz} |x_0+n\b|$ 
is the distance over the one-dimensional
torus of length $\b$ and
$|\vec x|_\L=\min_{\vec \ell\in\bbb}|\vec x+L\vec \ell|$ is the distance over the periodic lattice $\L$ 
(here $\BBBB$ is the triangular lattice defined after Eq.(\ref{1.1}));
see Appendix \ref{app0} for a proof. Moreover, 
$g^{(h)}(\xx)$ admits a Gram representation that, in notation analogous to Eq.(\ref{s5.8}), reads
\be g^{(h)}_{\r,\r'}(\xx-\yy)= \media{{\bf A}^{(h)}_{\xx,\r}, {\bf B}^{(h)}_{\yy,\r'}}\;,\label{B.3W}\ee
with
\bea && [{\bf A}^{(h)}_{\xx,\r}(\zz)]_i=\frac{1}{\b|\L|}\sum_{\kk\in\BBB^{(M)}_{\b,L}} \sqrt{f_{u.v.}(\kk)
H_h(k_0)}\frac{e^{-i\kk(\zz-\xx)}}{[k_0^2+v_0^2|\O(\vec k)|^2]^{1/4}}\d_{\r,i}\;,\label{B.4a}\\
&&[{\bf B}^{(h)}_{\xx,\r}(\zz)]_i=\frac{1}{\b|\L|}
\sum_{\kk\in\BBB^{(M)}_{\b,L}} \,\frac{\sqrt{f_{u.v.}(\kk)
H_h(k_0)}\,e^{-i\kk(\zz-\xx)}}{[k_0^2+v_0^2|\O(\vec k)|^2]^{3/4}}
\begin{pmatrix}i k_0 & -v_0\O^*(\vec k)\\
-v_0\O(\vec k)& i k_0\end{pmatrix}_{i,\r}\;,\nn\eea
and
\be ||{\bf A}^{(h)}_{\xx,\r}||^2=
||{\bf B}^{(h)}_{\xx,\r}||^2= \frac{1}{\b|\L|}\sum_{\kk\in\BBB^{(M)}_{\b,L}}
\frac{f_{u.v.}(\kk) H_h(k_0)}{[k_0^2+v_0^2|\O(\vec k)|^2]^{1/2}}
\le C\;,\label{B.5a}\ee
for a suitable constant $C$. Our goal is to compute
\bea
e^{-\b |\L|F_{0,M}-\VV_0(\psi^{(i.r)})}&=&\int P(d\psi^{[1,M]})
e^{-\VV(\psi^{(i.r.)}+\psi^{[1,M]})}\nn\\
&=& \exp\Big\{\log\int P(d\psi^{[1,M]})
e^{-\VV(\psi^{(i.r.)}+\psi^{[1,M]})}
\Big\}\;,\label{B.6}\eea
where $P(d\psi^{[1,M]})$ is the fermionic ``Gaussian integration''
associated with the propagator $\sum_{h=1}^{M}\hat g^{(h)}(\kk)$
(i.e., it is the same as $P(d\psi^{(u.v.)})$); moreover, we want to prove that 
$F_{0,M}$ and $\VV_0(\psi^{(i.r)})$ are uniformly convergent as $M\to\io$.
We perform the integration of (\ref{B.6}) in an iterative fashion: in fact,
we will inductively prove that for $1\le h\le M$, 
\be e^{-\b |\L|F_{0,M}-\VV_0(\psi^{(i.r)})}=e^{-\b|\L|F_h}\int P(d\psi^{[1,h]})
e^{-\VV^{(h)}(\psi^{(i.r.)}+\psi^{[1,h]})}\label{B.7}\ee
where $P(d\psi^{[1,h]})$  is the fermionic ``Gaussian integration''
associated with the propagator $\sum_{k=1}^{h}\hat g^{(k)}(\kk)$
and $\VV^{(M)}=\VV$; for $1\le h<M$,
\bea&& {\cal V}^{(h)}(\psi)
=\label{B.8}\\
&&\qquad=\sum_{n=1}^\io \sum_{\ul\r,\ul\s}\int\,d\xx_1\cdots d\xx_{2n}
\Big[\prod_{j=1}^n \psi^{+}_{\xx_{2j-1},\s_j,\r_{2j-1}}
\psi^{-}_{\xx_{2j},\s_j,\r_{2j}}\Big]
W_{M,2n,\ul\r}^{(h)}(\xx_1,\ldots,
\xx_{2n})\;.\nn\eea
In order to inductively prove (\ref{B.7})-(\ref{B.8})
we use the addition principle to rewrite
\be \int P(d\psi^{[1,h]})
e^{-\VV^{(h)}(\psi^{(i.r.)}+\psi^{[1,h]})}=\!\!
\int P(d\psi^{[1,h-1]})\!\!\int P(d\psi^{(h)})
e^{-\VV^{(h)}(\psi^{(i.r.)}+\psi^{[1,h-1]}+\psi^{(h)})}\label{B.9}\ee
where $P(d\psi^{(h)})$ is the fermionic Gaussian integration with
propagator $\hat g^{(h)}(\kk)$. After the integration of $\psi^{(h)}$
we define
\be e^{-\VV^{(h-1)}(\psi^{(i.r.)}+\psi^{[1,h-1]})
-\b|\L|\lis e_h}=\int P(d\psi^{(h)})
e^{-\VV^{(h)}(\psi^{(i.r.)}+\psi^{[1,h-1]}+\psi^{(h)})}\;,\label{B.10}\ee
which proves (\ref{B.7}) with 
\be F_h=\sum_{k=h+1}^M\lis e_k\;.\label{B.9a}\ee
Let $\EE^T_h$ be the truncated expectation associated 
to $P(d\psi^{(h)})$: then we have
\be \lis e_h+{\cal V}^{(h-1)}(\psi)=
\sum_{n\ge 1}\frac{1}{n!}(-1)^{n+1}\EE^T_h({\cal V}^{(h)}\big(
\psi+\psi^{(h)}\big);n)
\;.\label{B.11}\ee
Eq.(\ref{B.11}) can be graphically represented as in Fig.\ref{fig6.1}.
The tree in the l.h.s., consisting of a single horizontal branch, connecting the 
left node 
(called the {\it root} and associated to the {\it scale label} $h-1$)
with a big black dot on scale $h$, represents ${\cal V}^{(h-1)}(\psi)$. In the r.h.s., 
the term with $n$ final points represents the corresponding term in 
the r.h.s. of Eq.(\ref{B.11}): a scale label 
$h-1$ is attached to the leftmost node (the root); a scale label 
$h$ is attached to the central node (corresponding to the action of $\EE^T_h$);
a scale label $h+1$ is attached to the $n$ rightmost nodes with the big black dots 
(representing $\VV^{(h)}$).  Iterating the graphical equation in Fig.\ref{fig6.1} up to scale 
$M$, and representing the endpoints on scale $M+1$ as simple dots (rather than big black dots),
we end up with a graphical representation of $\VV^{(h)}$ in terms of {\it Gallavotti-Nicol\`o}
trees, see Fig.\ref{fig6.2}, defined in terms of the following features.
\begin{enumerate}
\item Let us consider the family of all trees which can be constructed
by joining a point $r$, the {\it root}, with an ordered set of $n\ge 1$
points, the {\it endpoints} of the {\it unlabeled tree},
so that $r$ is not a branching point. $n$ will be called the
{\it order} of the unlabeled tree and the branching points will be called
the {\it non trivial vertices}.
The unlabeled trees are partially ordered from the root
to the endpoints in the natural way; we shall use the symbol $<$
to denote the partial order.
Two unlabeled trees are identified if they can be superposed by a suitable
continuous deformation, so that the endpoints with the same index coincide.
It is then easy to see that the number of unlabeled trees with $n$ end-points
is bounded by $4^n$ (see, e.g., \cite[Appendix A.1.2]{GeM} for a proof of this fact).
We shall also consider the {\it labelled trees} (to be called
simply trees in the following); they are defined by associating
some labels with the unlabelled trees, as explained in the
following items.
\item We associate a label $0\le h\le M-1$ with the root and we denote by
$\widetilde\TT_{M;h,n}$ the corresponding set of labeled trees with $n$
endpoints. Moreover, we introduce a family of vertical lines,
labeled by an integer taking values in $[h,M+1]$, and we represent
any tree $\t\in\widetilde\TT_{M;h,n}$ so that, if $v$ is an endpoint, it is contained 
in the vertical line with index $h_v=M+1$, while if it is a non trivial vertex, 
it is contained in a vertical line with index
$h<h_v\le M$, to be called the {\it scale} of $v$; the root $r$ is on
the line with index $h$.
In general, the tree will intersect the vertical lines in set of
points different from the root, the endpoints and the branching
points; these points will be called {\it trivial vertices}.
The set of the {\it
vertices} will be the union of the endpoints, of the trivial
vertices and of the non trivial vertices; note that the root is not a vertex.
Every vertex $v$ of a
tree will be associated to its scale label $h_v$, defined, as
above, as the label of the vertical line whom $v$ belongs to. Note
that, if $v_1$ and $v_2$ are two vertices and $v_1<v_2$, then
$h_{v_1}<h_{v_2}$.
\item There is only one vertex immediately following
the root, called $v_0$ and with scale label equal to $h+1$.
\item Given a vertex $v$ of $\t\in\widetilde\TT_{M;h,n}$ that is not an endpoint,
we can consider the subtrees of $\t$ with root $v$, which correspond to the
connected components of the restriction of
$\t$ to the vertices $w\ge v$. If a subtree with root $v$ contains only
$v$ and one endpoint on scale $h_v+1$,
it will be called a {\it trivial subtree}.
\item With each endpoint $v$ we associate a factor $\VV(\psi^{(i.r.)}+\psi^{[1,M]})$
and a set $\xx_v$ of space-time
points (the corresponding integration variables in the $\xx$-space
representation).
\item We introduce a {\it field label} $f$ to distinguish the field variables
appearing in the factors $\VV(\psi^{(i.r.)}+\psi^{[1,M]})$ associated with the endpoints;
the set of field labels associated with the endpoint $v$ will be called $I_v$;
note that if $v$ is an endpoint $|I_v|=4$. Analogously, if $v$ is not an endpoint, we shall
call $I_v$ the set of field labels associated with the endpoints following
the vertex $v$; $\xx(f)$, $\e(f)$, $\s(f)$ and $\r(f)$ will denote the
space-time point, the $\e$ index, the $\s$ index and the $\r$ index, 
respectively, of the Grassmann field variable with label $f$.
\end{enumerate}

In terms of trees, the effective potential ${\cal V}^{(h)}$,
$0\le h\le M$ (with $\VV^{(0)}(\psi^{(i.r.)})$ identified with
$\VV_0(\psi^{(i.r.)})$), can be written as
\be {\cal V}^{(h)}(\psi^{(i.r.)}+\psi^{[1,h]}) + \b|\L| \lis e_{h+1}=
\sum_{n=1}^\io\sum_{\t\in\widetilde\TT_{M;h,n}}
\widetilde{\cal V}^{(h)}(\t,\psi^{(i.r.)}+\psi^{[1,h]})\;,\label{B.12}\ee
where, if $v_0$ is the first vertex of $\t$ and $\t_1,\ldots,\t_s$
($s=s_{v_0}$) are the subtrees of $\t$ with root $v_0$,
$\widetilde{\cal V}^{(h)}(\t,\psi^{(i.r.)}+\psi^{[1,h]})$
is defined inductively as:
\be \widetilde{\cal V}^{(h)}(\t,\psi^{[0,h]})=\frac{(-1)^{s+1}}{s!} \EE^T_{h+1}
\big[\widetilde{\cal V}^{(h+1)}(\t_1,\psi^{[0, h+1]});\ldots;\widetilde{\cal V}^{(h+1)}
(\t_{s},\psi^{[0,h+1]})\big]\;.\label{B.12a}\ee
where $\psi^{[0,h]}:=\psi^{(i.r.)}+\psi^{[1,h]}$ and, if $\t$ is a trivial subtree with root on scale $M$, then $\widetilde{\cal V}^{(M)}(\t,\psi^{[0,M]})={\cal V}(\psi^{[0,M]})$.

\begin{figure}[ht]
\includegraphics[height=.5\textwidth]{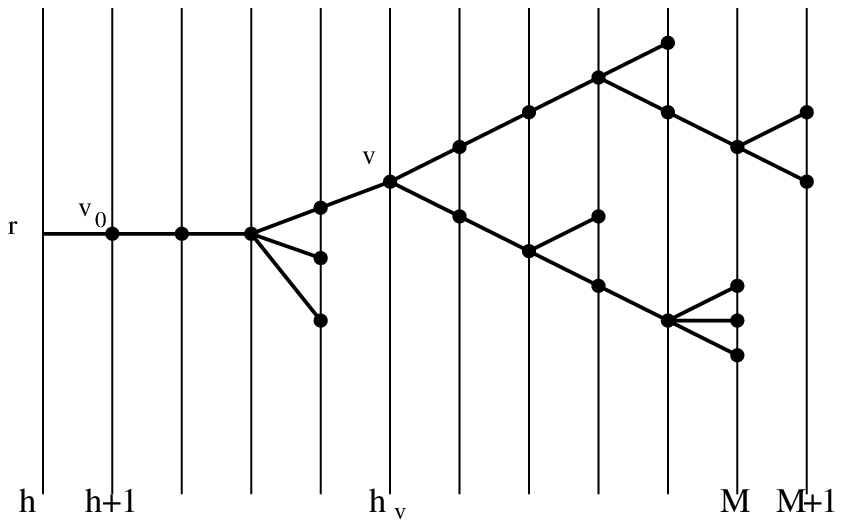}
\caption{A tree $\t\in\TT_{M;h,n}$ with $n=9$: the root is on scale $h$ and the 
endpoints are on scales $\le M+1$.\label{fig6.3}}
\end{figure}

For what follows, it is important to specify the action of the truncated expectations on the 
branches connecting any endpoint $v$ to the closest {\it non-trivial} vertex $v'$
preceding it. In fact, if $\t$ has only one end-point, 
it is convenient to rewrite $\widetilde\VV^{(h)}(\t,\psi^{[0,h]})=\EE^T_{h+1}\EE^T_{h+2}\cdots
\EE^T_M(\VV(\psi^{[0,M]}))$ as: 
\be \widetilde\VV^{(h)}(\t,\psi^{[0,h]})=\VV(\psi^{[0,h]})+
\EE^T_{h+1}\cdots\EE^T_M\big(\VV(\psi^{[0,M]})-
\VV(\psi^{[0,h]})\big)\;.\label{B.12b}\ee
Now, the key observation is that, since $\VV(\psi)$ is defined as in Eq.(\ref{s4.6}) and
since our explicit choice of the ultraviolet cutoff makes the tadpoles equal to zero 
(i.e., $[g^{(h)}(\V0)]_{\r\r}=0$), then the second term in the r.h.s. of Eq.(\ref{B.12b})
is identically zero:
\be \EE^T_{h+1}\cdots\EE^T_M\big(\VV(\psi^{[0,M]})-
\VV(\psi^{[0,h]})\big)=0\;,\label{B.12by}\ee
for all $0\le h<M$. Therefore, it is natural to shrink all the branches of $\t\in\widetilde\TT_{M;h,n}$
consisting of a subtree $\t'\subseteq\t$, having root $r'$ on scale $h'\in[h,M]$ 
and only one endpoint on scale $M+1$,  into a trivial subtree, rooted in $r'$ and associated 
to the factor $\VV(\psi^{[0,h']})$. By doing so, we end up with an alternative representation of the
effective potentials, which is based
on a slightly modified tree expansion. The set of modified trees with $n$ endpoints contributing 
to $\VV^{(h)}$ will be denoted by $\TT_{M;h,n}$; every $\t\in\TT_{M;h,n}$ is characterized
in the same way as the elements of $\widetilde\TT_{M;h,n}$, but for two features: 
(i) the endpoints of $\t\in\TT_{M;h,n}$ are not necessarily on scale $M+1$; (ii) every endpoint $v$ 
of $\t$ is attached to a non-trivial vertex on scale $h_v-1$ and is 
associated to the factor $\VV(\psi^{[0,h_{v}-1]})$. See Fig.\ref{fig6.3}.
In terms of these modified trees, we have:
\be {\cal V}^{(h)}(\psi^{[0,h]}) + \b|\L| \lis e_{h+1}=
\sum_{n=1}^\io\sum_{\t\in\TT_{M;h,n}}{\cal V}^{(h)}(\t,\psi^{[0,h]})\;,\label{B.12az}\ee
where
\be {\cal V}^{(h)}(\t,\psi^{[0,h]})=\frac{(-1)^{s+1}}{s!} \EE^T_{h+1}
\big[{\cal V}^{(h+1)}(\t_1,\psi^{[0, h+1]});\ldots;{\cal V}^{(h+1)}
(\t_{s},\psi^{[0,h+1]})\big]\label{B.13}\ee
and, if $\t$ is a trivial subtree with root on scale $k\in[h,M]$, then 
${\cal V}^{(k)}(\t,\psi^{[0,k]})={\cal V}(\psi^{[0,k]})$.

Using its inductive definition Eq.(\ref{B.13}), the right hand side of Eq.(\ref{B.12az}) can be
further expanded (it is a sum of several contributions, differing for the choices of the 
field labels contracted under the action of the truncated expectations $\EE^T_{h_v}$ 
associated with the vertices $v$ that are not endpoints), and in order to describe the resulting 
expansion we need some more definitions (allowing us to distinguish the fields that are 
contracted or not ``inside the vertex $v$").

We associate with any vertex $v$ of the tree a subset $P_v$ of $I_v$,
the {\it external fields} of $v$. These subsets must satisfy various
constraints. First of all, if $v$ is not an endpoint and $v_1,\ldots,v_{s_v}$
are the $s_v\ge 1$ vertices immediately following it, then
$P_v \subseteq \cup_i
P_{v_i}$; if $v$ is an endpoint, $P_v=I_v$.
If $v$ is not an endpoint, we shall denote by $Q_{v_i}$ the
intersection of $P_v$ and $P_{v_i}$; this definition implies that $P_v=\cup_i
Q_{v_i}$. The union ${\cal I}_v$ of the subsets $P_{v_i}\setminus Q_{v_i}$
is, by definition, the set of the {\it internal fields} of $v$,
and is non empty if $s_v>1$.
Given $\t\in\TT_{M;h,n}$, there are many possible choices of the
subsets $P_v$, $v\in\t$, compatible with all the constraints. We
shall denote by ${\cal P}_\t$ the family of all these choices and by ${\bf P}$
the elements of ${\cal P}_\t$. With these definitions, we can rewrite
${\cal V}^{(h)}(\t,\psi^{[0,h]})$ as:
\bea &&{\cal V}^{(h)}(\t,\psi^{[0,h]})=\sum_{{\bf P}\in{\cal P}_\t}
{\cal V}^{(h)}(\t,\PP)\;,\nn\\
&&{\cal V}^{(h)}(\t,\PP)=\int d\xx_{v_0}\psi^{[0,h]}_{P_{v_0}}
K_{\t,\PP}^{(h+1)}(\xx_{v_0})\;,\label{2.43}\eea
where $\xx_{v}=\cup_{f\in I_v}\{\xx_v\}$,
\be\psi^{[0,h]}_{P_{v}}=\prod_{f\in P_v}\psi^{[0,h]\,\e(f)}_{\xx(f),\s(f),\r(f)}\label{2.44}\ee
and $K_{\t,\PP}^{(h+1)}(\xx_{v_0})$ is defined inductively by
the equation, valid for any $v\in\t$ that is not an endpoint,
\be K_{\t,\PP}^{(h_v)}(\xx_v)=\frac{1}{s_v !}
\prod_{i=1}^{s_v} [K^{(h_v+1)}_{v_i}(\xx_{v_i})]\; \;\EE^T_{h_v}[\psi^{(h_v)}_{P_{v_1}\setminus 
Q_{v_1}},\ldots,\psi^{(h_v)}_{P_{v_{s_v}}\setminus
Q_{v_{s_v}}}]\;,\label{2.45}\ee
where $\psi^{(h_v)}_{P_{v_i}\setminus Q_{v_i}}$ has a definition
similar to Eq.(\ref{2.44}). Moreover, if $v_i$ is an endpoint
$K^{(h_v+1)}_{v_i}(\xx_{v_i})=U$; if $v_i$ is not an
endpoint, $K_{v_i}^{(h_v+1)}=K_{\t_i,\PP_i}^{(h_v+1)}$, 
where ${\bf P}_i=\{P_w, w\in\t_i\}$. Using in the r.h.s. of Eq.(\ref{2.45}) the
determinant representation of the truncated expectation discussed in the previous section,
we finally get:
\be {\cal V}^{(h)}(\t,\PP) = \sum_{T\in {\bf T}} \int d\xx_{v_0}
\psi^{[0,h]}_{P_{v_0}} W_{\t,\PP,T}^{(h)}(\xx_{v_0})
\= \sum_{T\in {\bf T}}
{\cal V}^{(h)}(\t,\PP,T)\;,\label{B.14}\ee
where
\bea&&
W_{\t,\PP, T}(\xx_{v_0}) =\label{B.16}\\
&&=U^n \Bigg\{\prod_{\substack{v\ {\rm not}\\ {\rm e.p.}}}\frac{1}{s_v!} \int
dP_{T_v}({\bf t}_v)\;{\rm det}\, G^{h_v,T_v}({\bf t}_v)\Biggl[
\prod_{\ell\in T_v} \d_{\s(\ell),\s'(\ell)}\,g_{\r(\ell),\r'(\ell)}^{(h_v)}(\xx(\ell)-\xx'(\ell))\,\Biggr]
\Bigg\}\nn\eea
and $G^{h_v,T_v}({\bf t}_v)$ is a matrix analogous to the one defined in previous section,
with $g$ replaced by $g^{(h)}$. Note that $W_{\t,\PP, T}$ and, therefore, $\VV^{(h)}(\t,\PP)$
do not depend on $M$: therefore, the effective potential $\VV^{(h)}(\psi)$ depends on $M$ only 
through the choice of the scale labels (i.e., the dependence on $M$ is all
encoded in $\TT_{M;h,n}$). Using Eqs.(\ref{B.14})-(\ref{B.16}), we finally get the bound:
\bea&& \frac1{\b|\L|}\int d\xx_1\cdots d\xx_{2l} |W^{(h)}_{M,2l,\ul\r}(\xx_1,
\ldots,\xx_{2l})|\le
\sum_{n\ge \max\{1,l-1\}}|U|^n\sum_{\t\in {\cal T}_{M;h,n}}
\sum_{\substack{\PP\in{\cal P}_\t\\ |P_{v_0}|=2l}}\sum_{T\in{\bf T}}\cdot
\nn\\
&&\cdot\int\prod_{\ell\in T}
d(\xx(\ell)-\xx'(\ell))\Bigg[\prod_{\substack{v\ {\rm not}\\ {\rm e.p.}}}\frac{1}{s_v!}
\max_{{\bf t}_v}\big|{\rm det}\, G^{h_v,T_v}({\bf t}_v)\big|
\prod_{\ell\in T_v}
\big|\big|g^{(h_v)}(\xx(\ell)-\xx'(\ell))\big|\big|\Bigg]\;.\nn\\&&\label{B.17}\eea
Now, an application of the Gram--Hadamard inequality Eq.(\ref{s5.6}), combined with
the representation Eq.(\ref{B.3W}) and the dimensional bounds Eq.(\ref{B.5a}), implies that
\be |{\rm det} G^{h_v,T_v}({\bf t}_v)| \le
(\const.)^{\sum_{i=1}^{s_v}|P_{v_i}|-|P_v|-2(s_v-1)}\;.\label{2.54}\ee
By the decay properties of $g^{(h)}(\xx)$ given by Eq.(\ref{B.3}), it also follows that
\be \prod_{\substack{v\ {\rm not}\\ {\rm e.p.}}}
\frac{1}{ s_v!}\int \prod_{\ell\in T_v} d(\xx(\ell)-\xx'(\ell))\,
||g^{(h_v)}(\xx(\ell)-\xx'(\ell))||\le (\const.)^n \prod_{\substack{v\ {\rm not}\\ {\rm e.p.}}}
\frac{1}{s_v!} 2^{-h_v(s_v-1)}\;.\label{2.55}\ee
Plugging Eqs.(\ref{2.54})-(\ref{2.55}) into Eq.(\ref{B.17}), we find that the l.h.s. of Eq.(\ref{B.17})
can be bounded from above by 
\be \sum_{n\ge \max\{1,l-1\}}\sum_{\t\in {\cal T}_{M;h,n}}
\sum_{\substack{\PP\in{\cal P}_\t\\ |P_{v_0}|=2l}}\sum_{T\in{\bf T}}
(\const.)^n|U|^n \Big[\prod_{\substack{v\ {\rm not}\\ {\rm e.p.}}} \frac{1}{s_v!}
2^{-h_v(s_v-1)}\Big]\;.\label{B.18}\ee
Using the following relation, which can be easily proved by induction,
\be\sum_{\substack{v\ {\rm not}\\ {\rm e.p.}}}h_v(s_v-1)=h(n-1)+
\sum_{\substack{v\ {\rm not}\\ {\rm e.p.}}}(h_v-h_{v'})(n(v)-1)\;,\label{B.18rt}\ee
where $v'$ is the vertex immediately preceding $v$ on $\t$
and $n(v)$ the number of endpoints following $v$ on $\t$, 
we find that Eq.(\ref{B.18}) can be rewritten as
\be \sum_{n\ge \max\{1,l-1\}}\sum_{\t\in {\cal T}_{M;h,n}}
\sum_{\substack{\PP\in{\cal P}_\t\\ |P_{v_0}|=2l}}\sum_{T\in{\bf T}}
(\const.)^n|U|^n 2^{-h(n-1)}\Big[\prod_{\substack{v\ {\rm not}\\ {\rm e.p.}}}
\frac{1}{s_v!}2^{-(h_v-h_{v'})(n(v)-1)}\Big]\;,
\label{B.18a}\ee
where, by construction, we have that $n(v)>1$ for any vertex $v$ of $\t\in\TT_{M;h,n}$
that is not an endpoint (simply because every endpoint $v$ 
of $\t$ is attached to a non-trivial vertex on scale $h_v-1$, see the discussion after 
Eq.(\ref{B.12by})). Now, the number of terms in $\sum_{T\in {\bf T}}$ can be bounded by
$(\const.)^n\prod_{v\ {\rm not}\ {\rm
e.p.}} s_v!$ (see \cite[Appendix A.3.3]{GeM}); moreover, $|P_v|\le 4n(v)$ and $n(v)-1\ge \max\{1,\frac{n(v)}2\}$, so that 
$n(v)-1\ge \frac12+\frac{|P_v|}{16}$. Therefore,
\bea&&  \frac1{\b|\L|}\int d\xx_1\cdots d\xx_{2l} |W^{(h)}_{M,2l,\ul\r}(\xx_1,
\ldots,\xx_{2l})|\le 
\sum_{n\ge \max\{1,l-1\}}(\const.)^n|U|^n2^{-h(n-1)}\cdot\nn\\
&&\cdot\sum_{\t\in {\cal T}_{M;h,n}} 
\big(\!\!\!\!\prod_{v\ {\rm not}\ {\rm e.p.}}2^{-\frac12(h_v-h_{v'})}\big)
\sum_{\substack{\PP\in{\cal P}_\t\\ |P_{v_0}|=2l}}\big(\prod_{\substack{v\ {\rm not}\\ {\rm e.p.}}}
2^{-|P_v|/16}\big)\;.\label{s6.boh} \eea
Now, the sum over $\PP$ can be bounded as follows: defining $p_v:=|P_v|$ 
(so that $p_v\le p_{v_1}+\cdots+p_{v_{s_v}}$),  then
\be\sum_{\substack{\PP\in{\cal P}_\t\\ |P_{v_0}|=2l}}\big(\prod_{\substack{v\ {\rm not}\\ {\rm e.p.}}}
2^{-|P_v|/16}\big)\le \prod_{\substack{v\ {\rm not}\\ {\rm e.p.}}}
\sum_{p_v}2^{-p_v/16}\begin{pmatrix}
p_{v_1}+\cdots+p_{v_{s_v}}\\ p_v\end{pmatrix} \;.\label{B.zero}\ee
The sum in the r.h.s. can be computed inductively, starting from the root and moving towards 
the endpoints; in fact, at the first step 
\be \sum_{p_{v_0}}2^{-p_{v_0}/16}\begin{pmatrix}
p_{v_{1}}+\cdots+p_{v_{s_{v_0}}}\\ p_{v_0}\end{pmatrix}=\prod_{j=1}^{s_{v_{0}}}
(1+2^{-1/16})^{p_{v_{j}}}\;;\label{B.first}\ee
at the second step, 
\be \prod_{j=1}^{s_{v_{0}}}
(1+2^{-1/16})^{p_{v_{j}}}\sum_{p_{v_{j}}}2^{-p_{v_{j}}/16}
\begin{pmatrix}
p_{v_{j,1}}+\cdots+p_{v_{j,s_{v_{j}}}}\\ p_{v_{j}}\end{pmatrix}=
 \prod_{j=1}^{s_{v_{0}}} \prod_{j'=1}^{s_{v_{j}}}(1+2^{-\frac1{16}}+2^{-\frac2{16}})^{p_{v_{j,j'}}}
 \label{B.second} \ee
so that, iterating,
\be \prod_{\substack{v\ {\rm not}\\ {\rm e.p.}}}
\sum_{p_v}2^{-p_v/16}\begin{pmatrix}
p_{v_1}+\cdots+p_{v_{s_v}}\\ p_v\end{pmatrix} = \prod_{v\ {\rm e.p.}}
\big[\sum_{k=1}^{L_v}2^{-\frac{k}{16}}\big]^4\label{B.third}\ee
where $L_v$ is the distance from $v$ to the root and we used the fact that $p_v=4$ for all 
endpoints $v$. Plugging Eq.(\ref{B.third}) into Eq.(\ref{B.zero}), we get
\be \sum_{\substack{\PP\in{\cal P}_\t\\ |P_{v_0}|=2l}}\big(\prod_{\substack{v\ {\rm not}\\ {\rm e.p.}}}
2^{-|P_v|/16}\big)\le \Big(\frac1{2^{1/16}-1}\Big)^n\;.\label{s6.boh2}\ee
Similarly, one can prove that 
\be \sum_{\t\in {\cal T}_{M;h,n}} \big(\prod_{\substack{v\ {\rm not}\\ {\rm e.p.}}}
2^{-\frac12(h_v-h_{v'})}\big)
\le \Big(\frac4{2^{1/2}-1}\Big)^n\;,\label{s6.boh3}\ee
uniformly in $M$ as $M\to\infty$, see \cite[Lemma A.2]{GeM}. Collecting all the previous bounds, we obtain
\be  \frac1{\b|\L|}\int d\xx_1\cdots d\xx_{2l} |W^{(h)}_{M,2l,\ul\r}(\xx_1,
\ldots,\xx_{2l})|\le \sum_{n\ge \max\{1,l-1\}}(\const.)^n|U|^n2^{-h(n-1)}\;,\label{2.61e}\ee
which implies Eq.(\ref{2.17}). The constant $\lis e_h$ can be bounded by the r.h.s.
of Eq.(\ref{B.17}) with $l=0$ and $n\ge2$ (because the contributions to $\lis e_h$
with $l=1$ are zero, by the condition that the tadpoles vanish), which implies
\be \lis e_h\le \sum_{n\ge 2}(\const.)^n|U|^n2^{-h(n-1)}\le (\const.)|U|^2 2^{-h}
\;.\label{2.61ee}\ee
Therefore, $F_{0,M}=\sum_{k=1}^M\lis e_k$ is given by an absolutely convergent 
power series in $U$, as desired. A critical analysis of the proof shows that all the bounds
are uniform in $M,\b,\L$ and all the expressions involved admit well-defined limits as 
$M,\b,|\L|\to\infty$. In particular, this implies that $F_0=
\lim_{M\to\infty}F_{0,M}$ is analytic in $U$ (and so is its $\b,|\L|\to\infty$ limit, which is reached 
uniformly) in the complex domain $|U|\le U_0$, for $U_0$ small enough. 
See \cite[Appendices C and D]{GM10} for details on 
these technical aspects. This concludes the proof of Lemma \ref{sec6}.1.\qed

\section{The multiscale integration: the infrared regime}\label{sec7}
\setcounter{equation}{0}
\renewcommand{\theequation}{\ref{sec7}.\arabic{equation}}

We are now left with computing 
\be \X_{M,\b,\L}=e^{-\b|\L|F_0}\int P(d\psi^{(i.r.)})e^{-\VV_0(\psi^{(i.r.)})}\;.\label{s7.1}\ee
We proceed in an iterative fashion, similar to the one described in the previous section 
for the integration of the large values of $k_0$. 
As a starting point, it is convenient to decompose the infrared
propagator as:
\be g^{(i.r.)}(\xx-\yy)=\sum_{\o=\pm} e^{-i \vec p_F^{\;\o}(\vec x-\vec
y)}g_{\o}^{(\le 0)}(\xx-\yy)\;,\label{2.22}\ee
where, if $\kk'=(k_0,\vec k')$,
\be g_{\o}^{(\le 0)}(\xx-\yy)=\frac1{\b|\L|}\sum_{\kk'\in\BBB^\o_{\b,L}}
\chi_0(|\kk'|)e^{-i\kk'(\xx-\yy)}\begin{pmatrix}-i k_0 & -v\O^*(\vec k'+
\vec p_F^{\;\o})
\\ -v\O(\vec k'+\vec p_F^{\;\o}) & -i k_0\end{pmatrix}^{-1}\label{2.23}\ee
and $\BBB^\o_{\b,L}=\BBB^{(M)}_\b\times\BBB^\o_L$, 
with $\BBB^\o_L=\{\frac{n_1}L\vec G_1
+\frac{n_2}L\vec G_2-\vec p_F^{\;\o}\;,\ 0\le n_1,n_2\le L-1\}$.
Correspondingly, we rewrite $\psi^{(i.r.)}$ as a sum of two independent
Grassmann fields:
\be \psi^{(i.r.)\pm}_{\xx,\s}=\sum_{\o=\pm}e^{i\vec p_F^{\;\o}\vec x}
\psi^{(\le 0)\pm}_{\xx,\s,\o}\label{2.24}\ee
and we rewrite Eq.(\ref{2.14}) in the form:
\be \Xi_{M,\b,\L}=e^{-\b |\L| F_{0}}\int P_{\c_0,A_0}(d\psi^{(\le 0)})
e^{-{\cal V}^{(0)}(\psi^{(\le 0)})}
\;,\label{2.25}\ee
where ${\cal V}^{(0)}(\psi^{(\le 0)})$ is equal to
${\cal V}_0(\psi^{(i.r.)})$, once $\psi^{(i.r.)}$ is rewritten as in
(\ref{2.24}), i.e.,
\bea && {\cal V}^{(0)}(\psi^{(\le 0)})=\label{V0}\\
&&=\sum_{n=1}^\io (\b|\L|)^{-2n}\!\!\!\sum_{\s_1,\ldots,\s_n=\uparrow
\downarrow}\;\sum_{\r_1,\ldots,\r_{2n}=1,2}^{\o_1,\ldots,\o_{2n}=\pm}\;
\sum_{\kk_1',\ldots,\kk_{2n}'}
\Big[\prod_{j=1}^n\hat \psi^{(\le 0)+}_{\kk_{2j-1}',\s_j,\r_{2j-1},\o_{2j-1}}
\hat \psi^{(\le 0)-}_{\kk_{2j}',\s_j,\r_{2j},\o_{2j}}\Big]\;\cdot\nn\\
&&\hskip3.8truecm\cdot\hat W_{2n,\ul\r,\ul\o}^{(0)}(\kk_1',\ldots,
\kk_{2n-1}')\;\d\big(\sum_{j=1}^{2n}(-1)^j(\pp_F^{\o_{j}}+\kk_{j}')\big)=\nn\\
&&=\sum_{n=1}^\io \sum_{\ul\s,\ul\r,\ul\o}
\int\,d\xx_1\cdots d\xx_{2n}
\Big[\prod_{j=1}^n \psi^{(\le 0)+}_{\xx_{2j-1},\s_j,\r_{2j-1},\o_{2j-1}}
\psi^{(\le 0)-}_{\xx_{2j},\s_j,\r_{2j},\o_{2j}}\Big]
W_{2n,\ul\r,\ul\o}^{(0)}(\xx_1,\ldots,
\xx_{2n})\nn\eea
with:\\
1) $\ul\o=(\o_1,\ldots,\o_{2n})$, $\ul\s=(\s_1,\ldots,\s_{n})$ and
$\pp_F^{\o}=(0,\vec p_F^{\;\o})$;\\
2) $\hat W_{2n,\ul\r,\ul\o}^{(0)}
(\kk_1',\ldots,\kk_{2n-1}')=\hat W_{M,2n,\ul\r}(\kk_1'+\pp_F^{\o_j},\ldots,
\kk_{2n-1}'+\pp_F^{\o_{2n-1}})$, see (\ref{2.16});\\
3) the kernels $W_{2n,\ul\r,\ul\o}^{(0)}(\xx_1,\ldots,
\xx_{2n})$ are defined as:
\bea &&W_{2n,\ul\r,\ul\o}^{(0)}(\xx_1,\ldots,
\xx_{2n})=\label{xspace}\\
&&=(\b|\L|)^{-2n}\sum_{\kk'_1,\ldots,\kk_{2n}'}
e^{i\sum_{j=1}^{2n}(-1)^j\kk_{j}\xx_{j}}
\hat W_{2n,\ul\r,\ul\o}^{(0)}(\kk_1',\ldots,
\kk_{2n-1}')\;\d\big(\sum_{j=1}^{2n}(-1)^j(\pp_F^{\o_{j}}+\kk_{j}')\big)
\;.\nn\eea
Moreover,
$P_{\c_0,A_0}(d\psi^{(\le 0)})$ is defined as
\bea &&P_{\c_0,A_0}(d\psi^{(\le 0)})={{\cal N}_0}^{-1}\Biggl[\;
\prod_{\kk'\in\BBB_{\b,L}^\o}^{\c_0(|\kk'|)>0}
\;\prod_{\s,\o,\r}
d\hat\psi^{(\le 0)+}_{\kk',\s,\r,\o}d\hat\psi^{(\le 0)-}_{\kk',\s,\r,\o}\Biggr]
\cdot\label{2.26}\\
&&\hskip2.truecm\cdot
\exp \Big\{-(\b|\L|)^{-1}\sum_{\o=\pm}^{\s=\uparrow\downarrow}
\sum_{\kk'\in\BBB_{\b,L}^\o}^{\c_0(|\kk'|)>0}\c_0^{-1}(|\kk'|)
\hat\psi^{(\le 0)+}_{\kk',\s,\cdot,\o}A_{0,\o}(\kk')\hat\psi^{(\le 0)-}_{
\kk',\s,\cdot,\o}\Big\}\;,\nn\eea
where:
\bea 
A_{0,\o}(\kk')&=&\begin{pmatrix}-i k_0 & -v_0\O^*(\vec k'+\vec p_F^{\;\o})
\\ -v_0\O(\vec k'+\vec p_F^{\;\o}) & -i k_0\end{pmatrix}=\nn\\
&=&\begin{pmatrix}-i \z_0 k_0 +s_{0}(\kk') & v_0(ik_1'-\o k_2')+t_{0,\o}(\kk')
\\ v_0(-ik_1'-\o k_2') +t_{0,\o}^*(\kk')& -i \z_0 k_0+s_{0}(\kk') \end{pmatrix}
\;,\nn\eea
${\cal N}_0$ is chosen in such a way that $\int P_{\c_0,A_0}
(d\psi^{(\le 0)})=1$, $\z_0=1$,
$s_{0}=0$ and $|t_{0,\o}(\kk')|\le C|\kk'|^2$.

It is apparent that the $\psi^{(\le 0)}$ field
has zero mass (i.e., its propagator
decays polynomially at large distances in $\xx$-space). Therefore,
its integration requires an infrared multiscale analysis. As in the analysis 
of the ultraviolet problem, we define a sequence of geometrically
decreasing momentum scales $2^h$, with $h=0,-1,-2,\ldots$ Correspondingly
we introduce compact support functions $f_h(\kk')=\c_0(2^{-h}|\kk'|)
-\c_0(2^{-h+1}|\kk'|)$ and we rewrite
\be \c_0(|\kk'|)=\sum_{h=h_\b}^0 f_h(\kk')\;,\label{2.27}\ee
with $h_\b:=\lfloor \log_2\big(\frac{3\p}{4\b}\big)\rfloor$ (the reason why the sum in Eq.(\ref{2.27})
runs over $h\ge h_\b$ is that, if $\kk'\in\BBB^\o_{\b,L}$ and $h<h_\b$,
then $f_h(\kk')$, simply because $|\kk'|\ge \p/\b$).
The purpose is to perform the integration of (\ref{2.25}) in an iterative way.
We step by step decompose the
propagator into a sum of two propagators, the first supported on
momenta $\sim 2^{h}$, $h\le 0$, the second supported
on momenta smaller than $2^h$. Correspondingly we rewrite the
Grassmann field as a sum of two independent fields: $\psi^{(\le
h)}=\psi^{(h)}+ \psi^{(\le h-1)}$ and we integrate the field
$\psi^{(h)}$. In this way
we inductively prove that, for any $h\le 0$,  Eq.\pref{2.25} can be
rewritten as
\be \Xi_{M,\b,\L}=e^{-\b|\L|F_h}\int P_{\c_h,A_h}(d\psi^{(\le h)})
e^{-{\cal V}^{(h)}(\psi^{(\le h)})}\;,\label{2.28}\ee
where $F_h, A_h, {\cal V}^{(h)}$ will be defined recursively,
$\c_h(|\kk'|)=
\sum_{k=h_\b}^h f_k(\kk')$ and $P_{\c_h,A_h}(d\psi^{(\le h)})$ is defined
in the same way as $P_{\c_0,A_0}(d\psi^{(\le 0)})$ with
$\psi^{(\le 0)}, \c_0$, $A_{0,\o}, \z_0$, $v_0, s_{0}, t_{0,\o}$ replaced
by $\psi^{(\le h)}, \c_h, A_{h,\o}, \z_h, v_h, s_h, t_{h,\o}$,
respectively. Moreover ${\cal V}^{(h)}(0)=0$ and
\bea && {\cal V}^{(h)}(\psi)=\sum_{n=1}^\io (\b|\L|)^{-2n}
\sum_{\ul\s,\ul\r,\ul\o}\;
\sum_{\kk_1',\ldots,\kk_{2n}'}
\Big[\prod_{j=1}^n\hat\psi^{(\le h)+}_{\kk_{2j-1}',\s_j,\r_{2j-1},\o_{2j-1}}
\hat \psi^{(\le h)-}_{\kk_{2j}',\s_j,\r_{2j},\o_{2j}}\Big]\;\cdot\nn\\
&&\hskip3.truecm\cdot\hat W_{2n,\ul\r,\ul\o}^{(h)}(\kk_1',\ldots,
\kk_{2n-1}')\;\d(\sum_{j=1}^{2n}(-1)^j(\pp_F^{\o_{j}}+\kk_{j}'))=\label{2.29}\\
&&=\sum_{\substack{n\ge 1\\\ul\s,\ul\r,\ul\o}}
\int\,d\xx_1\cdots d\xx_{2n}
\Big[\prod_{j=1}^n \psi^{(\le h)+}_{\xx_{2j-1},\s_j,\r_{2j-1},\o_{2j-1}}
\psi^{(\le h)-}_{\xx_{2j},\s_j,\r_{2j},\o_{2j}}\Big]
W_{2n,\ul\r,\ul\o}^{(h)}(\xx_1,\ldots,
\xx_{2n})\;.\nn\eea

Note that the field $\psi^{(\le h)}_{\kk',\s,\o}$,
whose propagator is given by $\c_h(|\kk'|)
[A^{(h)}_\o(\kk')]^{-1}$, has the same support as $\c_h$, that is on a
neighborood of size $2^h$ around the singularity $\kk'={\bf 0}$ (that, in the
original variables, corresponds to the Dirac point $\kk=\pp_F^\o$). It is
important for the following to think $\hat W^{(h)}_{2n,\ul\r,\ul\o}$, $h\le 0$,
as functions of the variables $\{\z_k,v_k\}_{h<k\le 0}$. The iterative
construction below will inductively imply that the dependence on these
variables is well defined. The iteration continues up to the scale $h_\b$
and the result of the last iteration is $\Xi_{M,\b,\L}=e^{-\b|\L|F_{\b,\L}^{(M)}}$.
\\
\\
{\it Localization and renormalization.}
In order to inductively prove Eq.\pref{2.28}
we write
\be {\cal V}^{(h)} =\LL{\cal V}^{(h)}+\RR{\cal V}^{(h)}\label{2.loc}
\ee
where
\be
\LL{\cal V}^{(h)}=
\frac{1}{\b|\L|}\sum_{\s=\uparrow
\downarrow}^{\o=\pm}\;
\sum_{\kk'}^{\c_h(|\kk'|)>0} \hat\psi^{(\le h)+}_{\kk',\s,\o}
\hat W_{2,(\o,\o)}^{(h)}(\kk')
\hat\psi^{(\le h)-}_{\kk',\s,\o}
\;,\label{2.30a}\ee
and $\RR{\cal V}^{(h)}$ is given by Eq.\pref{2.29} with
$\sum_{n=1}^\io$ replaced by $\sum_{n=2}^\io$, that is it contains
only the monomials with four fermionic fields or more. 
The symmetries of the action, which are described in Appendix \ref{app2} and
are preserved by the iterative integration procedure, imply that, in the zero 
temperature and thermodynamic limit, 
\be 
\hat W^{(h)}_{2,(\o,\o)}({\bf k'})=
\begin{pmatrix} -iz_hk_0 & \d_h(ik_1'-\o k_2') \\
\d_h(-ik_1'-\o k_2') & -iz_hk_0\end{pmatrix}+O(|\kk'|^2)\;, \label{2.31}\ee
for suitable real constants $z_h,\d_h$. Eq.(\ref{2.31}) is the analogue of Eq.(\ref{sym2}); its 
proof is completely analogous to the proof of Lemma \ref{sec6}.2 and will not be belabored here.

Once that the above definitions are given, we can describe our iterative
integration procedure for $h\le 0$. We start from Eq.(\ref{2.28})
and we rewrite it as
\be \int P_{\c_h,A_h}(d\psi^{(\le h)}) \, e^{-\LL{\cal V}^{(h)}
(\psi^{(\le h)})-\RR{\cal V}^{(h)}
(\psi^{(\le h)}) -\b|\L| F_h}\;,\label{2.32}\ee
with
\bea &&\LL{\cal V}^{(h)}(\psi^{(\le h)})=(\b|\L|)^{-1}\sum_{\o,\s}
\sum_{\kk'}^{\c_h(|\kk'|)>0}\cdot\label{2.33}\\
&&\hskip.5truecm\cdot
\hat\psi^{(\le h)+}_{\kk',\s,\o}\begin{pmatrix}
-iz_hk_0 +\s_h(\kk')& \d_h(ik_1'-\o k_2')+\t_{h,\o}(\kk') \\
\d_h(-ik_1'-\o k_2) +\t_{h,\o}^*(\kk')& -iz_hk_0+\s_h(\kk')\end{pmatrix}
\hat\psi^{(\le h)-}_{\kk',\s,\o}\;.\nn\eea
Then we include $\LL{\cal V}^{(h)}$
in the fermionic integration, so obtaining
\be
\int P_{\c_h,\lis A_{h-1}}(d\psi^{(\le h)}) \, e^{-
\RR{\cal V}^{(h)}
(\psi^{(\le h)}) -\b|\L| (F_h+e_h)}\;,\label{2.34}\ee
where 
\be e_h=\frac1{\b|\L|}\sum_{\o,\s}
\sum_{\kk'}\sum_{n\ge 1}\frac{(-1)^{n}}{n}
\Tr\Big\{\big[\c_h(\kk')A^{-1}_{h,\o}(\kk')
W^{(h)}_{2,(\o,\o)}(\kk')\big]^n\Big\}
\label{2.34z}\ee
is a constant taking into account the change in the
normalization factor of the measure and
\be \lis A_{h-1,\o}(\kk')=\begin{pmatrix}-i \lis \z_{h-1} k_0 +\lis s_{h-1}(\kk') &
\lis v_{h-1}(ik_1'-\o k_2')+\lis t_{h-1,\o}(\kk')
\\ \lis v_{h-1}(-ik_1'-\o k_2') +\lis t_{h-1,\o}^*(\kk')& -i \lis \z_{h-1} k_0
+\lis s_{h-1}(\kk')\end{pmatrix}\label{2.34aa}\ee
with:
\bea& \lis \z_{h-1}(\kk')= \z_h +z_h \c_h(\kk')\;,&\qquad
\lis v_{h-1}(\kk')= v_h +\d_h \c_h(\kk')\;,\label{2.34ab}\\
& \lis s_{h-1}(\kk')=s_{h}(\kk')+\s_h(\kk')\c_h(\kk')\;,&\qquad 
\lis t_{h-1,\o}(\kk')=t_{h,\o}(\kk')+\t_{h,\o}(\kk')\c_h(\kk')\;.
\nn\eea
Now we can perform the integration of the $\psi^{(h)}$ field.
We rewrite the Grassmann field $\psi^{(\le h)}$ as a sum of two independent
Grassmann fields $\psi^{(\le h-1)}+\psi^{(h)}$ and correspondingly we rewrite
Eq.(\ref{2.34}) as
\be e^{-\b|\L|(F_h+e_h)}\int
P_{\c_{h-1},A_{h-1}}(d\psi^{(\le h-1)}) \,
\int P_{f_h,\lis A_{h-1}}
(d\psi^{(h)})\, e^{-\RR{\cal V}^{(h)}(\psi^{(\le h-1)}+\psi^{(h)})}\;,
\label{2.35}\ee
where
\be A_{h-1,\o}(\kk')=\begin{pmatrix}-i\z_{h-1} k_0 +s_{h-1}(\kk') &
v_{h-1}(ik_1'-\o k_2')+t_{h-1,\o}(\kk')
\\ v_{h-1}(-ik_1'-\o k_2') +t_{h-1,\o}^*(\kk')& -i \z_{h-1} k_0
+s_{h-1}(\kk')\end{pmatrix}\label{2.34ac}\ee
with:
\bea& \z_{h-1}= \z_h +z_h\;,
&\qquad v_{h-1}= v_h +\d_h \;,\nn\\
& s_{h-1}(\kk')=s_{h}(\kk')+\s_h(\kk')\;,& \qquad t_{h-1,\o}(\kk')=t_{h,\o}(\kk')+\t_{h,\o}(\kk')\;.
\label{2.34ad}\eea

The single scale propagator is
\be \int P_{f_h,\lis A_{h-1}}(d \psi^{(h)})
\psi^{(h)-}_{\xx_1,\s_1,\o_1}
\psi^{(h)+}_{ \xx_2,\s_2,\o_2} = \d_{\s_1,\s_2}\d_{\o_1,\o_2}
g^{(h)}_\o(\xx_1-\xx_2)\;,\label{2.36}\ee
where
\be g^{(h)}_\o(\xx_1-\xx_2)=\frac1{\b|\L|}\sum_{\kk'\in\DD_{\b,L}^\o}
e^{-i\kk'(\xx_1-\xx_2)}f_h(\kk')\Big[\lis A_{h-1,\o}(\kk')\Big]^{-1}
\;.\label{2.37}\ee
After the integration of the field on scale $h$ we are left with an
integral involving the fields
$\psi^{(\le h-1)}$ and the new effective interaction
${\cal V}^{(h-1)}$, defined as
\be e^{-{\cal V}^{(h-1)}(\psi^{(\le h-1)})-\lis e_h \b|\L|}=
\int P_{f_h,\lis A_{h-1}}
(d\psi^{(h)})\, e^{-\RR{\cal V}^{(h)}(\psi^{(\le h-1)}+\psi^{(h)})}
\;,\label{2.39}\ee
with ${\cal V}^{(h-1)}(0)=0$. 
It is easy to see that ${\cal V}^{(h-1)}$ is of the form Eq.(\ref{2.29}) and that
$F_{h-1}=F_h+e_h+\lis e_h$. It is sufficient to use the identity
\be \lis e_h+{\cal V}^{(h-1)}(\psi^{(\le h-1)})=
\sum_{n\ge 1}\frac{1}{n!}(-1)^{n+1}\EE^T_h(\RR{\cal V}^{(h)}\big(
\psi^{(\le h-1)}
+\psi^{(h)}\big);n)
\;,\label{2.40}\ee
where $\EE^T_h(X(\psi^{(h)});n)$
is the truncated expectation of order $n$ w.r.t. the
propagator $g^{(h)}_\o$, which is the analogue of Eq.(\ref{2.15})
with $\psi^{(u.v.)}$ replaced by $\psi^{(h)}$ and with
$P(d\psi^{(u.v.)})$ replaced by $P_{f_h,\lis A_{h-1}}(d\psi^{(h)})$.

Note that the above procedure allows us to write the
{\it effective constants} $(\z_{h},v_{h})$, $h\le 0$,
in terms of $(\z_k,v_k)$, $h< k\le 0$, namely
$\z_{h-1}=\b_h^\z\big((\z_h,v_h),\ldots,(\z_0,v_0)\big)$
and $v_{h-1}=\b_h^v\big((\z_h,v_h),\ldots,(\z_0,v_0)\big)$,
where $\b_h^\#$ is the so--called {\it Beta function}.\\

An iterative implementation of Eq.(\ref{2.40}) leads to a representation of
${\cal V}^{(h)}$, $h<0$, in terms of a new tree expansion. The 
set of trees of order $n$ contributing to ${\cal V}^{(h)}(\psi^{(\le h)})$ is denoted by $\TT_{h,n}$.
The trees in $\TT_{h,n}$ are defined in a way very similar to 
those in $\widetilde \TT_{M;h,n}$, but for the following differences:
(i) all the endpoints are on scale $1$; 
(ii) the scale labels of the vertices of $\t\in\TT_{h,n}$ are between $h+1$ and $1$; 
(iii) with each endpoint $v$ we associate one of the monomials
with four or more Grassmann fields contributing to $\RR {\cal
V}^{(0)}(\psi^{(\le 0)})$, corresponding to the terms with
$n\ge 2$ in the r.h.s. of Eq.(\ref{V0}) (here $\RR$ is the linear operator acting as 
the identity on Grassmann monomials of order 4 or more, and acting as 0 on Grassmann 
monomials of order 0 or 2). In terms of these trees, the effective potential 
${\cal V}^{(h)}$, $h\le -1$, can be written as
\be {\cal V}^{(h)}(\psi^{(\le h)}) + \b|\L| \lis e_{k+1}=
\sum_{n=1}^\io\sum_{\t\in\TT_{h,n}}
{\cal V}^{(h)}(\t,\psi^{(\le h)})\;,\label{2.41}\ee
where, if $v_0$ is the first vertex of $\t$
and $\t_1,\ldots,\t_s$ ($s=s_{v_0}$)
are the subtrees of $\t$ with root $v_0$,
\be {\cal V}^{(h)}(\t,\psi^{(\le h)})=\frac{(-1)^{s+1}}{s!} \EE^T_{h+1}
\big[\RR{\cal V}^{(h+1)}(\t_1,\psi^{(\le h+1)});\ldots; \RR{\cal V}^{(h+1)}
(\t_{s},\psi^{(\le h+1)})\big]\;,\label{2.42}\ee
where, if $\t_i$ is trivial, then $h+1=0$ and $\RR{\cal V}^{(0)}(\t_i,\psi^{(\le h+1)})=
\RR{\cal V}^{(0)}(\psi^{(\le h+1)})$. 
Repeating step by step the discussion leading to Eqs.(\ref{2.43}),(\ref{B.14}) and (\ref{B.16}),
and using analogous definitions, we find that 
\be \VV^{(h)}(\t,\psi^{(\le h)})=\sum_{{\bf P}\in{\cal P}_\t}\sum_{T\in{\bf T}}
 \int d\xx_{v_0}\psi^{(\le h)}_{P_{v_0}} W_{\t,\PP,T}^{(h)}(\xx_{v_0})\;,\label{2.42A}\ee
with 
\bea &&
W_{\t,\PP, T}(\xx_{v_0}) =\left[\prod_{i=1}^n
K_{v_i^*}^{(h_i)} (\xx_{v_i^*})\right] \Bigg\{\prod_{\substack{v\ {\rm not}\\
{\rm e.p.}}}\frac{1}{s_v!} \int
dP_{T_v}({\bf t}_v)\;{\rm det}\, G^{h_v,T_v}({\bf t}_v)\cdot\nn\\ &&\qquad
\cdot\Biggl[
\prod_{\ell\in T_v} \d_{\o(\ell),\o'(\ell)} \d_{\s(\ell),\s'(\ell)}\,
\big[g^{(h_v)}_{\o(\ell)}(\xx(\ell)-\xx'(\ell))\big]_{\r(\ell),\r'(\ell)}\,\Biggr]
\Bigg\}\;.\label{2.50zz}\eea
and: $v_i^*$, $i=1,\ldots,n$, are the endpoints of $\t$; 
$K_{v_i^*}^{(h_i)} (\xx_{v_i^*})$ is the kernel of one of the monomials contributing to 
$\RR\VV^{(0)}(\psi^{(\le h_v)})$; $G^{h,T}$ is a matrix with elements 
\be G^{h,T}_{f,f'}=t_{i(f),i(f')}\d_{\o(f),\o(f')} \d_{\s(f),\s(f')}\,
\big[g^{(h)}_{\o(f)}(\xx(f)-\xx(f'))\big]_{\r(f),\r(f')}\;,\label{2.48}\ee
Once again, it is important to note that $G^{h,T}$ is a Gram matrix, i.e.,
defining ${\bf e}_+={\bf e}_\uparrow=(1,0)$ and
${\bf e}_-={\bf e}_\downarrow=(0,1)$,
the matrix elements in Eq.(\ref{2.48}), using a notation similar to Eq.(\ref{s5.8}),
can be written in terms of scalar products:
\bea&& G^{h,T}_{f,f'}=\label{2.48B}\\
&&=\media{{\bf u}_{i(f)}\otimes {\bf e}_{\o(f)}\otimes{\bf e}_{\s(f)}\otimes{\bf 
A}_{\xx(f),\r(f),\o(f)}, {\bf u}_{i(f')}\otimes {\bf e}_{\o(f')}\otimes
{\bf e}_{\s(f')}\otimes {\bf B}_{\xx(f'),\r(f'),\o(f')}}\;,\nn\eea
where
\bea &&[{\bf A}_{\xx,\r,\o}(\zz)]_i=\frac1{\b|\L|}\sum_{\kk'\in\BBB_{\b,L}^\o}
\frac{e^{-i\kk'(\zz-\xx)}\sqrt{f_h(\kk')}}{|\det \lis A_{h-1,\o}(\kk')|^{1/4}}\;\d_{\r,i}\;,\nn\\
&&[{\bf B}_{\xx,\r,\o}(\zz)]_i=\frac1{\b|\L|}\sum_{\kk'\in\BBB_{\b,L}^\o}
\frac{e^{-i\kk'(\zz-\xx)}\sqrt{f_h(\kk')}}{|\det \lis A_{h-1,\o}(\kk')|^{3/4}}\cdot\label{2.48b}\\
&&\hskip2.truecm\cdot\begin{pmatrix}i \lis \z_{h-1} k_0 -\lis s_{h-1}(\kk') &
\lis v_{h-1}(ik_1'-\o k_2')+\lis t_{h-1,\o}(\kk')
\\ \lis v_{h-1}(-ik_1'-\o k_2') +\lis t_{h-1,\o}^*(\kk')& i \lis \z_{h-1} k_0
-\lis s_{h-1}(\kk')\end{pmatrix}_{i,\r}\;.
\nn\eea
so that, using that $\lis s_{h-1}$ is purely imaginary,
\be ||{\bf A}_{\xx,\r,\o}||^2=||{\bf B}_{\xx,\r,\o}||^2=\frac1{\b|\L|}\sum_{\kk'\in\BBB^{\o}_{\b,L}}
\frac{f_h(\kk')}{|\det\lis A_{h-1,\o}(\kk')|^{1/2}}\le C 2^{2h}\;,\label{2.48B.5}\ee
for a suitable constant $C$. 

Using the representation Eqs.(\ref{2.41}),(\ref{2.42}),(\ref{2.50zz})
and proceeding as in the proof of Lemma \ref{sec6}.1, we can get a bound on the kernels of the 
effective potentials, which is the key ingredient for the proof of  Theorem \ref{thm1} and is 
summarized in the following theorem.

\begin{theorem}\label{thm2}
There exists a constants $U_0>0$, independent 
of $M$, $\b$ and $\L$, such that the kernels
$W^{(h)}_{2l,\ul\r,\ul\o}
(\xx_1,\ldots,\xx_{2l})$ in Eq.(\ref{2.29}), $h\le -1$,
are analytic functions of $U$
in the complex domain $|U|\le U_0$,
satisfying, for any $0\le \th<1$ and a suitable constant $C_\th>0$ 
(independent of $M,\b,\L$), the following estimates:
\be \frac1{\b|\L|}\int d\xx_1\cdots d\xx_{2l}|W^{(h)}_{2l,\ul\r,\ul\o}
(\xx_1,\ldots,\xx_{2l})|\le
2^{h (3-2l+\th)} \,(C_\th\,|U|)^{max(1,l-1)}\;.\label{2.52}\ee
Moreover, the constants $e_{h}$ and $\lis e_h$ defined by Eq.(\ref{2.34z})
and Eq.(\ref{2.40}) are analytic functions of $U$ in the same domain $|U|\le U_0$,
uniformly bounded as $|e_{h}|+|\lis e_{h}|\le C_\th |U| 2^{h(3+\th)}$. Both the kernels 
$W^{(h)}_{2l,\ul\r,\ul\o}(\xx_1,\ldots,\xx_{2l})$ and the constants $e_h,\lis e_h$ admit 
well defined limits as $M,\b,|\L|\to\infty$, which are reached uniformly 
(and, with some abuse of notation, will be denoted by the same symbols).
\end{theorem}
Before we present its proof, let us show how Theorem \ref{thm2} implies, as a corollary,
Theorem \ref{thm1}. It is enough to observe that, by Proposition \ref{prop1} and by the 
multiscale integration procedure described above,
$f_\b(U)=F_0+\sum_{h=h_\b}^0
(e_h+\lis e_h)$, with $F_0, e_h, \lis e_h$ analytic functions of $U$ for $U$ small enough
(see the discussion at the end of Section \ref{sec6} and the statement of Theorem \ref{thm2}). 
Since $|e_{h}|+|\lis e_{h}|\le C_\th |U| 2^{h(3+\th)}$, the sum $\sum_{h=h_\b}^0
(e_h+\lis e_h)$ is absolutely convergent, uniformly in $h_\b$; therefore, both 
$f_\b(U)$ and its $\b\to\infty$ limit are analytic functions of $U$ for $U$ small enough.
This concludes the proof of Theorem \ref{thm1}.\\

{\bf Proof of Theorem \ref{thm2}.}
Let us preliminarily assume that, for $h'\le h\le -1$,
and for suitable constants $c, c_n$, the corrections $z_h,\d_h,\s_h(\kk')$
and $\t_h(\kk')$ defined in Eq.\pref{2.31} and Eq.\pref{2.33}, satisfy the following
estimates:
\bea &&\hskip1.25truecm\max\ \{|z_h|, |\d_h|\}\le c |U|2^{\th h}
\;,\label{2.51}\\
&&\sup_{2^{h'-1}\le |\kk'|\le 2^{h'+1}}\{||\partial_{\kk'}^n
\s_h(\kk')||,||\partial_{\kk'}^n \t_{h,\o}(\kk')||\}\le c_n
|U|2^{2(h'-h)}2^{(1+\th-n)h}\;.\nn\eea
Using Eq.\pref{2.51} we inductively see that the running coupling functions
$\z_h$, $v_h$, $s_h(\kk')$ and $t_h(\kk')$ satisfy similar estimates:
\bea &&\hskip1.25truecm\max\ \{|\z_h-1|,|v_h-v_0|\}\le c |U|\;,\label{2.51a}\\
&&\sup_{2^{h'-1}\le |\kk'|\le 2^{h'+1}}\{||\partial_{\kk'}^n
s_h(\kk')||,||\partial_{\kk'}^n (t_{h,\o}(\kk')-t_{0,\o}(\kk'))||\}\le c_n
|U|2^{2(h'-h)}2^{(1+\th-n)h}\;.\nn\eea
Now, using the definition of $g_\o^{(h)}$, see Eq.(\ref{2.37}) and
Eq.(\ref{2.34aa}), and the bounds Eq.(\ref{2.51a}), 
we get (proceeding as in the proof of Eq.(\ref{2.38copy}) in Appendix \ref{app0}),
\be ||g^{(h)}_{\o}(\xx_1-\xx_2)||
\le C_{K}\frac{2^{2h}}{1+(2^{h}||\xx_1-\xx_2||)^K}\;,\label{2.38}\ee
for all $K\ge 0$ and a suitable constant $C_K$. Using the tree expansion described above and,
in particular, Eqs.(\ref{2.41}),(\ref{2.42A}),(\ref{2.50zz}), 
we find that the l.h.s. of Eq.(\ref{2.52}) can be bounded from above by
\bea && \sum_{n\ge 1}\sum_{\t\in {\cal T}_{h,n}}
\sum_{\substack{\PP\in{\cal P}_\t\\ |P_{v_0}|=2l}}\sum_{T\in{\bf T}}
\int\prod_{\ell\in T^*}
d(\xx(\ell)-\xx'(\ell)) \left[\prod_{i=1}^n|K_{v_i^*}^{(h_i)}(\xx_{v_i^*})|\right]\cdot
\label{2.53}\\
&&\hskip3.truecm\cdot\Bigg[\prod_{\substack{v\ {\rm not}\\{\rm e.p.}}}\frac{1}{s_v!}
\max_{{\bf t}_v}\big|{\rm det}\, G^{h_v,T_v}({\bf t}_v)\big|
\prod_{\ell\in T_v}
\big|\big|g^{(h_v)}_{\o(\ell)}(\xx(\ell)-\xx'(\ell))\big|\big|\Bigg]\nn\eea
where  $T^*$ is a tree graph
obtained from $T=\cup_vT_v$, by adding
in a suitable (obvious) way, for each endpoint $v_i^*$,
$i=1,\ldots,n$, one or more lines connecting the space-time points
belonging to $\xx_{v_i^*}$.

An application of the Gram--Hadamard inequality Eq.(\ref{s5.6}), combined with
the representation Eq.(\ref{2.48B}) and the dimensional bound Eq.(\ref{2.48B.5}),
implies that
\be |{\rm det} G^{h_v,T_v}({\bf t}_v)| \le
(\const.)^{\sum_{i=1}^{s_v}|P_{v_i}|-|P_v|-2(s_v-1)}\cdot\;
2^{{h_v}
\left(\sum_{i=1}^{s_v}|P_{v_i}|-|P_v|-2(s_v-1)\right)}\;.\label{2.54az}\ee
By the decay properties of $g^{(h)}_\o(\xx)$, Eq.(\ref{2.38}), it
also follows that
\be \prod_{\substack{v\ {\rm not}\\ {\rm e.p.}}}
\frac{1}{s_v!}\int \prod_{\ell\in T_v} d(\xx(\ell)-\xx'(\ell))\,
||g^{(h_v)}_{\o(\ell)}(\xx(\ell)-\xx'(\ell))||\le c^n \prod_{\substack{v\ {\rm not}\\ {\rm e.p.}}}
\frac{1}{s_v!} 2^{-h_v(s_v-1)}\;.\label{2.55az}\ee
The bound Eq.(\ref{2.17}) on the kernels produced by the ultraviolet integration
implies that
\be \int\prod_{\ell\in T^*\setminus\cup_v T_v}d(\xx(\ell)-\xx'(\ell))
\prod_{i=1}^n |K_{v_i^*}^{(h_i)}(\xx_{v_i^*})|\le \prod_{i=1}^n C^{p_i}
|U|^{\,\frac{p_i}2-1}\;,\label{2.56}\ee
where $p_i=|P_{v_i^*}|$. Combining the previous bounds, we find that
Eq.(\ref{2.53}) can be bounded from above by
\be  \sum_{n\ge 1}\sum_{\t\in {\cal T}_{h,n}}
\sum_{\substack{\PP\in{\cal P}_\t\\ |P_{v_0}|=2l}}\sum_{T\in{\bf T}}
C^n \Big[\prod_{\substack{v\ {\rm not}\\ {\rm e.p.}}} \frac{1}{s_v!}
2^{{h_v}\left(\sum_{i=1}^{s_v}|P_{v_i}|-|P_v|-3(s_v-1)\right)}\Big]
\Big[\prod_{i=1}^n C^{p_i}
|U|^{\,\frac{p_i}2-1}\Big]
\label{2.57}\ee
Let us recall that $n(v)=\sum_{i: v_i^*>v}\,1$ is the number of endpoints following
$v$ on $\t$, that $v'$ is the vertex immediately preceding $v$ on $\t$ and that $|I_v|$
is the number of field labels associated to the endpoints following $v$
on $\t$ (note that $|I_v|\ge 4 n(v)$).  Using the following relations, which can be easily proved by 
induction,
\bea && \sum_{v\ {\rm not}\ {\rm e.p.}}h_v\Big[\big(\sum_{i=1}^{s_v}
|P_{v_i}|\big)-|P_v|\Big]=h(|I_{v_0}|-|P_{v_0}|)+\sum_{v\ {\rm not}\ {\rm e.p.} }
(h_v-h_{v'})(|I_v|-|P_v|)\;,\nn\\
&&
\sum_{v\ {\rm not}\ {\rm e.p.}}h_v(s_v-1)=h(n-1)+\sum_{v\ {\rm not}\ {\rm e.p.}}
(h_v-h_{v'})(s_v-1)\;,\label{2.58}\eea
we find that Eq.(\ref{2.57}) can be bounded above by
\bea &&\sum_{n\ge 1}\sum_{\t\in {\cal T}_{h,n}}
\sum_{\substack{\PP\in{\cal P}_\t\\ |P_{v_0}|=2l}}\sum_{T\in{\bf T}}
C^n  2^{h(3-|P_{v_0}|+|I_{v_0}|-3n)}\cdot\nn\\
&&\hskip1.truecm \cdot
\Big[\prod_{\substack{v\ {\rm not}\\ {\rm e.p.}}} \frac{1}{s_v!}
2^{(h_v-h_{v'})(3-|P_v|+|I_v|-3n(v))}\Big]
\Big[\prod_{i=1}^n C^{p_i}
|U|^{\,\frac{p_i}2-1}\Big]\;.\label{2.59}\eea
Using the following identities
\bea &&2^{h  n}
\prod_{\substack{v\ {\rm not}\\ {\rm e.p.}}}
2^{(h_v-h_{v'}) n(v)}=\prod_{v\ {\rm e.p.}}
2^{h_{v'}}\;,\label{2.60Y}\\
&& 2^{h  |I_{v_0}|}
\prod_{\substack{v\ {\rm not}\\ {\rm e.p.}}}
2^{(h_v-h_{v'}) |I_v|}=\prod_{v\ {\rm e.p.}}
2^{h_{v'} |I_v|}\;,\label{2.60}\eea
combined with remark that all the endpoints are on scale $1$ (so that the right hand sides of 
Eqs.(\ref{2.60Y})-(\ref{2.60}) are equal to $1$),  we obtain
\bea&& \frac1{\b|\L|}\int d\xx_1\cdots d\xx_{2l}|W^{(h)}_{2l,\ul\r,\ul\o}
(\xx_1,\ldots,\xx_{2l})|\le
\sum_{n\ge 1}\sum_{\t\in {\cal T}_{h,n}}
\sum_{\substack{\PP\in{\cal P}_\t\\ |P_{v_0}|=2l}}\sum_{T\in{\bf T}}
C^n  2^{h(3-|P_{v_0}|)}\cdot\nn\\
&&\hskip2.2truecm \cdot \Big[\prod_{\substack{v\ {\rm not}\\ {\rm e.p.}}}
\frac{1}{s_v!} 2^{-(h_v-h_{v'})(|P_v|-3)}\Big]
\Big[\prod_{i=1}^n C^{p_i} |U|^{\,\frac{p_i}2-1}\Big]\;.\label{2.61} \eea
Note that, if $v$ is not an endpoint, $|P_v|-3\ge 1$ by the
definition of $\RR$. Now, note that
the number of terms in $\sum_{T\in {\bf T}}$ can be bounded by
$C^n\prod_{v\ {\rm not}\ {\rm
e.p.}} s_v!$. Using also that $|P_v|-3\ge 1$ and $|P_v|-3\ge|P_v|/4$,
we find that the l.h.s. of Eq.(\ref{2.61}) can be bounded as
\bea&& \frac1{\b|\L|}\int d\xx_1\cdots d\xx_{2l}|W^{(h)}_{2l,\ul\r,\ul\o}
(\xx_1,\ldots,\xx_{2l})|\le 2^{h(3-|P_{v_0}|)}
\sum_{n\ge 1}C^n\sum_{\t\in {\cal T}_{h,n}}\cdot\label{2.61b}\\
&&\cdot\big(\!\!
\prod_{\substack{v\ {\rm not}\\ {\rm e.p.}}}
2^{-\th(h_v-h_{v'})}2^{-(1-\th)(h_v-h_{v'})/2}\big)
\sum_{\substack{\PP\in{\cal P}_\t\\ |P_{v_0}|=2l}}\big(\prod_{\substack{v\ {\rm not}\\ {\rm e.p.}}}
2^{-(1-\th)|P_v|/8}\big)\prod_{i=1}^n
C^{p_i} |U|^{\,\frac{p_i}2-1}\;.\nn \eea
Proceeding as in the previous section, we get the analogue of Eq.(\ref{s6.boh2}):
$$\sum_{\substack{\PP\in{\cal P}_\t\\
|P_{v_0}|=2l}}\big(\prod_{\substack{v\ {\rm not}\\ {\rm e.p.}}}
2^{-(1-\th)|P_v|/8}\big)\prod_{i=1}^n
C^{p_i} |U|^{\,\frac{p_i}2-1}\le C_\th^n|U|^n\;.$$
Finally, using that $\prod_{v\ {\rm not}\ {\rm e.p.}}
2^{-\th(h_v-h_{v'})}\le 2^{\th h}$ and that, for $0<\th<1$ (in analogy with Eq.(\ref{s6.boh3})),
$$\sum_{\t\in {\cal T}_{h,n}}
\prod_{v\ {\rm not}\ {\rm e.p.}}
2^{-(1-\th)(h_v-h_{v'})/2}\le C^n\;;$$
and collecting all the previous bounds, we obtain
\be \frac1{\b|\L|}\int d\xx_1\cdots d\xx_{2l}|W^{(h)}_{2l,\ul\r,\ul\o}
(\xx_1,\ldots,\xx_{2l})|\le 2^{h(3-|P_{v_0}|+\th)}
\sum_{n\ge 1}C^n |U|^n\;,\label{2.61ez}\ee
which is the desired result.

We now need to prove the assumption Eq.\pref{2.51}. We proceed by induction.
It is easy to see that the assumption is valid for $h=0$; in fact, 
\bea && |z_0|=
|\dpr_{k_0}\hat W^{(0)}_{2,(1,1),(\o,\o)}(\pp_F^\o)|\;,\qquad 
|\d_0|=|\dpr_{k_1}\hat W_{2,(1,2),(\o,\o)}(\pp_F^\o)|\;,\nn\\
&&|\s_0(\kk')|=
\Big|\sum_{\m,\n=0}^2\int_0^1ds\int_0^s ds' k_\m'k_\n'\dpr_\m\dpr_\n W_{2,(1,1),(\o,\o)}(s'\kk')
\Big|\;,\label{d.tr}\\
&&|\t_0(\kk')|=
\Big|\sum_{\m,\n=0^2}\int_0^1ds\int_0^s ds' k_\m'k_\n'\dpr_\m\dpr_\n W_{2,(1,2),(\o,\o)}(s'\kk')
\Big|\;,\nn\eea
so that, 
\bea &&
\max\{|z_0|,|\d_0|\}\le \frac1{\b|\L|}\int d\xx d\yy\, ||\xx-\yy||\, ||W_{2,(\o,\o)}(\xx-\yy)||\;,\label{ci.ut}\\
&& \sup_{|\kk'|\sim 2^{h'}} \{||\dpr^n_{\kk'}\s_0(\kk')||,||\dpr^n_{\kk'}\t_0(\kk')||\}\le 
\frac{C2^{2h'}}{\b|\L|}\int d\xx d\yy\, ||\xx-\yy||^{n+2}\, ||W_{2,(\o,\o)}(\xx-\yy)||\;,\nn\eea
which can be bounded via the same strategy as the proof of Eq.(\ref{2.17}), yielding
Eq.(\ref{2.51}) for $h=0$. Similarly,
assuming that Eq.\pref{2.51} is valid for all $h\ge k+1$, the quantities of interest for $h=k$
can be bounded by 
\be \max\{|z_k|,|\d_k|\}\le \frac1{\b|\L|}\int d\xx d\yy\,||\xx-\yy||\,||W^{(k)}_{2,(\o,\o)}(\xx,\yy)||\;,\label{2.61c}\ee
and
\bea &&\sup_{2^{h'-1}\le |\kk'|\le 2^{h'+1}}\{||\partial_{\kk'}^n
\s_k(\kk')||,||\partial_{\kk'}^n \t_{k,\o}(\kk')||\}\le \nn\\
&&\qquad\le
\frac{C2^{2h'}}{\b|\L|}\int d\xx d\yy\,
||\xx-\yy||^{n+2}||W^{(k)}_{2,(\o,\o)}(\xx,\yy)||\;.\label{2.61d}\eea
The same proof leading to Eq.(\ref{2.61ez}) shows that the r.h.s. of Eq.(\ref{2.61c})
can be bounded by the r.h.s. of Eq.(\ref{2.61ez}) times $2^{-k}$ (that is
the dimensional estimate for $||\xx-\yy||$), and that
the r.h.s. of Eq.(\ref{2.61c})
can be bounded by the r.h.s. of Eq.(\ref{2.61ez}) times $(\const.)^n2^{2h'}2^{-(n+2)k}$
(where $2^{-k(n+2)}$ is the dimensional estimate for $||\xx-\yy||^{n+2}$).

It remains to prove the estimates on $e_h,\lis e_h$. The bound on $\lis e_h$
is an immediate corollary of the discussion above, simply because $\lis e_h$
can be bounded by Eq.(\ref{2.53}) with $l=0$. Finally, remember that 
$e_h$ is given by Eq.(\ref{2.34z}): an explicit computation of $A^{-1}_{h,\o}
(\kk')W^{(h)}_{2,(\o,\o)}(\kk')$ and the use of 
Eqs.(\ref{2.51})-(\ref{2.51a}) imply that $||A^{-1}_{h,\o}
(\kk')W^{(h)}_{2,(\o,\o)}(\kk')||\le C|U|2^{\th h}$, from which: 
$|e_h|\le C'2^{3h}\sum_{n\ge 1}(C|U|2^{\th h})^n$, as desired. This concludes the proof of Theorem \ref{thm2} and, therefore, as discussed after the statement of Theorem \ref{thm2},
it also concludes the proof of analyticity of the specific free energy and ground state energy.
\qed

\section{Conclusions}\label{sec8}
\setcounter{equation}{0}
\renewcommand{\theequation}{\ref{sec8}.\arabic{equation}}

In conclusion, I presented a self-contained proof of the analyticity of the specific 
free energy and ground state energy of the 2D Hubbard model on the honeycomb lattice,
at half-filling and weak coupling. The proof is based on rigorous
fermionic RG methods and can be extended to the construction of the interacting correlations,
i.e., the off-diagonal elements of the reduced density matrices of the system \cite{GM10},
and to the computation of the universal optical conductivity \cite{GMP3}. Such construction
shows that the interacting correlations decay to zero at infinity with the same decay exponents as
those of the non-interacting case. The ``only" effect of the interactions is to change by a finite
amount the quasi-particle weight $Z^{-1}$ at the Fermi surface and the Fermi velocity $v$. 

The example presented here is the only known example of a realistic 2D interacting Fermi system 
for which the ground state (including the correlations) can be constructed. The main difference
with respect to other more standard 2D Fermi systems is the fact that here, at half-filling, the 
Fermi surface reduces to a set of two isolated points. This fact dramatically improves the infrared 
scaling properties of the theory: the four-fermions interaction, rather than being marginal, 
as in many other similar cases, is irrelevant; this is the technical point that makes the construction 
of the ground state possible and ``relatively easy". 

It is natural to ask how the system behaves in the presence of electromagnetic interactions among 
the electrons, which is the case of interest for applications to clean graphene samples. 
In this case, the system has many analogies with $(2+1)$-dimensional QED. 
The four-fermions interaction, rather than being irrelevant, is marginal, and the fixed point of the
theory is expected to be non-trivial. The long distance decay of correlations is expected to be
described in terms of anomalous critical exponents and the effective Fermi velocity is expected to 
grow up to the maximal possible value, i.e., the speed of light. The specific values of 
the critical exponents suggest that local distortions of the lattice (in the form 
of the so-called Kekul\'e pattern \cite{HCM}) are amplified by electromagnetic interactions: 
this led us to propose a possible mechanism for the spontaneous formation of the 
Kekul\'e pattern, via a mechanism analogous to the 1D Peierls' instability \cite{GMP2}.
All these claims have been proved so far only order by order in renormalized perturbation 
theory \cite{GMP1,GMP2}; proving them in a non-perturbative fashion is an important open problem,
whose solution would represent a corner stone in the mathematical theory of quantum Coulomb systems.

\appendix

\section{Dimensional estimates of the propagators}\label{app0}
\setcounter{equation}{0}
\renewcommand{\theequation}{\ref{app0}.\arabic{equation}}

In this Appendix we prove the dimensional bounds on the propagators
used in Sections \ref{sec4}-\ref{sec5}-\ref{sec6}-\ref{sec7} and, in particular, the bounds 
$||g||_{\infty}\le CM$ and $||g||_1\le C\b$, stated right before Eq.(\ref{s4.16}), and the estimates 
Eqs.(\ref{B.3}),(\ref{2.38}). The basic idea is to decompose the propagator in Eq.(\ref{s4.4a}),
\be g(\xx)=\frac{1}{\b|\L|}
\sum_{\kk\in\BBB^{(M)}_{\b,L}}e^{-i\kk\xx}\frac{\c_0(2^{-M}|k_0|)}{k_0^2+v_0^2
|\O(\vec k)|^2} \begin{pmatrix}i k_0 & -v_0\O^*(\vec k) \\ -v_0\O(\vec k) &
ik_0\end{pmatrix}\;,\label{0.1}\ee
 as a sum of single scale propagators: 
\be g(\xx)=\sum_{h=1}^M g^{(h)}(\xx)+\sum_{\o=\pm}
\sum_{h=h_\b}^0 \lis g^{(h)}_\o(\xx)\;,\label{0.2}\ee
where: (i) if $1\le h\le M$, then 
$g^{(h)}(\xx)$ is defined as in Eq.(\ref{B.2}); (ii) if $h_\b:=\lfloor
\log_2\big(\frac{3\p}{4\b}\big)\rfloor$ and $h_{\b}\le h\le 0$, then  
\be \lis g^{(h)}_\o(\xx)=\frac{1}{\b|\L|}
\sum_{\kk\in\BBB^{(M)}_{\b,L}}e^{-i\kk\xx}\frac{f_h(k_0, \vec k-\vec p_F^{\,\o})}{k_0^2+v_0^2
|\O(\vec k)|^2} \begin{pmatrix}i k_0 & -v_0\O^*(\vec k) \\ -v_0\O(\vec k)& ik_0\end{pmatrix}\;,\label{0.3}\ee
with $f_h(\kk')=\c_0(2^{-h}|\kk'|)-\c_0(2^{-h+1}|\kk'|)$. The main issue is to prove Eq.(\ref{B.3}), i.e., 
\be ||g^{(h)}(\xx)||\le \frac{C_K}{1+ (2^h|x_0|_\b+|\vec x|_\L)^K}
\;,\label{B.3copy}\ee
and the analogue of Eq.(\ref{2.38}), i.e., 
\be ||g^{(h)}_{\o}(\xx)||\le \frac{C_{K}2^{2h}}{1+(2^{h}||\xx||)^K}\;,\label{2.38copy}\ee
where $||\xx||^2=|x_0|_\b^2+|\vec x|_\L^2$, for all $K\ge 0$. 
Note that $||g||_{\infty}\le CM$ and $||g||_1\le C\b$ are immediate corollaries of 
Eqs.(\ref{B.3copy})-(\ref{2.38copy}). In fact, plugging these bounds into Eq.(\ref{0.2}),
we find:
\bea  ||g||_{\infty} &\le& \sum_{h=1}^M C_0+\sum_{\o=\pm}
\sum_{h=h_\b}^0 C_0 2^{2h}\le C_0(M+\frac83)\;,\label{0.4}\\
 ||g||_{1} &\le&  C_4 \sum_{h=1}^M \int_{(\b,\L)}\hskip-.5truecm
 d\xx\frac{1}{1+ (2^h|x_0|_\b+|\vec x|_\L)^K}+ 2C_4 
 \sum_{h=h_\b}^0 \int_{(\b,\L)}\hskip-.5truecm
 d\xx\frac{2^{2h}}{1+ (2^h||\xx||)^4}
 \le\nn\\
 &\le &  C_4'\sum_{h=h_\b}^M 2^{-h}\le C_4' 2^{-h_\b+1}\;,\label{0.4bb}\eea
which are the desired bounds. 

So, let us start by proving Eq.(\ref{B.3copy}). We denote by $\dpr_\kk$ the discrete derivative, and 
by $\tilde \dpr_\m=\tilde e_\m\cdot \dpr_\kk$, with $\m\in\{0,1,2\}$,
the discrete derivative in the direction $\tilde e_\m$ (here $\tilde e_0=(1,0,0)$, 
$\tilde e_1=(0,\frac12,\frac{\sqrt3}{2})$ and $\tilde e_2=(0,\frac12,-\frac{\sqrt3}{2})$),
defined as follows: given a compact support function $\hat F(\kk)$ with 
$\kk\in {\cal B}_{\b,L}^{(\infty)}$, we let
\bea && \tilde\dpr_{\m} \hat F(\kk) =\tilde e_\m\cdot\dpr_\kk\hat F(\kk)=
\frac{\hat F(\kk+\tilde e_\m \D k_\m)-\hat F(\kk)}{\D k_\m} 
\;,\label{0.8}\eea
with $\D k_0 = \frac{2\p}{\b}$ and $\D k_1 = \D k_2=\frac{4\p}{3L}$.
%
Note that, if
\be F(\xx) = \sum_{\kk\in {\cal B}_{\b,L}^{(\infty)}} e^{-i\kk\cdot\xx}
\hat F(\kk) \; , \label{0.9}\ee
then, defining $\tilde x_\m=\xx\cdot\tilde e_\m$,
\be \sum_{\kk\in {\cal B}_{\b,L}^{(\infty)}} e^{-i\kk\cdot\xx} \tilde\dpr_{\m}
\hat F(\kk) = \left( \frac{e^{i\D k_\m \tilde x_\m} - 1}{\D k_\m} \right)
\sum_{\kk\in {\cal B}_{\b,L}^{(\infty)}} e^{-i\kk\cdot\xx} \hat F(\kk)
\; , \label{0.10}\ee
so that, using the fact that $|x_0|_\b\le \frac{\p}2 d_\b(x_0)$ and $|\vec x|_\L\le \frac{\p}{\sqrt2}
(d_L^2(\tilde x_1)+d_L^2(\tilde x_2))^{1/2}$ (here $d_\b(x_0)=\frac{\sin(\p x_0/\b)}{\p/\b}$ and
$d_L(\tilde x_i)=\frac{\sin(2\p \tilde x_i/3L)}{2\p/3L}$), we find
\bea  |x_0|_\b^2 |F(\xx)|
&\le& \frac{\p^2}4d_\b^2(x_0)|F(\xx)|=
\frac{\p^2}4\Big|\frac{e^{i\D k_0 x_0} - 1}{\D k_0}\Big|^2\cdot\Big|
\sum_{\kk\in {\cal B}_{\b,L}^{(\infty)}} e^{-i\kk\cdot\xx} \hat F(\kk)\Big|=\nn\\
&=&\frac{\p^2}4\Big|
\sum_{\kk\in {\cal B}_{\b,L}^{(\infty)}} e^{-i\kk\cdot\xx} \dpr_{k_0}^2\hat F(\kk)\Big|
\le \frac{\p^2}{4}\sum_{\kk\in {\cal B}_{\b,L}^{(\infty)}} \big| \dpr_{k_0}^2\hat F(\kk)\big|
\; , \label{0.11}\eea
and, similarly,
 \bea |\vec x|_{\L}^2 |F(\xx)|
&\le& \frac{\p^2}2\Big(d_L^2(\tilde x_1)+d_L^2(\tilde x_2)\Big)|F(\xx)|=\nn\\
&=&\frac{\p^2}2\sum_{\m=1}^2\Big|\frac{e^{i\D k_\m \tilde x_\m} - 1}{\D k_\m}\Big|^2\cdot\Big|
\sum_{\kk\in {\cal B}_{\b,L}^{(\infty)}} e^{-i\kk\cdot\xx} \hat F(\kk)\Big|=\label{0.11as}\\
&=&\frac{\p^2}2\sum_{\m=1}^2\Big|
\sum_{\kk\in {\cal B}_{\b,L}^{(\infty)}} e^{-i\kk\cdot\xx} \tilde\dpr_\m^2\hat F(\kk)\Big|
\le \frac{3\p^2}{8}\sum_{\kk\in {\cal B}_{\b,L}^{(\infty)}} \big| (\dpr_{k_1}^2+\dpr_{k_2}^2)
\hat F(\kk)\big| \; .\nn\eea
Coming back  to Eq.(\ref{B.3copy}), recalling that 
\be g^{(h)}(\xx)=\frac{1}{\b|\L|}\sum_{\kk\in\BBB^{(M)}_{\b,L}} e^{-i\kk\xx}\,
\frac{f_{u.v.}(\kk)H_h(k_0)}{k_0^2+v_0^2 |\O(\vec k)|^2} \begin{pmatrix}i k_0 & 
-v_0\O^*(\vec k) \\ -v_0\O(\vec k) & ik_0\end{pmatrix}\;,\qquad h\ge 2\;.\label{0.12z}\ee
Therefore, 
\be ||g^{(h)}(\xx)||\le \frac{1}{\b|\L|}
\sum_{\kk\in\BBB^{(M)}_{\b,L}} 
\frac{f_{u.v.}(\kk)H_h(k_0)}{\big[k_0^2+v_0^2 |\O(\vec k)|^2\big]^{1/2}}\le (\const.)2^h\, 2^{-h}\;,
\label{0.12ed}\ee
where in the last inequality we used the fact that, on the support of $f_{u.v.}(\kk)H_h(k_0)$, 
$k_0^2+v_0^2|\O(\vec k)|^2\sim 2^{2h}$ 
(here $\sim$ means that the ratio of the two sides is bounded 
above and below by universal constants) and that, moreover, the measure of the support of 
$f_{u.v.}(\kk)H_h(k_0)$ is itself of order $2^h$, that is $(\b|\L|)^{-1}\sum_\kk f_{u.v.}(\kk)
H_h(k_0)\sim 2^h$. 

Moreover, using Eqs.(\ref{0.11})-(\ref{0.11as}), we have that 
for all $N\ge 0$ and a suitable constant $C$,
\bea && |x_0|_\b^{2N}||g^{(h)}(\xx)||\le \frac{C^N}{\b|\L|}
\sum_{\kk\in\BBB^{(M)}_{\b,L}} \Big|\!\Big|\dpr_{k_0}^{2N}
\Big[\frac{f_{u.v.}(\kk)H_h(k_0)}{k_0^2+v_0^2 |\O(\vec k)|^2} \begin{pmatrix}i k_0 & 
-v_0\O^*(\vec k) \\ -v_0\O(\vec k) & ik_0\end{pmatrix}\Big]\Big|\!\Big|\;,\nn\\
&&|\vec x|_\L^{2N}||g^{(h)}(\xx)||\le \frac{C^N}{\b|\L|}\!\!
\sum_{\kk\in\BBB^{(M)}_{\b,L}} \Big|\!\Big|(\vec \dpr_{\vec k}\cdot \vec\dpr_{\vec k})^{N}
\Big[\frac{f_{u.v.}(\kk)H_h(k_0)}{k_0^2+v_0^2 |\O(\vec k)|^2} \begin{pmatrix}i k_0 & 
-v_0\O^*(\vec k) \\ -v_0\O(\vec k) & ik_0\end{pmatrix}\Big]\Big|\!\Big|\nn\eea
Now, in the first line, every derivative with respect to $k_0$ can act either 
on the denominator $k_0^2+v_0^2|\O(\vec k)|^2$, or on 
$f_{u.v.}(\kk)H_h(k_0)$, or on the diagonal elements of the matrix,
$ik_0$; in all these cases, recalling that, on the support of $f_{u.v.}H_h$, $|k_0|\sim 2^h$, every 
derivative $\dpr_{k_0}$ can be estimated dimensionally by a factor proportional to $2^{-h}$. 
This leads to the bound
\be  |x_0|_\b^{2N}||g^{(h)}(\xx)||\le (\const.)^N 2^h\, 2^{-2Nh}\,2^{-h}\;,\label{0.12dr}\ee
which differs from Eq.(\ref{0.12ed}) precisely by the factor $2^{-2Nh}$. Similarly, in the second line,
the derivative with respect to $\vec k$ can act either on the denominator $k_0^2+
v_0^2|\O(\vec k)|^2$, or on 
$f_{u.v.}(\kk)H_h(k_0)$, or on the off-diagonal elements of the matrix, $-v_0\O(\vec k)$ or $-v_0\O^*(\vec k)$;
in the first case,  $\vec \dpr_{\vec k}$ can be estimated dimensionally by a factor proportional to 
$2^{-h}$ while, in the other cases, by an order 1 factor. This leads to the bound 
\be  |\vec x|_\L^{2N}||g^{(h)}(\xx)||\le (\const.)^N 2^h\,2^{-h}\;,\label{0.12dp}\ee
which, if combined with Eqs.(\ref{0.12ed})-(\ref{0.12dr}), finally implies Eq.(\ref{B.3copy}).

The proof of Eq.(\ref{B.3copy}) is very similar. In fact, for all $N\ge 0$,
\be ||\xx||^{2N}\,||\lis g_\o^{(h)}(\xx)||\le \frac{C^N}{\b|\L|}
\sum_{\kk\in\BBB^{(M)}_{\b,L}} \Big|\!\Big|\dpr_{\kk}^{2N}
\Big[\frac{f_h(k_0, \vec k-\vec p_F^{\,\o})}{k_0^2+v_0^2 |\O(\vec k)|^2} \begin{pmatrix}i k_0 & 
-v_0\O^*(\vec k) \\ -v_0\O(\vec k) & ik_0\end{pmatrix}\Big]\Big|\!\Big|\label{0.14}\ee
Now, using the fact that, on the support of $f_h(k_0, \vec k-\vec p_F^{\,\o})$, 
$|k_0|,|\vec k-\vec p_F^\o|,|\O(\vec k)|\sim 2^h$, we see that 
every derivative $\dpr_{\kk}$, when acting either on $f_h$, or on the denominator $k_0^2+v_0^2|\O(\vec k)|^2$, or on the elements of the matrix, can be estimated dimensionally by a factor proportional to 
$2^{-h}$; moreover, the measure of the support of 
$f_{h}$ is itself of order $2^{3h}$, that is $(\b|\L|)^{-1}\sum_\kk f_{h}(k_0,\vec k-\vec p_F^\o)\sim 2^{3h}$. This leads to the bound
\be  ||\xx||^{2N}\,||\lis g_\o^{(h)}(\xx)||\le (\const.)^N 2^{3h}\, 2^{-2Nh}\, 2^{-h}\;,\label{0.15}\ee
which finally implies Eq.(\ref{B.3copy}).\qed

\section{Truncated expectations and determinants}\label{app1}
\setcounter{equation}{0}
\renewcommand{\theequation}{\ref{app1}.\arabic{equation}}
 
\begin{figure}[hbtp]
\centering
\includegraphics[height=0.35\textwidth]{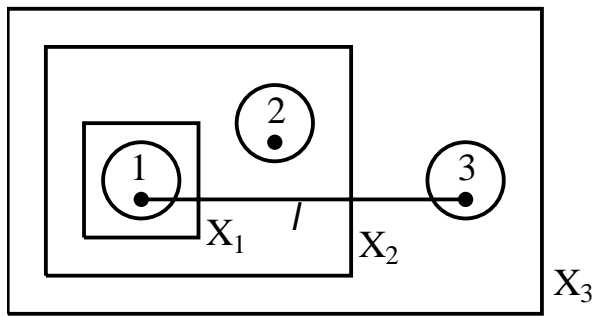}
\caption{Graphical representation of the sets $X_{k}$,
$ k=1,2,3$. In the example $X_{1}=\{1\}$, $X_{2}=\{1,2\}$, $X_{3}=\{1,2,3\}$ and $\ell=(1, 3)$.
The line $\ell$ intersects both the boundary of 
$X_1$ and of $X_2$, i.e., $\ell\sim\dpr X_1$ and $\ell\sim\dpr X_2$.}
\label{figa1.1}
\end{figure}

In this Appendix we prove Eq.(\ref{s5.3}), following \cite[Appendix A.3.2]{GeM}.
Given $s$ set of indices $P_{1},\ldots,P_{s}$, consider the quantity 
$\EE^T(\psi_{P_1},\ldots,\psi_{P_s})$. Define 
\be P_j^\pm=\{f\in P_j\ :\ \e(f)=\pm\}\label{a1.1}\ee
and set $f=(j,i)$ for $f\in P_j^\pm$, with $i=1,\ldots,|P_j^\pm|$. Note that 
$\sum_{j=1}^s|P_j^+|=\sum_{j=1}^s|P_j^-|$, otherwise the considered truncated expectation is 
vanishing. Define
\be {\cal D}\psi=\prod_{j=1}^{s}
\Big[\prod_{f\in P_{j}^+} \der\psi^{+}_{\xx(f),\s(f),\r(f)} \Big]\Big[\prod_{f\in P_{j}^-} 
\der\psi^{-}_{\xx(f),\s(f),\r(f)} \Big]\;,\label{a1.2}\ee
\be ( \psi^+,G\psi^- ) =\sum_{j,j'=1}^s\sum_{i=1}^{|P_j^-|}\sum_{i'=1}^{|P_{j'}^+|}
\psi^+_{(j',i')}G_{(j,i),(j',i')}\psi^-_{(j,i)}\;,\label{a1.3FG}\ee
where $\psi^\pm_{(j,i)}:=\psi^\pm_{\xx(j,i),\s(j,i),\r(j,i)}$ and, if 
$ n = \sum_{j=1}^{s} \left| P_{j}^+ \right|=\sum_{j=1}^{s} \left| P_{j}^-\right|$,
then $G$ is the $n\times n$ matrix with entries
\be G_{(j,i),(j',i')}:= \d_{\s(j,i),\s(j',i')}g_{\r(j,i),\r(j',i')}(\xx(j,i)-\xx(j',i'))\;,\label{a1.4JU}\ee
so that 
\be \EE \left( \prod_{j=1}^{s} \psi_{P_{j}} \right)
= \det G = \int {\cal D}\psi \, \exp\left[
-\left( \psi^+,G\psi^- \right) \right] \; . \label{a1.6}\ee
Setting $X\=\{1,\ldots,s\}$ and
\be \overline V_{jj'}
= \sum_{i=1}^{|P_j^-|}\sum_{i'=1}^{|P_{j'}^+|}
\psi^+_{(j',i')}G_{(j,i),(j',i')}\psi^-_{(j,i)}\;,\label{a1.7}\ee
we write
\be V(X) = \sum_{j,j'\in X} \overline V_{jj'} = \sum_{j \le j'}
V_{jj'} \; , \label{a1.8}\ee
where 
\be V_{jj'} = \begin{cases}
\overline V_{jj'} \;, & {\rm if}\  j=j' \; ,  \\
\overline V_{jj'} + \overline V_{j'j} \; , & {\rm if}\ j<j' \; .
\end{cases} \label{a1.9}\ee
In terms of these definitions, Eq.(\ref{a1.6}) can be rewritten as
\be \EE \left( \prod_{j=1}^{s} \psi_{P_{j}} \right)
= \det G = \int {\cal D}\psi \, 
e^{-V(X)}\;. \label{a1.10}\ee
We now want to express the last expression in terms 
of the functions $W_{X}$, defined as follows:
\be W_{X}(X_{1},\ldots,X_{r};t_{1},\ldots,t_{r}) =
\sum_{\ell\in L(X)} \prod_{k=1}^{r} t_{k}(\ell) \, V_{\ell} \; , \label{a1.11}\ee
where:\begin{enumerate}
\item $X_{k}$ are subsets of $X$ with $|X_{k}|=k$ and such that 
\be \begin{cases}
X_{1} = \{1\} \; , & \\
X_{k+1} \supset X_{k} \; ; \end{cases} \label{a1.12}\ee
\item $L(X)$ is the set of unordered pairs in $X$, i.e., the set of pairs $(j,j')$ with 
$j,j'\in X$, such that $(j,j')$ is identified with $(j',j)$ and, possibly, $j=j'$;
we shall say that $\ell=(j,j')$ is {\it non tivial} if $j\neq j'$, and {\it trivial} otherwise;
\item the functions $t_{k}(\ell)$ are defined as follows:
\be t_{k}(\ell) = \begin{cases}
t_{k} \; , & {\rm if} \quad \ell \sim \dpr X_{k} \; ,  \\
1 \; , & {\rm otherwise} \; , \end{cases} \label{a1.13}\ee
where $\ell\sim X_{k}$ means that $\ell=(j,j')$ ``intersects the boundary'' of 
$X_{k}$, i.e., it connects $j \in X_{k}$ with  $j'\notin X_{k}$, or viceversa.
See Fig. \ref{figa1.1}.
\end{enumerate}

Let us show how to re-express $e^{-V(X)}$ in terms of the $W_X$'s. The basic step is the following: using the definition
Eq.(\ref{a1.11}), we rewrite
\be W_{X}(X_{1};t_{1}) =
 t_{1} V(X)+(1-t_{1}) \left[ V(X_{1}) + V(X \bs X_{1}) \right]
 \;;\label{a1.14}\ee
that is, we recognize that $ W_{X}(X_{1};t_{1}) $ interpolates between the full $V(X)$ and two
of its ``proper subsets", $V(X_1)$, $V(X\bs X_1)$. In this way,
\bea e^{-V(X)} & =& e^{-W_{X}(X_{1};0) } +\int_{0}^{1} \der t_{1} \left[
\frac{\partial}{\partial t_{1} }\, e^{-W_{X}(X_{1};t_{1}) }
\right] \nn\\
& =&e^{-W_{X}(X_{1};0) } - \sum_{ \ell_{1} \sim \partial X_{1} } V_{\ell_{1}} \,
\int_{0}^{1} \der t_{1} \, e^{-W_{X}(X_{1};t_{1}) } \; . \label{a1.15}\eea
Let us now iterate this construction: let us consider one of the terms in the summation in 
the r.h.s. of Eq.(\ref{a1.15}); if $\ell_1=(1,j^*)$, we let  $X_{2} \= X_{1} \cup \{j^*\}$ and we 
note that, by definition,
\be
W_{X}( X_{1}, X_{2}; t_{1}, t_{2} )  =
t_{2} W_{X} ( X_{1};t_{1} ) + 
(1-t_{2}) \left[ W_{X_{2}} ( X_{1};t_{1} ) +
V ( X \bs X_{2} ) \right]\;,\label{a1.16}\ee
so that 
\bea 
e^{ - W_{X} ( X_{1};t_{1} ) } & =&e^{-W_{X}(X_{1},X_{2};t_{1},0)}+
\int_{0}^{1} \der t_{2} \, \left[ \frac{\partial}{\partial t_{2}}
\, e^{-W_{X}(X_{1},X_{2};t_{1},t_{2}) } \right] 
 \label{a1.17}\\
& =& e^{-W_{X}(X_{1},X_{2};t_{1},0)}- \sum_{\ell_{2} \sim \partial X_{2} } 
V_{\ell_{2}} \int_{0}^{1} \der t_2 \, t_{1} (\ell_{2})
\, e^{-W_{X}(X_{1},X_{2};t_{1},t_{2}) }\; . \nn\eea
Substituting Eq.(\ref{a1.17}) into Eq.(\ref{a1.15}) we get:
\bea  e^{-V(X)}
& = &e^{-W_{X}(X_{1};0)}+ \sum_{ \ell_{1} \sim \partial X_{1} } 
\int_{0}^{1} \der t_{1} \, (-1) \, V_{\ell_{1}} \,
e^{-W_{X}(X_{1},X_{2};t_{1},0) }  +\label{a1.18}\\
&+&\sum_{ \ell_{1} \sim \partial X_{1} }
\sum_{ \ell_{2} \sim \partial X_{2} }
\int_{0}^{1} \der t_{1}
\int_{0}^{1} \der t_{2} \,
(-1)^{2} \, V_{\ell_{1}} \, V_{\ell_{2}} \,
t_{1}(\ell_{2}) \, e^{-W_{X}(X_{1},X_{2};t_{1},t_{2}) } \; . \nn\eea
Using the fact that 
\bea && W_X(X_1,\ldots, X_{p+1}; t_1,\ldots, t_{p+1})= 
t_{p+1}W_X(X_1,\ldots, X_{p}; t_1,\ldots, t_{p})+\nn\\
&& (1-t_{p+1})\left[
W_{X_{p+1}}(X_1,\ldots, X_{p}; t_1,\ldots, t_{p})+V(X\setminus X_{p+1})\right]
\label{a1.19}\eea
and iterating we find
\bea e^{-V(X)}
& = &\sum_{r=0}^{s-1} \sum_{ \ell_{1} \sim \partial X_{1} }
\ldots
\sum_{ \ell_{r} \sim \partial X_{r} }
\int_{0}^{1} \der t_{1} \ldots
\int_{0}^{1} \der t_{r} \, 
(-1)^{r} \,
V_{\ell_{1}} \ldots V_{\ell_{r}} \nn\\
&& \left(
\prod_{k=1}^{r-1} t_{1}(\ell_{k+1}) \ldots
t_{k}(\ell_{k+1}) \right)
e^{-W_{X}(X_{1}, \ldots , X_{r+1} ; 
t_{1}, \ldots, t_{r}, 0 ) } \; , \label{a1.21}\eea
where, if $r=0$, the summand should be interpreted as equal to $e^{-W_X(X_1;0)}$.
Moreover, by the definition  of the $W_X$'s, if $r>1$,
\be
W_{X} ( X_{1}, \ldots , X_{r} ; t_{1}, \ldots, t_{r-1}, 0 ) =
W_{X_{r}} ( X_{1}, \ldots , X_{r-1} ; t_{1}, \ldots, t_{r-1} )
+ V ( X \setminus X_{r} ) \;.\label{a1.22}\ee
%
Using this remark and the notion of {\it anchored tree}, which will be defined in a moment,
Eq.(\ref{a1.21}) can be rewritten in a more suggestive and convenient way.
Let $\TT(X)$ be the set of tree graphs on $X$, i.e., the set of $(|X|-1)$-ples of lines in $L(X)$
connecting (in a minimal way) all the elements of $X$. Given a sequence of subsets 
$X_1\subset\cdots\subset X_r$ as above, we shall say that $T\in\TT(X_r)$ is an {\it anchored tree
on $(X_1,\ldots,X_r)$} if its lines can be ordered in such a way that 
$\ell_{1}\sim \dpr X_{1}, \ell_{2}\sim \dpr X_{2},\ldots, \ell_{r-1}\sim\dpr X_{r-1}$.
Moreover, given a sequence $X_1\subset\cdots\subset X_s$ as above and a non-trivial line 
$\ell\in L(X)$, we let
\be n(\ell) = \max \{ k : \ell \sim \partial X_{k} \} \; , \qquad
n'(\ell) = \min \{ k : \ell \sim \partial X_{k} \} \;; \label{a1.24}\ee
if $\ell$ is trivial, we let $n(\ell)=n'(\ell)=0$. Using these definitions, we rewrite Eq.(\ref{a1.21}) as:
\bea &&e^{-V(X)}  = \sum_{r=1}^{s} 
\sum_{X_{r} \subset X } \; 
\sum_{ X_{2}, \ldots, X_{r-1} } \;
\sum_{T \; {\rm on} \; (X_1,\ldots,X_{r})}
(-1)^{r-1} \prod_{\ell \in T } V_{\ell}\cdot  \label{a1.25}\\
&&\qquad\cdot \int_{0}^{1} \der t_{1} \ldots \int_{0}^{1} \der t_{r-1}
\left( \prod_{\ell \in T}
\frac{ \prod_{k=1}^{r-1} t_{k}(\ell) }{
t_{n(\ell)} } \right) 
e^{-W_{X_{r}} (X_{1}, \ldots , X_{r-1} ; t_{1}, \ldots, t_{r-1} ) } 
\, e^{-V(X \setminus X_{r}) }\;,\nn\eea
where ``$T \; {\rm on} \; (X_1,\ldots,X_{r})$'' means that $T$ is an anchored tree on 
$(X_1,\ldots,X_r)$. Defining
\bea K(X_{r}) & = &
\sum_{ X_{2},\ldots, X_{r-1} } \;
\sum_{T \; {\rm on} \; (X_1,\ldots,X_{r})}
\prod_{\ell \in T } V_{\ell} \label{a1.26}\\
&& \int_{0}^{1} \der t_{1} \ldots
\int_{0}^{1} \der t_{r-1} \left(
\prod_{\ell \in T}
\frac{ \prod_{k=1}^{r-1} t_{k}(\ell)}{
t_{n(\ell)} } \right)
e^{-W_{X_{r}} (X_{1}, \ldots , X_{r-1} ; t_{1}, \ldots, t_{r-1} ) } 
\; , \nn\eea
Eq.(\ref{a1.25}) can be further rewritten as
\be e^{-V(X)} = \sum_{\substack{Y \subset X \\Y \ni \{ 1 \}} }
(-1)^{|Y|-1} \, K(Y) \, e^{-V(X \setminus Y)} \; , \label{a1.27}\ee
and, iterating,
\be e^{-V(X)} = \sum_{m=1}^{s}\sum_{(Y^1,\ldots,Y^m)}
(-1)^{s} \,
(-1)^{m} \prod_{i=1}^{m} \, K(Y^i) \; , \label{a1.28}\ee
where the second summation runs over partitions of $X$ of multiplicity $m$, i.e., 
over $m$-ples of disjoint sets $Y^{1},\ldots,Y^{m}$
such that $\cup_{i=1}^mY^i=X$. Plugging Eq.(\ref{a1.28}) into Eq.(\ref{a1.10}) gives
\be\EE \left( \prod_{j=1}^{s} \psi_{P_{j}} \right) =
 \sum_{m=1}^{s}\sum_{(Y^1,\ldots,Y^m)}
(-1)^{s+m}(-1)^\s\int {\cal D}\psi_{Y^i} \prod_{i=1}^m K(Y^i)\;,\label{a1.31}\ee
where ${\cal D}\psi_{Y^i}=\prod_{j\in Y^i}
\Big[\prod_{f\in P_{j}^+} \der\psi^{+}_{\xx(f),\s(f),\r(f)} \Big]\Big[\prod_{f\in P_{j}^-} 
\der\psi^{-}_{\xx(f),\s(f),\r(f)} \Big]$ and $(-1)^\s$ is the sign of the permutation leading from the
ordering of the fields in $\DD\psi$ to the ones in $\prod_i\DD\psi_{Y^i}$. In Eq.(\ref{a1.31}),
each factor $K(Y^i)$, after small manipulations of its definition, Eq.(\ref{a1.26}), can be rewritten as
\bea 
K(Y^i)&=&\sum_{T\in\TT(Y^i)}
\sum_{\substack{Y^i_2,\ldots, Y^i_{|Y^i|-1}\\{\rm compatible}\ {\rm with}\  T}}
\prod_{\ell\in T}V_\ell\int_0^1 d t_1\cdots\int_0^1 d t_{|Y^i|-1}\cdot\nn\\
&&\cdot\prod_{\ell\in T}(t_{n'(\ell)}\cdots t_{n(\ell)-1})e^{-\sum_{\ell\in L(Y^i)}
t_{n'(\ell)}\cdots t_{n(\ell)}V_\ell}\;,\label{a1.30}\eea
where: (i) $Y^i_1:=\{\min\{ j: j\in Y^i\}\}$; (ii) the second summations is over sequences of subsets
$Y^i_q$ of $Y^i$ with $|Y^i_q|=q$ and 
such that $Y^i_1\subset Y^i_2\subset\cdots\subset Y^i_{|Y^i|-1}\subset Y^i_{|Y^i|}\equiv Y^i$
and  $T$ is anchored on $(Y^i_1,\ldots,Y^i_{|Y^i|})$; (iii) in the second line, if $n(\ell)=n'(\ell)$, 
the factor $t_{n'(\ell)}\cdots t_{n(\ell)-1}$ should be interpreted as equal to $1$; (iv)
in the exponent, if $\ell$ is trivial (and, therefore, $n'(\ell)=n(\ell)=0$), $t_0$ should be interpreted
as equal to $1$. Now, using the analogue of Eq.(\ref{s4.12}) in a case, like the one at hand, 
where the monomials $\psi_{}$ do not necessarily all commute among each other, 
we have that (denoting the elements of $Y^i$ as $Y^i=\{j^i_1,\ldots,j^i_{|Y_i|}\}$)
\be \EE \left( \prod_{j=1}^{s} 
\psi_{P_j}\right) =\sum_{m=1}^s\sum_{(Y^1,\ldots, Y^m)}(-1)^{\s}
\EE^T(\psi_{P_{j^1_1}},\ldots,\psi_{P_{j^1_{|Y^1|}}})\cdots
\EE^T(\psi_{P_{j^m_1}},\ldots,\psi_{P_{j^m_{|Y^m|}}})
\label{a1.32}\ee
where $(-1)^\s$ is the parity of 
the permutation leading to the ordering on the r.h.s. from the one
on the l.h.s.; note that $\s$ is the same as in eqn(\ref{a1.31}). Comparing Eq.(\ref{a1.32}) with 
Eq.(\ref{a1.31}) and using Eq.(\ref{a1.30}), we get:
\be\EE^T(\psi_{P_1},\ldots,\psi_{P_s})=(-1)^{s+1}
\sum_{T\in\TT(X)}
\int {\cal D}\psi
\prod_{\ell\in T}V_\ell \int dP_T(\tt)e^{-V(\tt)}\;,\label{a1.33}\ee
where we defined:
\be dP_T(\tt)=\sum_{\substack{X_2,\ldots, X_{s-1}\\
{\rm compatible}\ {\rm with}\  T}}\prod_{\ell \in T}
\left( t_{n' ( \ell ) } \ldots t_{n( \ell )-1 } \right)
\prod_{q=1}^{s-1} \der t_{q}\label{a1.34}\ee
and 
\be V(\tt) \equiv \sum_{\ell \in L(X)}t_{n' ( \ell ) } \ldots t_{n( \ell )} \, V_{\ell}\;.\label{a1.35}\ee
Finally, if we use the definitions Eqs.(\ref{a1.7}),(\ref{a1.9}) and explicitly integrate the Grassmann 
variables along the lines of the anchored tree, we end up with 
\be \EE^{T} \left(\psi_{P_{1}}, \ldots,
\psi_{P_{s}} \right) = \sum_{T\in {\bf T({\cal P})}}\a_T
\prod_{\ell \in T } G_{f^1_\ell,f^2_\ell}
\int {\cal D}^*(\der\psi)\int \der P_{T}(\tt) \, e^{-V^*(\tt)} \; , \label{a1.37}\ee
where ${\cal P}=\{P_1,\ldots,P_s\}$ and ${\bf T({\cal P})}$ is the set of anchored trees
between the ``boxes" $P_1,\ldots,P_s$, i.e., the graphs that become trees if one identifies 
all the points in the same ``clusters" $P_i$ (note that now the lines of an anchored tree $T
\in{\bf T}(\PPP)$ are pairs of field variables $(f,f')$, rather than pairs of indices $(j,j')$). Moreover, 
$\a_T$ is a sign (irrelevant to the purpose of the bounds performed in this paper),
$f^1_\ell$, $f^2_\ell$ are the two field labels associated to the two (entering and exiting)
half-lines contracted into $\ell$, and
\be {\cal D}^*(\der\psi)=\prod_{\substack{f\in P\\ f\not\in T}}d \psi^{\e(f)}_{\xx(f),\s(f),\r(f)}
\;,\qquad V^*(\tt)=\sum_{\ell \in L(X)}
t_{n' ( \ell ) } \ldots t_{n( \ell )} \, V^T_{\ell}\label{a1.38}\ee
where 
\be V^T_{jj'} = \begin{cases}
\overline V^T_{jj'} \;, & {\rm if}\  j=j' \; ,  \\
\overline V^T_{jj'} + \overline V^T_{j'j} \; , & {\rm if}\ j<j' \; .
\end{cases} \label{a1.9T}\ee
and, if $\big((j,i),(j',i')\big)$ is the line obtained by contracting 
$\psi^-_{(j,i)}$ with $\psi^+_{(j',i')}$, 
\be \overline V^T_{jj'}
= \sum_{i=1}^{|P_j^-|}\sum_{i'=1}^{|P_{j'}^+|}
\psi^+_{(j',i')}G_{(j,i),(j',i')}\psi^-_{(j,i)}\,\c\Big(\big((j,i),(j',i')\big)\not\in T\Big)\;.\label{a1.7T}\ee
The term $\int {\cal D}^*(\der\psi) \, e^{-V^*(\tt)}$
in Eq.(\ref{a1.37}) is the determinant of the $(n-s+1)\times (n-s+1)$ matrix
$G^{T}(\tt)$ (here $2n=\sum_{i=1}^s|P_{i}|$), with elements $G^{T}_{f,f'}(\tt)=t_{j(f),j(f')} G_{f,f'}$,
where $f, f'\not\in T$, $j(f)\in\{1,\ldots,s\}$ and $t_{j,j'}=
t_{n'((j,j'))} \ldots t_{n((j,j'))}$: 
\be  \int {\cal D}^*(\der\psi) \, e^{-V^*(\tt)} =\det G^T(\tt)\;.\label{a1.39}\ee
Plugging this into Eq.(\ref{a1.37}) finally gives Eq.(\ref{s5.3}). In order to complete the proof of 
the claims following Eq.(\ref{s5.3}) we are left with proving the following Lemma.

\begin{lemma}\label{lemA1.1}
$dP_T(\tt)$ is a normalized, positive and $\s$--additive 
measure on the natural $\s$--algebra of $[0,1]^{s-1}$. Moreover 
there exists a set of unit vectors ${\bf u}_j\in\RRR^s$, $j=1,\ldots,s$, 
such that $t_{j,j'}={\bf u}_j\cdot{\bf u}_{j'}$.\end{lemma}

\begin{figure}[hbtp]
\centering
\includegraphics[height=.7\textwidth]{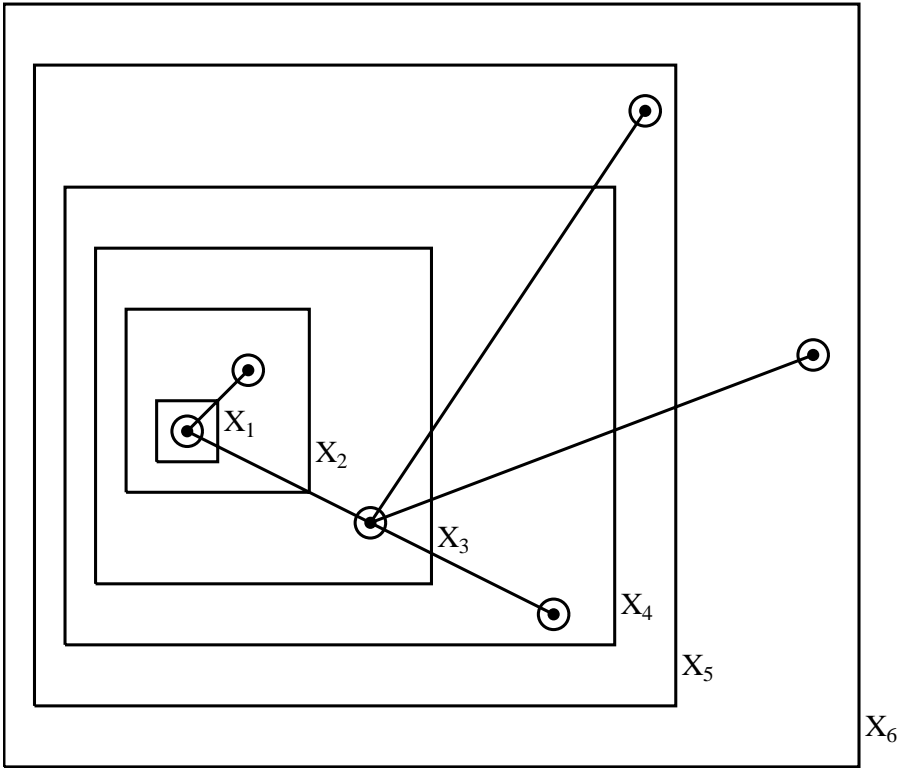}
\caption{The sets $X_{1},\ldots,X_{6}$, the anchored tree
$T$ and the lines $\ell_1,\ldots,\ell_5$ belonging to $T$. In the example, 
the coefficients $b_1,\ldots, b_5$ are respectively 
equal to: $2,1,3,2,1$.}\label{figa1.2}
\end{figure}

{\bf Proof.} Let us denote by $b_k$ the number of lines $\ell\in T$ exiting from
the points $x(j,i)$, $j\in X_k$, such that $\ell\sim X_k$. Let us consider
the integral 
\be \sum_{\substack{X_2,\ldots, X_{s-1}\\
{\rm compatible}\ {\rm with}\  T}}
\int_{0}^{1} \der t_{1} \ldots
\int_{0}^{1} \der t_{s-1} \,
\prod_{\ell \in T}
\left( t_{n' ( \ell ) } \ldots
t_{n( \ell )-1 } \right) = 1 \; , \label{a1.41}\ee
and note that, by construction, the 
parameter $t_k$ inside the integral in the l.h.s. appears at the 
power $b_k-1$. In fact any line intersecting $\dpr X_k$ contributes 
by a factor $t_k$, except for the line connecting $X_k$ with the point 
in $X_{k+1}\bs X_k$. See Fig. \ref{figa1.2}.

Then
\be \prod_{\ell \in T}
\left( t_{n'(\ell)} \ldots t_{n( \ell )-1 } \right) =
\prod_{k=1}^{s-1} t_{k}^{b_{k}-1} \; , \label{a1.42}\ee
and in Eq.(\ref{a1.41}) the $s-1$ integrations are independent.
One has: 
\be \int_{0}^{1} \der t_{1} \ldots
\int_{0}^{1} \der t_{s-1} \,
\prod_{\ell \in T}
\left( t_{n' ( \ell ) } \ldots
t_{n( \ell )-1 } \right) =\prod_{k=1}^{s-1} \left(
\int_{0}^{1} \der t_{k} \, t_{k}^{b_{k}-1} \right)  =
\prod_{k=1}^{s-1} \frac{1}{b_{k}} \; , \label{a1.43}\ee
which is well defined, since $b_{k}\ge 1$. 
Moreover we can write:
\be   \sum_{\substack{X_2,\ldots, X_{s-1}\\
{\rm compatible}\ {\rm with}\  T}}=
\sum_{\substack{X_2\\ X_1\ {\rm fixed}}}^*\ \
\sum_{\substack{ X_{3} \\  X_{1},X_{2}\;{\rm fixed}} }^* \ldots
\sum_{\substack{ X_{s-1} \\  X_{1},\ldots,X_{s-2}\;{\rm fixed} }}^*
\; , \label{a1.44}\ee
where the $*$'s over the sums remind that all the summations are subject to the constraint
that the subsets $X_1,X_2,\ldots,X_s$ must be compatible with the structure of $T$. Now, 
the number of terms in the sum over $X_k$, once that
$T$ and the sets $X_1,\ldots, X_{k-1}$ are fixed, is exactly $b_{k-1}$, so that 
\be   \sum_{\substack{X_2,\ldots, X_{s-1}\\
{\rm compatible}\ {\rm with}\  T}}1=b_{1} \ldots b_{s-2} \; , \label{a1.45}\ee
and, recalling that $b_{s-1}=1$,
\be   \sum_{\substack{X_2,\ldots, X_{s-1}\\
{\rm compatible}\ {\rm with}\  T}}
\int_{0}^{1} \der t_{1} \ldots
\int_{0}^{1} \der t_{s-1} \,
\prod_{\ell \in T}
\left( t_{n' ( \ell ) } \ldots
t_{n( \ell )-1 } \right) =
\prod_{k=1}^{s-2} \frac{b_{k}}{b_{k} } \; , \label{a1.46}\ee
yielding to $\int d P_T(\tt)=1$. The positivity and $\s$--addivity of 
$d P_T(\tt)$ is obvious by definition. 

We are left with proving that, for any given sequence of subsets $X_1\subset X_2\subset 
\cdots\subset X_s$ compatible with $T$, we can find unit vectors 
${\bf u}_j\in\RRR^s$ such that $t_{j,j'}={\bf u}_j\cdot{\bf u}_{j'}$. With no losso of generality,
we can assume that $X_1=\{ 1\}$, $X_2=\{ 1,2\}$, $\ldots$, 
$X_{s-1}=\{ 1,\ldots, s-1\}$. We introduce a family of unit vectors in $\RRR^s$ defined as follows: 
\be \begin{cases}
\uu_{1} = \vv_{1} \; , & \\
\uu_{j} = t_{j-1} \uu_{j-1} + \vv_{j}
\sqrt{1-t_{j-1}^2} \; , & j=2,\ldots,s \; ,  \end{cases} \label{a1.47}\ee
where $\{ {\bf v}_i\}_{i=1}^s$ is an orthonormal basis. From this equation, as desired:
\be \uu_{j} \cdot \uu_{j'} = t_{j} \ldots t_{j'-1}
\;. \label{a1.49}\ee
This concludes the proof of the determinant formula.\qed

\section{Symmetry properties}\label{app2}
\setcounter{equation}{0}
\renewcommand{\theequation}{\ref{app2}.\arabic{equation}}

In this Appendix we prove Lemma \ref{sec6}.2. The key remark underlying the nice properties
stated in the Lemma is that both the Gaussian integration $P_M(d\psi)$ and the
interaction $\VV(\psi)$ are invariant under the action of a number
of remarkable symmetry transformations, which are also preserved by the multiscale integration. 
In the following, we denote by $\s_1,\s_2,\s_3$ the standard Pauli matrices:
\be \s_1=\begin{pmatrix} 0 & 1\\ 1& 0\end{pmatrix}\;,\qquad 
\s_2=\begin{pmatrix} 0 & -i\\ i& 0\end{pmatrix}\;,\qquad 
\s_1=\begin{pmatrix} 1 & 0\\ 0& -1\end{pmatrix}\;.\label{app.pauli}\ee
Moreover, in order to avoid confusion with the Pauli matrices, we shall use the symbol $\t$ as spin
index rather than $\s$. As a preliminary result, let us start by listing all the transformation 
properties under which our theory is invariant.

\begin{lemma}\label{lem2.4}
For any choice of $M,\b,L$, the fermionic Gaussian
integration $P(d\psi)$ and the interaction $\VV(\psi)$
are separately invariant under the following transformations:
\begin{enumerate}
\item[(1)] \underline{Spin flip}:
$\hat\psi^{\varepsilon}_{\kk,\t}\leftrightarrow
\hat\psi^{\varepsilon}_{\kk,-\t}$;
\item[(2)] \underline{Global $U(1)$}:
$\hat\psi^{\varepsilon}_{\kk,\t}\rightarrow e^{ i
\varepsilon \a_{\t}}\hat\psi^{\varepsilon}_{\kk,\t}$, with
$\a_{\t}\in \RRR/2\p\ZZZ$ independent of $\kk$;
\item[(3)] \underline{Spin $SO(2)$}:
$\Biggl(\begin{matrix}
\hat\psi^{\varepsilon}_{\kk,\uparrow,\cdot} \\
\hat\psi^{\varepsilon}_{\kk,\downarrow,\cdot} \end{matrix}\Biggr)
\rightarrow e^{-i\th\s_2}\Biggl(\begin{matrix}
\hat\psi^{\varepsilon}_{\kk,\uparrow,\cdot} \\
\hat\psi^{\varepsilon}_{\kk,\downarrow,\cdot} \end{matrix}\Biggr)$, with
$\th\in\! \RRR/2\p\ZZZ$ independent of $\kk$;
\item[(4)] \underline{Discrete rotations}:
$\hat \psi_{\kk,\t}^-\to e^{-i\vec k(\vec \d_3
-\vec \d_1)\frac{\s_3}2}\hat \psi_{T\kk,\t}^-$ and
$\hat\psi_{\kk,\t}^+\to\hat\psi_{T\kk,\t}^+e^{i\vec k(\vec \d_3
-\vec \d_1)\frac{\s_3}2}$, with $T\kk=(k_0,e^{-i\frac{2\p}{3}\s_2}\vec k)$;
\item[(5)] \underline{Complex conjugation}:
$\hat\psi^{\varepsilon}_{\kk,\t}\rightarrow
\hat\psi^{\varepsilon}_{-\kk,\t}$ and
$c\rightarrow c^{*}$, where $c$ is generic
constant appearing in $P(d\Psi)$ or in $\VV(\Psi)$;
\item[(6.a)] \underline{Horizontal reflections}:
$\hat\psi^{-}_{\kk,\t}\to \s_1\hat\psi^-_{R_h\kk,\t}$ and
$\hat\psi^{+}_{\kk,\t}\to \hat\psi^+_{R_h\kk,\t}\s_1$,
with $R_h\kk=(k_0,-k_1,k_2)$;
\item[(6.b)] \underline{Vertical reflections}:
$\hat\psi^{\varepsilon}_{\kk,\t}\rightarrow
\hat\psi^{\varepsilon}_{R_v\kk,\t}$, 
with $R_v\kk=(k_0,k_1,-k_2)$;
\item[(7)] \underline{Particle-hole}:
$\hat\psi^{-}_{\kk,\t}\to i\hat\psi^{+,T}_{P\kk,\t}$,
$\hat\psi^{+}_{\kk,\t}\to i\hat\psi^{-,T}_{P\kk,\t}$,
with $P\kk=(k_0,-k_1,-k_2)$;
\item[(8)] \underline{Inversion}:
$\hat\psi^{-}_{\kk,\t}\to -i\s_3\hat\psi^-_{I\kk,\t}$,
$\hat\psi^{+}_{\kk,\t}\to -i\hat\psi^+_{I\kk,\t}\s_3$,
with $I\kk=(-k_0,k_1,k_2)$.
\end{enumerate}
\end{lemma}
\noindent{\bf Proof of Lemma \ref{lem2.4}.} 
Let us first recall the definitions of $P_M(d\psi)$ and $\VV(\psi)$:
\bea&& P_M(d\psi) = \frac1{\NN} \prod_{\kk\in\BBB_{\b,L}^{(M)}}^{\t=\uparrow\downarrow}
d\hat\psi_{\kk,\t,1}^+d\hat\psi_{\kk,\t,1}^-d\hat\psi_{\kk,\t,2}^+
d\hat\psi_{\kk,\t,2}^-\, e^{-\frac1{\b|\L|}\hat \psi^+_{\kk,\t}\hat g^{-1}_\kk\hat \psi^-_{\kk,\t}}\;,
\nn\\
&& \VV(\psi)
=\frac{U}{(\b |\L|)^3}\sum_{\kk,\kk',\pp}^{\a=\pm}\, \big(\hat \psi^+_{\kk+\pp,\uparrow}n_\a
\hat\psi^-_{\kk,\uparrow}\big)\big(\hat\psi^+_{\kk'-\pp,\downarrow}n_\a
\hat\psi^-_{\kk',\downarrow}\big)\;,\nn\eea
where $\NN$ is the normalization constant in Eq.(\ref{2.3}) and $n_\pm=(1\pm\s_3)/2$. 
Given these definitions and recalling the definition of $\hat g^{-1}_\kk$, 
we see that the invariance of $P_M(d\psi)$ and $\VV(\psi)$ are 
equivalent to the invariance of the following combinations:
\bea (*)&:=&\sum_{\kk,\t}\hat\psi^+_{\kk,\t}
\begin{pmatrix} i k_0 & v_0\O^*(\vec k)\\ v_0\O(\vec k)& ik_0\end{pmatrix}\hat\psi^-_{\kk,\t}\;,\label{a1.000}\\
(**)&:=&\sum_{\kk,\kk',\pp}^{\a=\pm}\, \big(\hat \psi^+_{\kk+\pp,\uparrow}n_\a
\hat\psi^-_{\kk,\uparrow}\big)\big(\hat\psi^+_{\kk'-\pp,\downarrow}n_\a
\hat\psi^-_{\kk',\downarrow}\big)\;.\label{a1.001}\eea
Now, the invariance of $(*)$ and $(**)$ under symmetries (1) and (2) is completely apparent. 
Let us check one by one the invariance under the other symmetries.\\

\noindent{\it Symmetry (3).} Note that $(*)$ and $(**)$ can be rewritten as
\bea &&(*)=\sum_{\kk,\r,\r'}(\hat g^{-1}_\kk)_{\r\r'}\sum_\t \hat\psi^+_{\kk,\t,\r}\hat \psi^-_{\kk',\t,\r'}\;,
\label{a1.00}
\\
&& (**)=\frac12\sum_{\substack{\kk,\kk',\pp\\ \a,\r}}(n_\a)_{\r\r}\sum_{\t,\t'}
\big(\hat\psi^+_{\kk+\pp,\t,\r} \hat\psi^-_{\kk,\t,\r}\big)\big(\hat\psi^+_{\kk'-\pp,\t',\r} 
\hat\psi^-_{\kk',\t',\r}\big)\label{a1.01}\eea
Invariance of these expression under symmetry (3) follows from the invariance of the combination
$\sum_{\t}\hat\psi^+_{\kk,\t,\r}\hat \psi^-_{\kk',\t,\r'}$: in fact, 
denoting by $R^\th$ the matrix $e^{-i\th\s_2}$, we see that under (3)
\be \sum_{\t}\hat\psi^+_{\kk,\t,\r}\hat \psi^-_{\kk',\t,\r'}\to 
\sum_{\t,\t_1,\t_2}\hat\psi^+_{\kk,\t_1,\r}R^{\th,T}_{\t_1,\t}R^{\th}_{\t,\t_2}\hat \psi^-_{\kk',\t_2,\r'}\;.\label{a1.2F}\ee
which is invariant, simply by the fact that $R^\th$ is orthogonal.\\

\noindent{\it Symmetry (4).} Under the action of (4), 
\be (*) \to \sum_{\kk,\t}\hat\psi^+_{T\kk,\t}e^{i\vec k(\vec \d_3
-\vec \d_1)\frac{\s_3}2}\begin{pmatrix} i k_0 & v_0\O^*(\vec k)\\ v_0\O(\vec k)& ik_0\end{pmatrix}e^{-i\vec k(\vec \d_3
-\vec \d_1)\frac{\s_3}2}\hat\psi^-_{T\kk,\t}\;,\label{a1.3}\ee
with $T\kk=(k_0,R^{\frac{2\p}3}\vec k)$. The r.h.s. of Eq.(\ref{a1.3}) can be rewritten as
\be \sum_{\kk,\t}\hat\psi^+_{T\kk,\t}\begin{pmatrix} i k_0 & v_0\O^*(\vec k)e^{i\vec k(\vec \d_3
-\vec \d_1)}\\ v_0\O(\vec k)e^{-i\vec k(\vec \d_3
-\vec \d_1)}& ik_0\end{pmatrix}\hat\psi^-_{T\kk,\t}\;,\label{a1.4}\ee
which is the same as $(*)$, as it follows by the remark that $\O(\vec k)e^{-i\vec k(\vec \d_3
-\vec \d_1)}=\O(R^{\frac{2\p}{3}}\vec k)$. Regarding the interaction term, note that 
\be \big[e^{i(\vec k+\vec p)(\vec\d_3-\vec\d_1)\frac{\s_3}2}
n_\a e^{-i\vec k(\vec\d_3-\vec\d_1)\frac{\s_3}2}\big]=
e^{i\a\frac{\vec p}2(\vec\d_3-\vec\d_1)}n_\a\;,\label{A1.3}\ee
which immediately shows that $(**)$ is invariant under (4). \\

\noindent{\it Symmetry (5).}
The term $(*)$ is changed under (5) as:
\be (*)\to \sum_{\kk,\t}\hat\psi^+_{-\kk,\t}
\begin{pmatrix} -i k_0 & v_0\O(\vec k)\\ v_0\O^*(\vec k)& -ik_0\end{pmatrix}\hat\psi^-_{-\kk,\t}\;,
\label{A1.005}\ee
which is the same as $(*)$ (simply because $\O^*(\vec k)=\O(-\vec k)$). Similarly, the term $(**)$ is changed under (5) as:
\be (**)\to \sum_{\kk,\kk',\pp}^{\a=\pm}\, \big(\hat \psi^+_{-\kk-\pp,\uparrow}n_\a
\hat\psi^-_{-\kk,\uparrow}\big)\big(\hat\psi^+_{-\kk'+\pp,\downarrow}n_\a
\hat\psi^-_{-\kk',\downarrow}\big)\;,
\label{A1.006}\ee
which is the same as $(**)$.\\

\noindent{\it Symmetry (6.a).}
The term $(*)$ is changed under (6.a) as:
\bea (*)&\to& \sum_{\kk,\t}\hat\psi^+_{R_h\kk,\t}\s_1
\begin{pmatrix} i k_0 & v_0\O^*(\vec k)\\ v_0\O(\vec k)& ik_0\end{pmatrix}\s_1\hat\psi^-_{R_h\kk,\t}
=\nn\\
&=&\sum_{\kk,\t}\hat\psi^+_{R_h\kk,\t}
\begin{pmatrix} i k_0 & v_0\O(\vec k)\\ v_0\O^*(\vec k)& ik_0\end{pmatrix}\hat\psi^-_{R_h\kk,\t}
\;.\label{A1.007}\eea
Using the fact that $\O^*(\vec k)=\O((-k_1,k_2))$, we see that this term is
invariant under (6.a). Regarding the interaction term, if we note that $\s_1 n_\a\s_1=n_{-\a}$,
we immediately see that it is invariant under (6.a).\\

\noindent{\it Symmetry (6.b).}
Invariance of the term $(*)$ follows from the remark that $\O^*(\vec k)=\O((k_1,-k_2))$; 
invariance of $(**)$ is trivial.\\

\noindent{\it Symmetry (7).}
The term $(*)$ is changed under (7) as:
\be (*)\to \sum_{\kk,\t}\hat\psi^+_{P\kk,\t}
\begin{pmatrix} i k_0 & v_0\O^*(\vec k)\\ v_0\O(\vec k)& ik_0\end{pmatrix}^{\!T}\hat\psi^-_{P\kk,\t}\;,
\label{a1.005}\ee
which is the same as $(*)$ (simply because $\O^*(\vec k)=\O(-\vec k)$). Similarly, using the fact 
that $n_\a^T=n_\a$, the term $(**)$ is changed under (7) as:
\be (**)\to \sum_{\kk,\kk',\pp}^{\a=\pm}\, \big(\hat \psi^+_{P\kk,\uparrow}n_\a
\hat\psi^-_{P(\kk+\pp),\uparrow}\big)\big(\hat\psi^+_{P\kk',\downarrow}n_\a
\hat\psi^-_{P(\kk'-\pp),\downarrow}\big)\;,
\label{a1.006}\ee
which is the same as $(**)$.\\

\noindent{\it Symmetry (8).}
The term $(*)$ is changed under (8) as:
\bea (*)&\to& -\sum_{\kk,\t}\hat\psi^+_{I\kk,\t}\s_3
\begin{pmatrix} i k_0 & v_0\O^*(\vec k)\\ v_0\O(\vec k)& ik_0\end{pmatrix}\s_3\hat\psi^-_{I\kk,\t}=\nn
\\
&=& \sum_{\kk,\t}\hat\psi^+_{I\kk,\t}\s_3
\begin{pmatrix} -i k_0 & v_0\O^*(\vec k)\\ v_0\O(\vec k)& -ik_0\end{pmatrix}\s_3\hat\psi^-_{I\kk,\t}\;,
\label{a1.008}\eea
which is the same as $(*)$. Moreover, note that $\s_3 n_\a\s_3=n_\a$, so that the term $(**)$ is
obviously invariant under (8). This concludes the proof of Lemma \ref{lem2.4}.\qed
\vskip.4truecm
Now, we are ready to prove Lemma \ref{sec6}.2.\\

{\bf Proof of Lemma \ref{sec6}.2.} The key remark is that, in addition to the fact that $P_M(d\psi)$
is invariant under the transformations (1)--(8), the ultraviolet and infrared integrations
$P(d\psi^{(u.v.)})$ and $P(d\psi^{(i.r.)})$ are separately invariant under the analogous 
transformations of the fields $\psi^{(u.v.)}$ and $\psi^{(i.r.)}$. Therefore, the effective potential 
$\VV_0(\psi^{(i.r.)})$ is also invariant under the same transformations. Let us restrict our attention 
to the quadratic contribution to $\VV_0$, i.e., 
\be \frac1{\b|\L|}\sum_{\kk,\t}\hat \psi^{(i.r.)+}_{\kk,\t}\hat W_{2}(\kk)\hat \psi^{(i.r.)-}_{\kk,\t}\;,
\label{a1.91}\ee
which, as we said, must be invariant under the symmetries (1)--(8) listed in Lemma \ref{lem2.4}.
The very fact that we can write the quadratic term in the form Eq.(\ref{a1.91}), with 
$\hat W_{2}(\kk)$ independent of the spin index $\t$, is a consequence of symmetries (1)--(3). 
It is straightforward to check that Eq.(\ref{sym}) follows from the invariance of Eq.(\ref{a1.91})
under (4)--(8). 

So, we are left with proving Eq.(\ref{sym2}). Let us start by showing that 
$\hat W(\pp_F^\o)=0$. Writing $\hat W(\pp_F^\o)=a^\o_0I+
a^\o_1\s_1+a^\o_2\s_2+a^\o_3\s_3$ and using the fact that, 
by Eq.(\ref{sym}), $\hat W(\pp_F^\o)
=\s_1\hat W(\pp_F^\o)\s_1=-\s_3\hat W(\pp_F^\o)\s_3$, we immediately see that $a^\o_0=a^\o_2=
a^\o_3=0$. Moreover, using the fact that (still by Eq.(\ref{sym}))
$\hat W(\pp_F^\o)=e^{i\o\frac{2\p}{3}\frac{\s_3}2}\hat W(\pp_F^\o)e^{-i\o\frac{2\p}{3}\frac{\s_3}2}$,
we get $a^\o_1\s_1=a^\o_1(\cos(2\p/3)\s_1-\o\sin(2\p/3)\s_2)$, which implies $a^\o_1=0$.

Let us now look at $A^\o_\m:=\dpr_{k_\m}\hat W(\pp_F^\o)$. Writing $A^\o_0=b_0^\o I+
b^\o_1\s_1+b^\o_2\s_2+b^\o_3\s_3$ and using that, by Eq.(\ref{sym}), 
$A^\o_0=-(A^{-\o}_0)^*=A^{-\o}_0=\s_1A^\o_0\s_1=\s_3A^\o_0\s_3$, we see that 
$A^\o_0=-iz_0I$, with $z_0$ real and independent of $\o$. In a similar way, writing 
$A^\o_1=c_0^\o I+ c^\o_1\s_1+c^\o_2\s_2+c^\o_3\s_3$ and $A^\o_2=d_0^\o I+
d^\o_1\s_1+d^\o_2\s_2+d^\o_3\s_3$, using the fact that 
$A^\o_1=-(A_1^{-\o})^*=A^{-\o}_1=-\s_1A^\o_1\s_1=-(A^{-\o}_1)^T=-\s_3A^\o_1\s_3$ 
and $A^\o_2=-(A_2^{-\o})^*=-A^{-\o}_2=\s_1A^\o_2\s_1=-(A^{-\o}_2)^T=-\s_3A^\o_2\s_3$,
we see that $A^\o_1=c_2\s_2$ and $A^\o_2=\o d_1\s_1$, with $c_2$ and $d_1$ real and 
independent of $\o$. Finally, using the fact that, again by Eq.(\ref{sym}), 
$A^\o_i=\sum_{j=1}^2 e^{i\o\frac{2\p}{3}\s_3} (R^{\frac{2\p}3})_{ij}A^\o_j$, we get $c_2=d_1
=:-\d_0$ that, if combined with our previous findings, implies Eq.(\ref{sym2}). This concludes the 
proof of Lemma \ref{sec6}.2. \qed

\vskip.2truecm {\bf Acknowledgements.} A particular thank goes to 
Vieri Mastropietro: we started the research project on graphene 
together and the material reviewed in these lecture notes is based on 
joint work with him. I also gratefully acknowledge financial support 
from the ERC Starting Grant CoMBoS-239694.


\thebibliography{0}
\bibitem{footnote} Of course, there is some arbitrariness 
in the definition of $\hat a^\pm_{\vec k,\s}, 
\hat b^\pm_{\vec k,\s}$: we could change the two sets of operators by multiplying them by 
two $\vec k$-dependent
phase factors, $e^{\mp i\vec k\vec A}, e^{\mp i\vec k\vec B}$, with $\vec A,\vec B$ two arbitrary 
constant vectors, and get a completely equivalent 
theory in momentum space.
The freedom in the choice of these phase factors corresponds to the freedom in the choice
of the origins of the two sublattices $\L_A,\L_B$; this symmetry is sometimes referred to as 
Berry-gauge invariance and it can be thought as a local gauge symmetry in momentum space. 
Note that, by changing the Berry phase, the boundary conditions on $\hat a^\pm_{\vec k,\s}, 
\hat b^\pm_{\vec k,\s}$ at the boundaries of the first Brillouin zone change; our explicit 
choice of the Berry phase has the ``advantage" of making $\hat a^\pm_{\vec k,\s}, 
\hat b^\pm_{\vec k,\s}$ periodic over $\L^*$ rather than quasi-periodic.
\endthebibliography
\end{document}